\documentclass[aps,prx,onecolumn,english,balance,superscriptaddress,floats,showpacs,letter,twocolumn]{revtex4-2}
\usepackage[T1]{fontenc}
\usepackage[latin9]{inputenc}
\usepackage{amsmath}
\usepackage{amssymb}
\usepackage{stmaryrd}
\usepackage{graphicx}
\usepackage{esint}
\usepackage{subfigure}
\usepackage{multirow}
\usepackage{xcolor}
\usepackage{ulem}
\usepackage{accents}
\usepackage{soul}

\makeatletter

\newcommand{\beq}{\begin{equation}}
	\newcommand{\eeq}{\end{equation}}
\newcommand{\bea}{\begin{eqnarray}}
	\newcommand{\eea}{\end{eqnarray}}
\newcommand{\bwt}{\begin{widetext}}
	\newcommand{\ewt}{\end{widetext}}
\@ifundefined{textcolor}{}
{%
	\definecolor{BLACK}{gray}{0}
	\definecolor{WHITE}{gray}{1}
	\definecolor{RED}{rgb}{1,0,0}
	\definecolor{GREEN}{rgb}{0,1,0}
	\definecolor{BLUE}{rgb}{0,0,1}
	\definecolor{CYAN}{cmyk}{1,0,0,0}
	\definecolor{MAGENTA}{cmyk}{0,1,0,0}
	\definecolor{YELLOW}{cmyk}{0,0,1,0}
}

\newcommand{\mG}{\mathcal{G}}
\newcommand{\mK}{\mathcal{K}}
\newcommand{\mM}{\mathcal{M}}
\newcommand{\mO}{\mathcal{O}}
\newcommand{\mR}{\mathcal{R}}
\newcommand{\mT}{\mathcal{T}}

\newcommand{\fvec}[1]{\boldsymbol{#1}}

\newcommand{\half}{\frac{1}{2}}
\newcommand{\pd}{\partial}
\newcommand{\rmd}{{\rm d}}

\makeatletter
\newcommand{\doublewidetilde}[1]{{%
  \mathpalette\double@widetilde{#1}%
}}
\newcommand{\double@widetilde}[2]{%
  \sbox\z@{$\m@th#1\widetilde{#2}$}%
  \ht\z@=.9\ht\z@
  \widetilde{\box\z@}%
}
\makeatother

\DeclareMathOperator{\Tr}{Tr}

\usepackage{babel}

\makeatother

\usepackage{babel}

\begin{document}

	\title{Analytical solution for the relaxed atomic configuration of twisted bilayer graphene including heterostrain}

	\author{Jian Kang}
	\email{kangjian@shanghaitech.edu.cn}
	\affiliation{State Key Laboratory of Quantum Functional Materials, School of Physical Science and Technology, ShanghaiTech University, Shanghai 201210, China}
    
	\author{Oskar Vafek}
	\email{vafek@magnet.fsu.edu}
	\affiliation{National High Magnetic Field Laboratory, Tallahassee, Florida, 32310, USA}
	\affiliation{Department of Physics, Florida State University, Tallahassee, Florida 32306, USA}
	
	\begin{abstract}
    Continuum atomic relaxation models for twisted bilayer graphene involve minimization of the sum of intralayer elastic energy and interlayer adhesion energy. The elastic energy favors a rigid twist i.e. no distortion in the twisted honeycomb lattices, while the adhesion energy favors Bernal stacking and breaking the relaxation into triangular AB and BA stacked domains.
	We compare the results of two relaxation models with the published Bragg interferometry data, finding good agreement with one of the models. We then provide a method for finding a highly accurate approximation to the solution of this model which holds above the twist angle of $\approx0.7^\circ$ and thus covers the first magic angle. We find closed form expressions in the absence, as well as in the presence, of external heterostrain. These expressions are not written as a Taylor series in the ratio of adhesion and elastic energy, because, as we show, the radius of convergence of such a series is too small to access the first magic angle.
	\end{abstract}
	
	\maketitle

\section{Introduction}
It became apparent fairly shortly after the discovery of correlated electron phenomena in the magic angle twisted bilayer graphene~\cite{Cao2018Insulator, Cao2018SC} that in order to explain the observed gap between the narrow and remote bands, it is necessary to model the structure of the bilayer beyond a simple rigid twist~\cite{BMModel} and to account for atomic lattice relaxation~\cite{KaxirasTBM16,KaxirasPRB18,LiangPRX,LiangStrain,KangPRX18,JKPRB23,LuskinPRB22,KoshinoPRB17,*KoshinoPRB17Erratum}. The lattice relaxation originates in the difference between the energy of Bernal stacking compared to other stackings, favoring the growth of the AB and BA stacked regions relative to the AA stacking, thus deforming the atomic configuration of the bilayer compared to the rigid twist. Such deformation results in strains within each monolayer that cost elastic energy. The relaxed configuration is understood as a result of a balance between these two effects.

\begin{figure}[t] 
	\centering
	\subfigure[\label{Fig:Relax:UnStrained}]{\includegraphics[width=0.9\columnwidth]{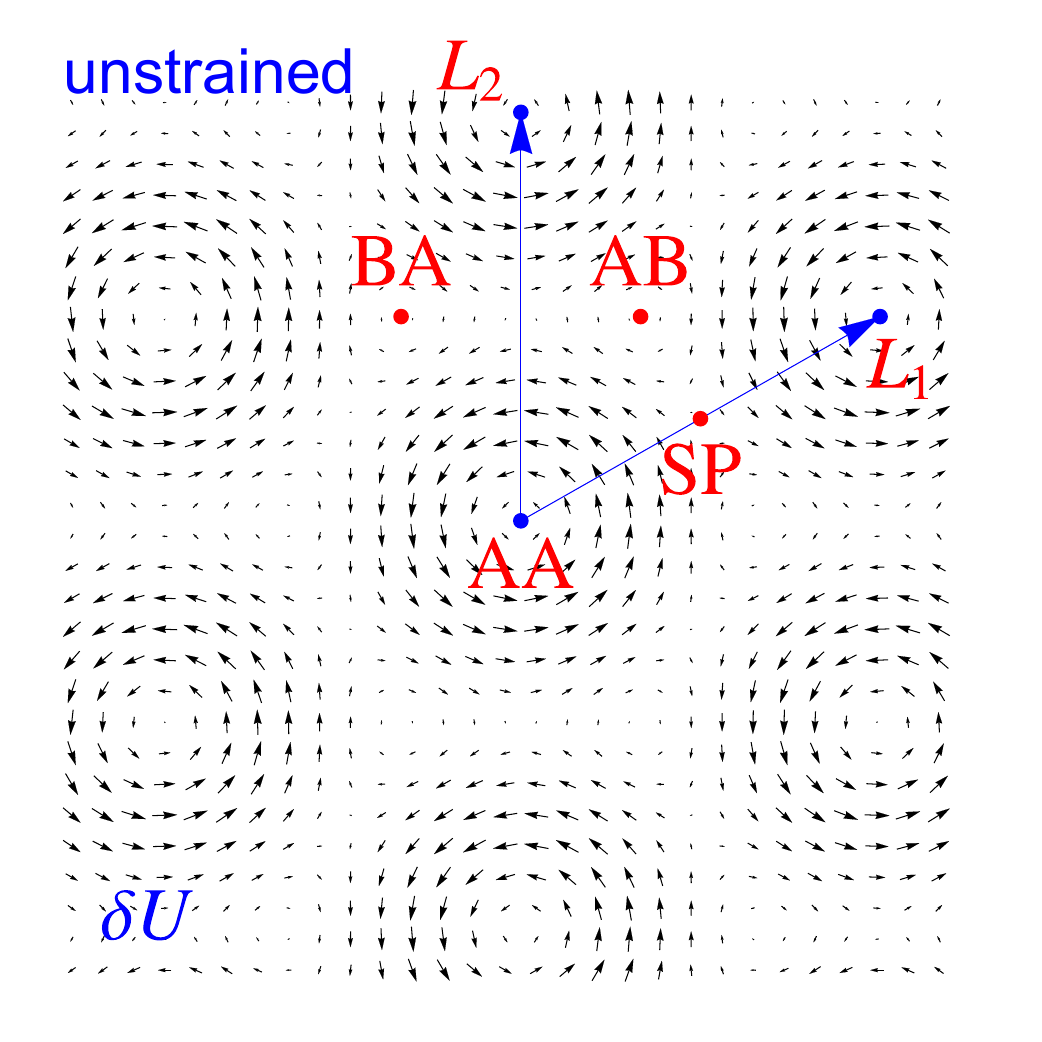}}
	\subfigure[\label{Fig:Relax:Strained}]{\includegraphics[width=0.9\columnwidth]{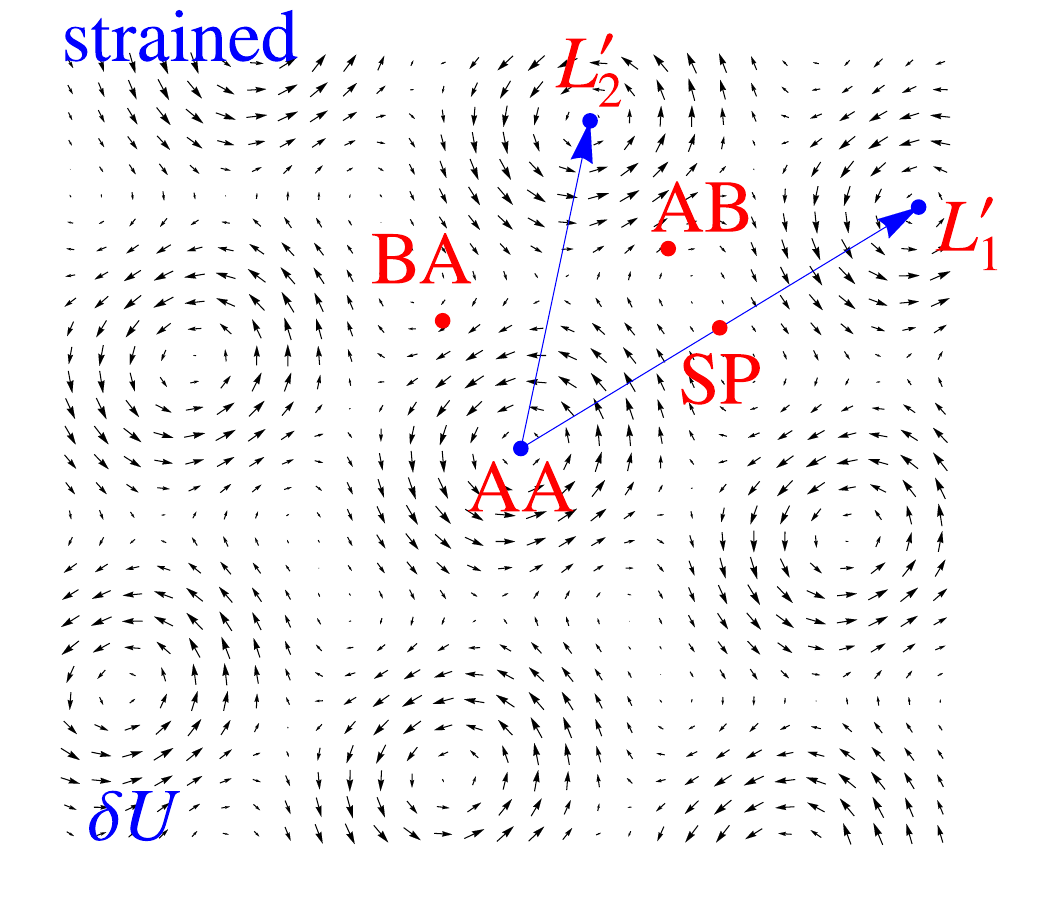}}
	\caption{Lattice relaxation for the twist angle $\theta = 1.05^{\circ}$ with (a) no external heterostrain and (b) $\epsilon_1 = 0.5\times 10^{-2}$, $\epsilon_2 = -0.8 \epsilon_1$, and $\phi = \pi/12$. The moire lattice vectors as well as various stacking points are also marked.}
	\label{Fig:Relax}
\end{figure}

More recently, it also became apparent that an unintentional external heterostrain, i.e. the layer antisymmetric component of strain, has a significant effect on the energy width of the magic angle twisted bilayer graphene narrow bands, and can qualitatively modify the nature of the correlated insulator states~\cite{NickIKSPRX,NickStrainCNPPRL}. Even for twist angles which, while small, are larger than the magic angle, the moire pattern magnifies the effect of strain~\cite{XiaoyuPNAS}. For example, at $\sim 1.38^\circ$ twist, a $\sim 0.3\%$ heterostrain can lead to $\sim 12\%$ change in the real space moire lattice vectors. We illustrate this effect at the magic angle $1.05^\circ$ in the Fig.~\ref{Fig:Relax} showing the difference between the unstrained moire unit cell lattice vectors $\fvec L_j$ and their heterostrained counterparts $\fvec L'_j$. Such a large change in atomic lattice constants would be difficult to achieve in a regular material due to the limits on its stability, but are readily observed in twisted bilayer graphene~\cite{RenardPRL18,Pasupathy19,XieNature2019,WongNature2020,BediakoNM21}. This phenomenon opens an avenue towards strain engineering the electronic properties of moire materials. For example, Ref.~\cite{XiaoyuPNAS} emphasized that even a small, $\sim0.2\%$, heterostrain can split the energetic degeneracy of the three Van Hove singularities in, say, the conduction moire band and introduce open Fermi surfaces within a finite energy window of the band spectrum, capable of accommodating more than one electron per moire unit cell. The magnetoresistance of the resulting electronic system with open Fermi surfaces can be very large and non-saturating~\cite{XiaoyuPNAS}. In contrast, in the absence of the external heterostrain, the Van Hove singularities are degenerate and the window of electron filling with open Fermi surfaces shrinks to a point.
For these and other reasons, we consider it important to understand the lattice relaxation in the presence of realistic heterostrain. 

The purpose of this paper is to provide analytical results which can be used as an input to model the electronic properties of twisted bilayer graphene, with or without heterostrain.
In order to have some confidence in the applicability of the theoretical model whose output  is the relaxation profile, we first compare the results obtained numerically with the experimentally measured relaxation and strain profiles of twisted bilayer graphene using Bragg interferometry\cite{BediakoNM21} as a function of the twist angle within the range from $0.2^\circ$ to $1.4^\circ$. We do this comparison for two theoretical models, Ref.\cite{KaxirasPRB18}
and Ref.\cite{KoshinoPRB17}, and conclude that the results of the model from Ref.\cite{KaxirasPRB18} follows the experiment much more closely. We therefore focus our attention on obtaining the analytical expressions for the relaxed configuration using the model of Ref.\cite{KaxirasPRB18}.

To this end, we introduce a dimensionless parameter
$\lambda$ that quantifies the ratio between the interlayer adhesion energy and the intralayer elastic energy at a given twist angle $\theta$ (it is precisely defined in the Eq.\ref{Eq:lambdaDefinition}). For smaller $\theta$, $\lambda$ is larger due to relative growth of the adhesion energy compared to the elastic energy and leading to a stronger lattice relaxation. At the first magic angle of $\theta\approx 1.05^\circ$, the value of $\lambda$ reaches $\approx 0.256$. Naively, one may expect that such a small value would allow finding the relaxed configuration by Taylor expanding it in powers of $\lambda$, an approach adopted recently in the Ref.\cite{ShaffiquePRL24}. Interestingly, we find that this approach encounters shortcomings just above the first magic angle. In this regard, we are able to precisely show that although the exact solution for the lattice relaxation is an analytic function of $\lambda$ near the origin, the distance to the nearest point of non-analyticity in the complex $\lambda$ plane is such that the radius of convergence of the Taylor series is only $\approx0.217$. This means that including more terms in the Taylor series does not necessarily improve the accuracy of the solution near the magic angle. Instead, we introduce a different method, which allows us to obtain a simple explicit form for the relaxation as a function of $\lambda$, which is highly accurate at, and below, the first magic angle. This approach does not suffer from the problems encountered by the series solution method. Moreover, it can be systematically improved and extended to include external heterostrain. It is sufficiently general to allow for uniaxial and/or biaxial heterostrain that can be treated on equal footing. The analytic formulas we present are accurate to at least $1\%$ heterostrain, and thus within the range of the vast majority of the experimentally examined twisted bilayer graphene devices.

The rest of the paper is organized as follows: in the next subsection we summarize the main result so that the reader can readily find it without needing to go over the details of the derivation. In the section \ref{Sec:LatticeRelax} we set up the formalism to describe twisted and strained honeycomb bilayer, review the expressions for the intralayer elastic energy and the interlayer adhesion energy, and compare the numerical solution for the relaxed configuration with the available experimental data. In the subsection \ref{Sec:AnalyticalNoStrain} we present analytical formulas for the relaxed configuration in the absence of external heterostrain, while in the subsection \ref{Sec:radius of convergence} we calculate the radius of convergence of the series expansion in $\lambda$ solution by studying its analytic properties in the complex $\lambda$ plane and finding the branch points. In the section \ref{Sec:relaxation with strain} we solve for the relaxed configuration in the presence of external heterostrain. In the section \ref{Sec:LatCorrugation} we discuss the out-of-plane corrugation. The last section contains the discussion.

\subsection{Summary of main results} 

The main result of this paper is the Eq.~\ref{Eqn:main formula} which gives a formula for the in-plane lattice relaxation of a twisted bilayer graphene subject to an external heterostrain in the Eq.\ref{Eqn:StrainMat}. The lattice distortion is defined via the Eq.~\ref{Eqn:LatDis}, the layer antisymmetric combination in Eq.~\ref{Eq:UminusDef}, and the Eq.~\ref{Eq:Udef}. The three monolayer graphene reciprocal lattice vectors $\fvec G_a$ are defined in the Eqs.~\ref{Eqn:GVectors} and \ref{Eqn:G3Vector}, and are unaffected by the strain. This is unlike the three moire reciprocal lattice vectors $\fvec g_a$ in the Eq. \ref{Eqn:gdef}. The explicit formulas for the Fourier amplitudes $\zeta_{1,2,3}$ are given in the Eqs.\ref{Eqn:Zeta1Appr}, \ref{Eqn:Zeta2Appr} and \ref{Eqn:Zeta3Appr}, while the additional strain correction is given in the Eq.~\ref{Eqn:DWAppr}, in terms of Eqs.~\ref{Eqn:uForm}, \ref{Eqn:MForm}, \ref{Eqn:DiagTAppr} and \ref{Eqn:OffTAppr}. The values of the elastic constants and the adhesion energy parameters are provided in the Table~\ref{Tab:RelaxPara}.

\section{Lattice Distortion} 
\label{Sec:LatticeRelax}

\begin{figure}[htb] 
	\centering
	\includegraphics[width=0.9\columnwidth]{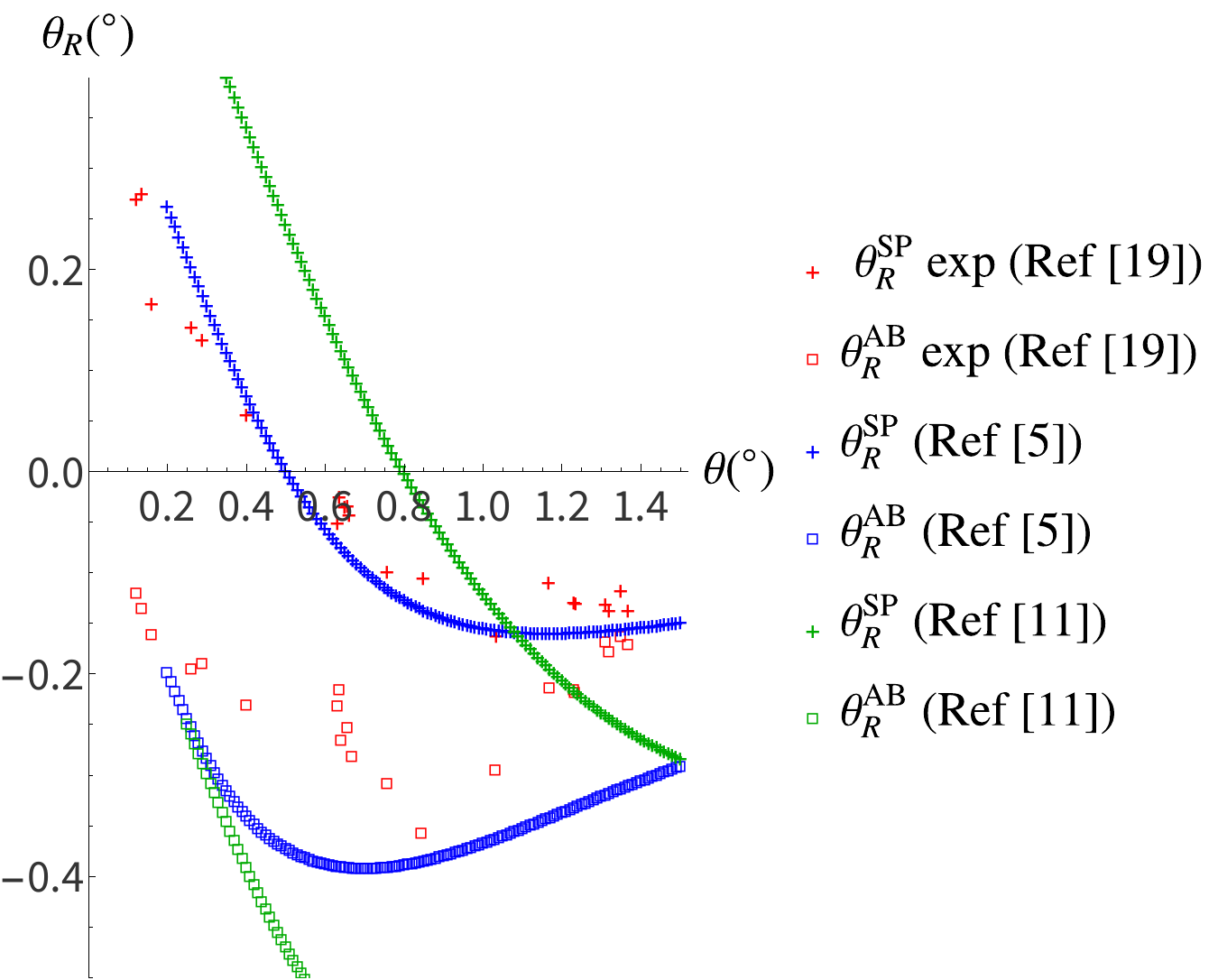}
	\caption{$\theta_R = \half \left( \partial_x \delta U_y - \partial_y \delta U_x \right)$ at two positions: SP stacking (``+'') at $\fvec x = \frac12 \fvec L_1$   and AB stacking (empty square)  at $\fvec x = \frac13(\fvec L_1 + \fvec L_2)$, when the heterostrain is absent. The three pairs of plots show the experimental measurements (red), the results of the model in Ref.~\cite{KoshinoPRB17,*KoshinoPRB17Erratum} (green)~\cite{KangLatticeRelax2025}, and the results of the model in Ref.~\cite{KaxirasPRB18} (blue)~\cite{KangLatticeRelax2025}. The experimental data is more consistent with the model in Ref.~\cite{KaxirasPRB18}. More distortion parameters are plotted and compared in Fig.~\ref{FigS:RelaxComp}.}
	\label{Fig:ThetaRSPAB}
\end{figure}

\begin{figure}[htb] 
	\centering
	\subfigure[\label{Fig:Schematic:GgMap}]{\includegraphics[width=0.9\columnwidth]{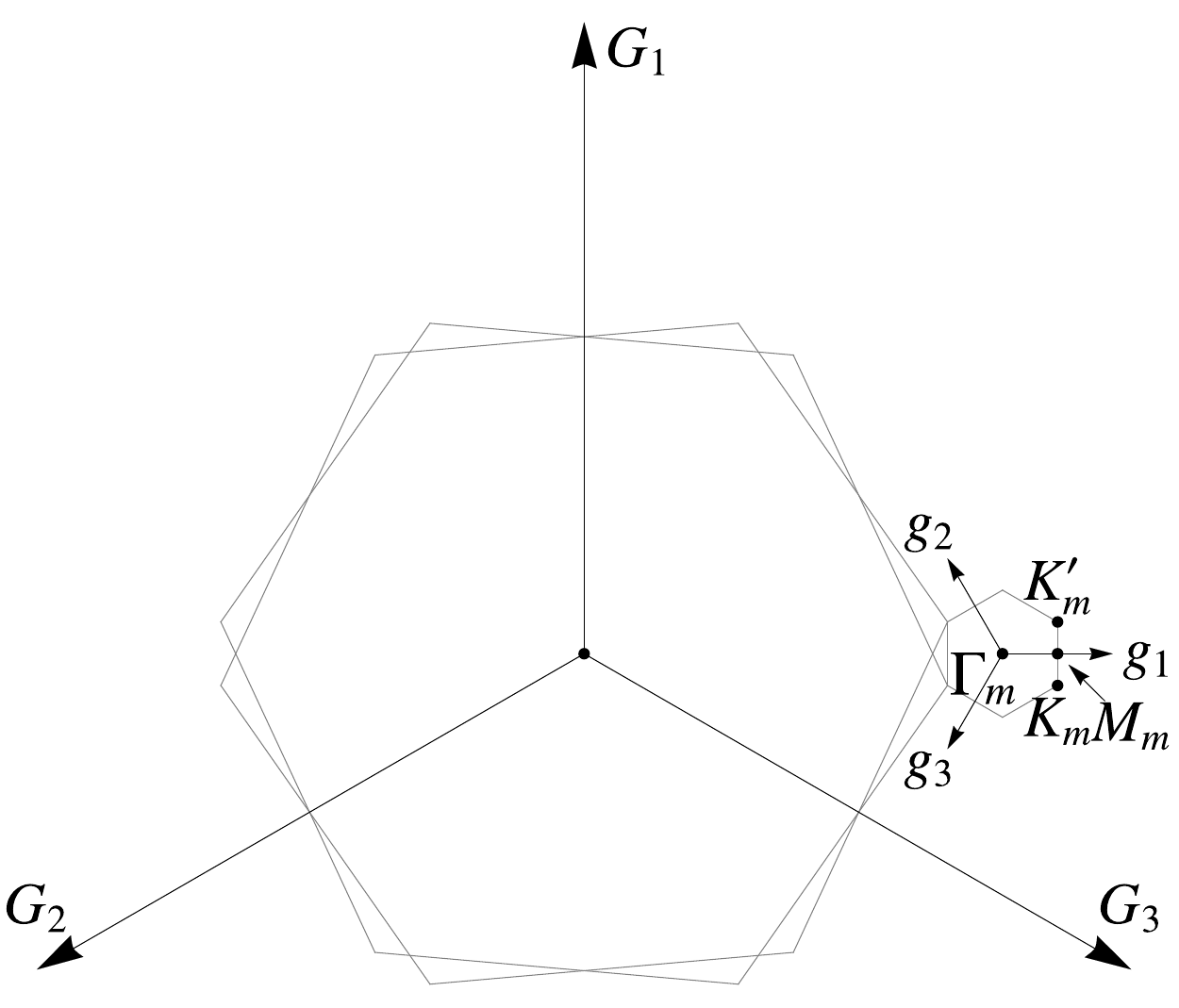}}
    \subfigure[\label{Fig:Schematic:GShell}]{\includegraphics[width=0.49\columnwidth]{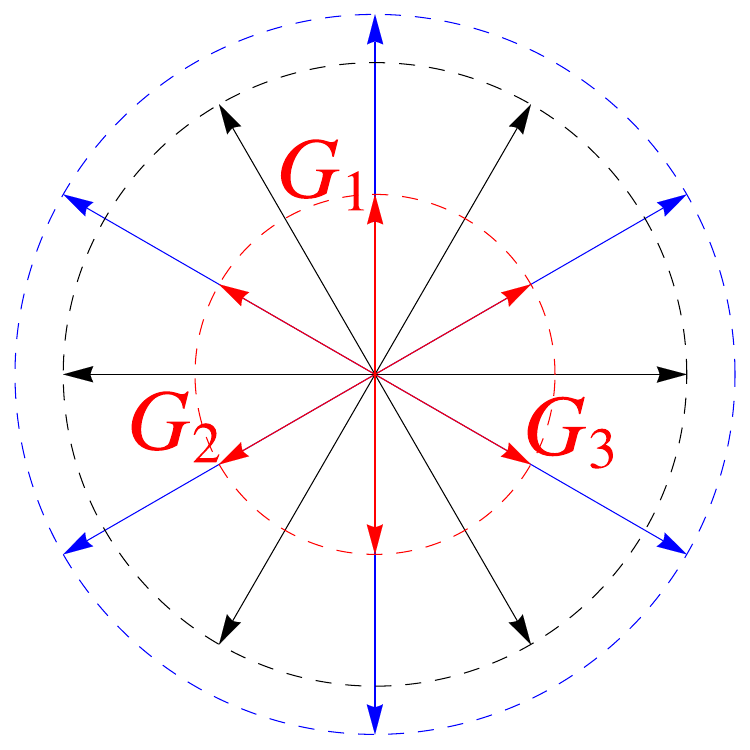}}
    \subfigure[\label{Fig:Schematic:gShell}]{\includegraphics[width=0.49\columnwidth]{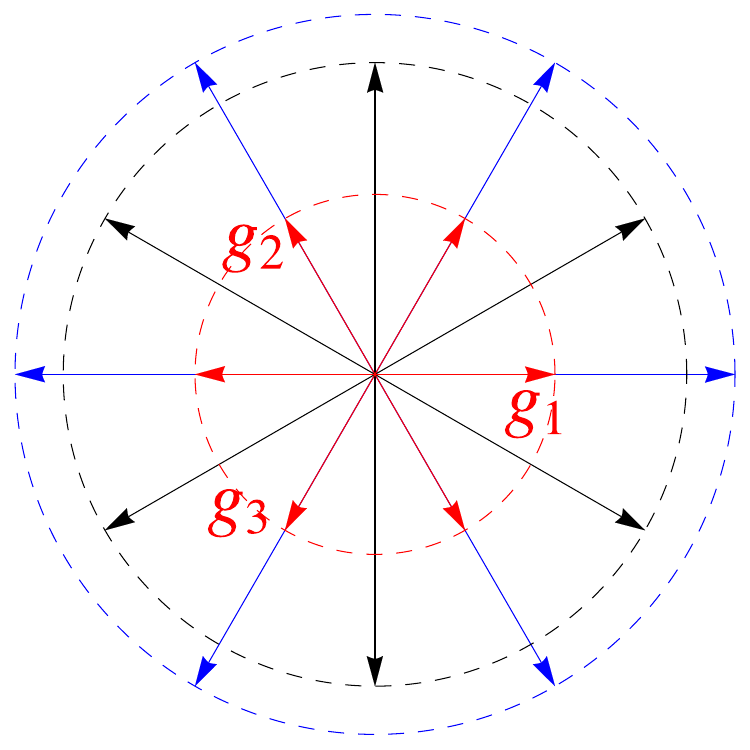}}
	\caption{(a,b) $\fvec G_i$ are the reciprocal lattice vectors of the undistorted monolayer graphene, with $\fvec G_1$, $\fvec G_2$ defined by the Eq.~\ref{Eqn:GVectors} and $\fvec G_3 = -(\fvec G_1 + \fvec G_2)$. (c) $\fvec g_i$ ($i =1$, $2$, $3$) are the reciprocal vectors of the moire lattice given by Eq.~\ref{Eqn:gdef}; in the (a) and (c) for we did not include heterostrain, only the twist. (b) the first three $\fvec G$-shells marked by red (the first shell), black (the second shell), and blue (the third shell) colored arrows. Note that in our definition the $\fvec G$ vectors are unaffected by strain, unlike the $\fvec g$ vectors. (c) the first three $\fvec g$-shells marked by the same colored arrows as for $\fvec G$-shells. Unlike in (a), in the schematics (b) and (c) we do not distinguish between different sizes of $\fvec G$ and $\fvec g$ vectors.}
	\label{Fig:Schematic}
\end{figure}
We start by focusing on two different theoretical models for lattice relaxation proposed in Refs.~\cite{KoshinoPRB17,*KoshinoPRB17Erratum} and \cite{KaxirasPRB18}, and compare the distortion obtained numerically from these two models with the experimental measurements of Ref.~\cite{BediakoNM21}.
The experimental data cover the range of twist angle $\theta$ between $0.2^{\circ}$ and $1.4^{\circ}$.
As shown in the Fig.~\ref{Fig:ThetaRSPAB}, we find a closer match for the distortion obtained using the model in Ref.~\cite{KaxirasPRB18} than in Ref.~\cite{KoshinoPRB17,*KoshinoPRB17Erratum}. Therefore, in later sections, we choose the model of Ref.~\cite{KaxirasPRB18} to obtain the lattice distortion, not just numerically but also analytically. 

Before proceeding with the details of the derivation, we introduce two unit cell vectors of monolayer graphene
\begin{align}
	\fvec a_1 = a \left( - 1/2,\ \sqrt{3}/2  \right) \quad \mbox{and} \quad \fvec a_2 = a \left( -1,\ 0 \right)  \ ,  \label{Eqn:MLGLatVecs}
\end{align} 
where $a\approx 0.246$nm is the magnitude of the monolayer graphene lattice constant. The corresponding reciprocal lattice vectors are
\begin{align}
	\fvec G_1 & = \frac{2\pi}a \left(0,\ \frac2{\sqrt{3}}\right) \ , \quad        \fvec G_2  = -\frac{2\pi}a \left( 1, \ \frac{1}{\sqrt{3}} \right) \ . \label{Eqn:GVectors}
\end{align}
For notational convenience, we also introduce 
\begin{equation}\fvec G_3 = - (\fvec G_1 + \fvec G_2) = \frac{2\pi}a \left( 1, \ - \frac1{\sqrt{3}} \right),
\label{Eqn:G3Vector}
\end{equation} shown in the Fig.~\ref{Fig:Schematic:GgMap}.


The heterostrain is described by a $2 \times 2$ symmetric matrix $S^{\epsilon}$. It can always be diagonalized by an orthogonal matrix $R(\phi)$~\cite{Pasupathy19, LiangStrain,XiaoyuPNAS}:
\begin{align}
	S^{\epsilon} =   R(-\phi) \begin{pmatrix} \epsilon_1 & 0\\ 0 & \epsilon_2 \end{pmatrix} R(\phi)  
 \label{Eqn:StrainMat}
\end{align}
where $R(\phi) = \begin{pmatrix} \cos\phi & - \sin\phi \\ \sin\phi & \cos\phi \end{pmatrix}$ is the two-dimensional rotational matrix. Note that this rotation has nothing to do with the twist. Rather, the angle $\phi$ determines the orientation of the two principal axes of the strain tensor. $\epsilon_{1,2}$  are the magnitudes of the strain along the two principal axes. In the case of uniaxial strain along the first principal axis, $\epsilon_2 = - \nu \epsilon_1$ where $\nu \approx 0.16$ is the Poisson ratio~\cite{LiangStrain}. In this work, we also allow for the biaxial strain, therefore, both $\epsilon_1$ and $\epsilon_2$ are treated as external parameters and we will not assume any specific relation between them. In the presence of experimentally relevant small twist and small heterostrain, the moire lattice vectors $\fvec L_j$, and the corresponding moire reciprocal lattice vectors $\fvec g_j$, can be safely expanded~\cite{LiangStrain} to the linear order in the twist angle $\theta$ and $\epsilon_{1,2}$ as
\begin{align}
    L_{j, \mu} & =  \big( S^{\epsilon} - i \theta \sigma^2 \big)^{-1}_{\mu\nu} a_{j, \nu}  \label{Eqn:LiVec} \\
   g_{j, \mu} & = \big( S^{\epsilon} + i \theta \sigma^2  \big)_{\mu\nu} G_{j, \nu}  \quad \mbox{for}\ j = 1,\ 2, \  3 \ . \label{Eqn:gdef}  
\end{align}
The Pauli matrix $i\sigma^2=\left(\begin{array}{cc} 0 & 1\\ -1 & 0\end{array}\right)$ and $S^{\epsilon}$ is the strain matrix defined in Eq.~\ref{Eqn:StrainMat}. The full set of the moire reciprocal lattice vectors is given by \begin{equation}
    \fvec g = n_1 \fvec g_1 + n_2 \fvec g_2,
\end{equation} with arbitrary integers $n_1$ and $n_2$. Note that $S^{\epsilon} + i \theta \sigma^2$ is a non-singular matrix unless the strain and the twist angle are fine-tuned~\cite{PacoPRL23}. Thus, we can set up a one-to-one mapping between the set of monolayer graphene reciprocal lattice vectors, $\fvec G$, and the set of $\fvec g$'s as
\begin{align}
    g_{\fvec G, \mu} = \left(S^{\epsilon}+i \theta \sigma^2   \right)_{\mu\nu} G_{\nu} \ .  \label{Eqn:gGMap}
\end{align}
We illustrate this in the Fig.~\ref{Fig:Schematic:GShell} and \ref{Fig:Schematic:gShell} in the absence of heterostrain for the first three shells.

As discussed in the Ref.~\cite{JKPRB23}, under the assumptions that the lattice distortion on both layers can be approximated by a continuous function whose variation is small on the length scale of $a$, and that the lattice distortion is independent of the graphene sublattice, we can use the Eulerian coordinates to express the distorted in-plane position $\fvec X_{j}$, on the layer $j=t$ or $b$, via a continuous function of the undistorted position $\fvec r$ as
\begin{align}
	\fvec X_j = \fvec r + \fvec U_{j}(\fvec X_{j}) \ . \label{Eqn:LatDis}
\end{align} 
Here $\fvec U_{j}$ gives the distortion of the carbon atom due to the twist, the strain, and the relaxation. We start by considering only the (dominant) in-plane lattice relaxation and address the out-of-plane corrugation in Sec.~\ref{Sec:LatCorrugation}. To proceed, we also introduce the layer symmetric and the layer antisymmetric combination of the in-plane distortions
\begin{align}
	\fvec U^+ & = \half \big( \fvec U_t + \fvec U_b  \big) \ , \quad
	\fvec U^-  = \fvec U_t - \fvec U_b \ .
    \label{Eq:UminusDef}
\end{align}

\begin{table}
    \centering
    \begin{tabular}{|c|c|c|} \hline
        $U_E$ &  $\mathcal{K}$ & $ \mathcal{G}$   \\ \hline
        Ref.~\cite{KoshinoPRB17,*KoshinoPRB17Erratum} &  $12.82$eV/\AA$^2$ &  $9.57$eV/\AA$^2$ \\ \hline
        Ref.~\cite{KaxirasPRB18} & $13.265$eV/\AA$^2$ & $9.035$eV/\AA$^2$  \\ \hline
   \end{tabular} 
\newline
\vspace*{0.1cm}
\newline   
   \begin{tabular}{|c|c|c|c|}    \hline
    $U_B$ &  $c_1$ & $c_2$ & $c_3$   \\ \hline
    Ref.~\cite{KoshinoPRB17,*KoshinoPRB17Erratum} &  $3.206$meV/\AA$^2$ & $0$ & $0$ \\ \hline
    Ref.~\cite{KaxirasPRB18} &  $0.775$meV/\AA$^2$ & $-0.071$meV/\AA$^2$ & $-0.018$meV/\AA$^2$ \\ \hline
   \end{tabular}
   \caption{The elastic parameters $\mathcal{K}$ and $\mathcal{G}$, and the parameters of the interlayer adhesion potential for two models proposed in Refs.~\cite{KoshinoPRB17} and \cite{KaxirasPRB18}.  }
   \label{Tab:RelaxPara}
\end{table}

Although the first-principles approaches have been used to obtain the lattice relaxation~\cite{Guinea19, Lucignano20, JungPRB22, Yazyev17}, the vector fields $\fvec U^{\pm}(\fvec x)$ can be conveniently obtained by minimizing the sum of the elastic energy, $U_E$, and the interlayer adhesion energy, $U_B$,
\begin{equation}
    U_{\text{tot}}=U_E+U_B,
\end{equation}
for example, as done in Refs.~\cite{KoshinoPRB17,KaxirasPRB18,Cazeaux2023}.
The elastic energy can be expressed in terms of the sum of the layer symmetric and layer anti-symmetric terms
\begin{equation}
U_E  =  U_E^+ + U_E^-.
\end{equation}
The layer anti-symmetric term is a functional of $\fvec{U}^-(\fvec{x})$ alone
\begin{eqnarray}
&&U_E^-[\fvec{U}^-] = \frac14 \int\rmd^2\fvec x\ \left[  \mathcal{K} \left( \partial_x U_{x}^-  + \partial_y U_{y}^- \right)^2 + \right. \nonumber \\
	&& \left. \mathcal{G} \left( ( \partial_x U_{x}^- - \partial_y U_{y}^- )^2 + ( \partial_x U_{y}^- + \partial_y U_{x}^- )^2  \right)   \right]  \  , \label{Eqn:ElasticEne}
\end{eqnarray}
and the layer symmetric functional $U_E^+$, which depends only on $\fvec{U}^+(\fvec{x})$
can be obtained using
\begin{equation}
         U_E^+[\fvec{U}^+] = 4 U_E^-\left[\fvec{U}^+ \right].\label{Eqn:ElasticEnePlus}
    \end{equation}
The elastic moduli $\mathcal{K}$ and $\mathcal{G}$ are listed in the Table \ref{Tab:RelaxPara}.
The interlayer adhesion term is a functional of $\fvec{U}^-(\fvec{x})$ alone. It is given by
\begin{align}
        & U_B  = \half \sum_{\fvec G} V_{\fvec G}\int\rmd^2\fvec x\   \cos(\fvec G \cdot \fvec U^-(\fvec x)),  \label{Eqn:AdhesionEne}
\end{align}
where $\sum_{\fvec G}$ sums over all the reciprocal lattice vectors of monolayer graphene.
Note that $V_{\fvec G}$ must be symmetric under various transformations, including the six-fold rotation $C_{6z}$ and the mirror reflection about the $xz$-plane, $\mathcal{R}_y$, leading to the following constraints 
\begin{align}
    & V_{\pm \fvec G_i} = c_1, \quad  V_{\pm 2\fvec G_i} = c_3, \qquad  i = 1,\ 2,\ 3\\
    & V_{\pm (\fvec G_1 - \fvec G_2)} = V_{\pm (\fvec G_2 - \fvec G_3)} = V_{\pm (\fvec G_3 - \fvec G_1)} = c_2 \ .
\end{align}
In most theoretical models, $V_{\fvec G}$ is negligible for  $|\fvec G| > 2|\fvec G_1|$ and thus can be safely ignored outside the first three shells. The interlayer adhesion potential parameters on the first shell, $c_1$, the second shell, $c_2$, and the third shell, $c_3$, for two models proposed in Ref.~\cite{KaxirasPRB18} and Ref.~\cite{KoshinoPRB17,*KoshinoPRB17Erratum} can be found in the Table~\ref{Tab:RelaxPara}.

Thus, $\fvec U^{+}(\fvec x)$ decouples from $\fvec U^-(\fvec x)$ (see e.g. Ref.~\cite{JKPRB23}). Therefore, minimizing $U_E^+$ with respect to $\fvec U^+(\fvec x)$ leads to $\partial_x U_x^+ = \partial_y U_y^+ = 0$ and $\partial_x U_y^+ + \partial_y U_x^+ = 0$. This implies that $\fvec U^+(\fvec x) = \vartheta_0 \hat z \times \fvec x + \fvec d_0$, where $\fvec d_0$ is a constant two-component vector and $\vartheta_0$ is a constant. Thus, $\fvec U^+(\fvec x)$ simply represents an overall rotation of both layers by the same angle of $\vartheta_0$ followed by an in-plane translation by the same vector $\fvec d_0$. Such a transformation does not induce any change of the elastic energy and electronic spectrum (unlike the relative twist), and thus we can safely set $\fvec U^+(\fvec x)=0$.

In the presence of a twist by a small angle $\theta$ and a small external heterostrain, $\fvec U^-(\fvec x)$ can be written as the sum of two terms
\begin{align}
    U^-_{\mu}(\fvec x) =  \big(S^{\epsilon} -i \theta \sigma^2  \big)_{\mu\nu} x_{\nu} + \delta U_{\mu}(\fvec x)   \  ,  \label{Eq:Udef}
\end{align}
where the last term $\delta \fvec U$ gives the lattice relaxation; $S^\epsilon$ is given by Eq.~\ref{Eqn:StrainMat}. For a uniform twist and an external heterostrain, we seek a moire lattice periodic $\delta \fvec U(\fvec x)$, with the lattice vectors $\fvec L_i$ defined in Eq.~\ref{Eqn:LiVec}, i.e. we impose the constraint $\delta \fvec U(\fvec x) = \delta \fvec U(\fvec x + \fvec L_i)$ when minimizing $U^{-}_E+U_B$. 
We are thus tacitly assuming that the substrate stabilizes the ``flat'' structure and prevents the buckling transition, with bending stiffness collapse and moire translation symmetry breaking, found in the theoretical study of freestanding twisted bilayer graphene in Ref.~\cite{TosattiPRB23}; our assumption is supported by the data of Ref.\cite{BediakoNM21}.

As explained in Ref.~\cite{JKPRB23}, the average of $\delta \fvec U(\fvec x)$ over the space can be absorbed by redefining the origin of $\fvec x$ and thus the average can be set to $0$ without loss of generality. Then, both the elastic and interlayer adhesion potential can be shown to be invariant under space inversion~\cite{JKPRB23}. Although the global minima may break this symmetry, numerical calculations haven't found such cases for $\theta \geq 0.2^{\circ}$ and $|\epsilon_{1,2}| < 1.5\%$. Thus, in the rest of this work, we assume that the lattice relaxation is $C_{2z}$ odd, i.e.~$\delta \fvec U(\fvec x) = - \delta \fvec U(-\fvec x)$, leading to its Fourier series 
\begin{align}
    & \delta \fvec U(\fvec x)  = \sum_{\fvec g} \fvec W(\fvec g) \sin(\fvec g \cdot \fvec x)   \label{Eqn:WgDef}
\end{align}
where $\sum_{\fvec g}$ sums over all the moire reciprocal lattice vectors and $\fvec W(\fvec g)$ are real and odd under $\fvec g \rightarrow - \fvec g$. Since Eq.~\ref{Eqn:gGMap} constructs a one-to-one mapping between the moire reciprocal lattice vectors $\fvec g$ and the reciprocal vectors $\fvec G$ of the undistorted monolayer graphene lattice, it is sometimes convenient to label the Fourier components $\fvec W(\fvec g)$ using $\fvec G$ vectors. To this end, we introduce $\fvec W_{\fvec G} \equiv \fvec W(\fvec g_{\fvec G})$, and the lattice relaxation will also be interchangeably written as
$\delta \fvec U(\fvec x)= \sum_{\fvec G} \fvec W_{\fvec G} \sin(\fvec g_{\fvec G} \cdot \fvec x)$.

Minimizing the sum of $U^-_E$ and $U_B$ leads to the self-consistent equation for $\fvec W_{\fvec G}$ (when $\fvec G \neq 0$):
\begin{align}
    & \sum_{\nu} M_{\mu\nu}(\fvec g_{\fvec G}) W_{\fvec G, \nu} = \sum_{\fvec G'} V_{\fvec G'} f_{\fvec G'}\left( \fvec g_{\fvec G},\ \{ \fvec W_{\fvec G} \} \right) G'_{\mu}, \label{Eqn:LatticeUqRelax}
\end{align}
where $M(\fvec g)$ is a $2 \times 2$ matrix that can be written as
\begin{align}
    M_{\mu\nu}(\fvec g) = \mathcal{G} |\fvec g|^2 \delta_{\mu\nu} + \mathcal{K} g_{\mu} g_{\nu} \ ,  \label{Eqn:MMatrix}
\end{align}
and $f_{\fvec G'}$ is the Fourier series amplitude defined via 
\begin{eqnarray}
\sin(\fvec G' \cdot \fvec U^-(\fvec x)) &=&
\sin(\fvec g_{\fvec G'} \cdot \fvec x + \fvec G' \cdot \delta \fvec U(\fvec x))\nonumber\\
&=& \sum_{\fvec g} f_{\fvec G'}(\fvec g,\{ \fvec W_{\fvec G} \}) \sin( \fvec g \cdot \fvec x) .
\end{eqnarray} 
The Eq.~\ref{Eqn:LatticeUqRelax} is a nonlinear equation that can be solved numerically by the iteration method~\cite{KoshinoPRB17,*KoshinoPRB17Erratum, KaxirasPRB18, OchoaPRB19, JKPRB23, LiuPRB23}. Tab.~\ref{Tab:RelaxPara} lists the parameters for two different models proposed in Ref.~\cite{KoshinoPRB17,*KoshinoPRB17Erratum} and \cite{KaxirasPRB18} that produce markedly different lattice distortions due to the significant difference in the parameter $c_1$. Based on these two models, we calculate their local twist $\theta_R = \half \left( \partial_x \delta U_y - \partial_y \delta  U_x \right)$ produced by the lattice relaxation at two positions: saddle point (SP) stacking at $\fvec x = \fvec L_1/2$, and AB stacking at $\fvec x = \frac13 (\fvec L_1 + \fvec L_2)$, see Fig.\ref{Fig:Relax:UnStrained}, with the twist angle between $0.2^{\circ}$ and $1.4^{\circ}$. The numerical results for $\theta_R$ are compared with the experimental measurements of Ref.~\cite{BediakoNM21} in the  Fig.~\ref{Fig:ThetaRSPAB}. The measured $\theta_R$ at the SP point follows the prediction of the model in Ref.~\cite{KaxirasPRB18}, but significantly deviates from the prediction of the model in Ref.~\cite{KoshinoPRB17,*KoshinoPRB17Erratum}.
Although the calculation in the Fig.~\ref{Fig:ThetaRSPAB} is done in the absence of an external strain, including the experimentally estimated $0.2\%$ heterostrain when $\theta \approx 1.3^{\circ}$ for either model does not result in a significant change.
The agreement with the data at the AB point is also significantly better for the model of Ref.~\cite{KaxirasPRB18}, although it seems to somewhat overestimate $\theta_R$ at the AB point.
This disparity may be related to the experimental measurement, which inherently averages over a particular region around a special point, such as SP, AB, and AA. Our numerical calculations shown in the Fig.\ref{Fig:ThetaRSPAB} report the lattice distortion exactly at these special points, without further averaging, and thus always produce stronger distortions than the reported measurements. Therefore, precise details of the inherent averaging in experiments would be needed to further judge the accuracy of the model in Ref.~\cite{KaxirasPRB18}.
Comparisons for other distortion parameters are presented in the appendix. Although the mentioned overestimates also occur for other distortion parameters, the overall consistency between the experiments and the model in Ref.~\cite{KaxirasPRB18} persists. In the rest of this manuscript, we will therefore focus on this lattice relaxation model.

\subsection{Analytical formula for in-plane relaxation in the absence of external strain}
\label{Sec:AnalyticalNoStrain}
\begin{figure}[htb] 
	\centering
	\subfigure[\label{Fig:LatRelaxNoStrain:Inner}]{\includegraphics[width=0.85\columnwidth]{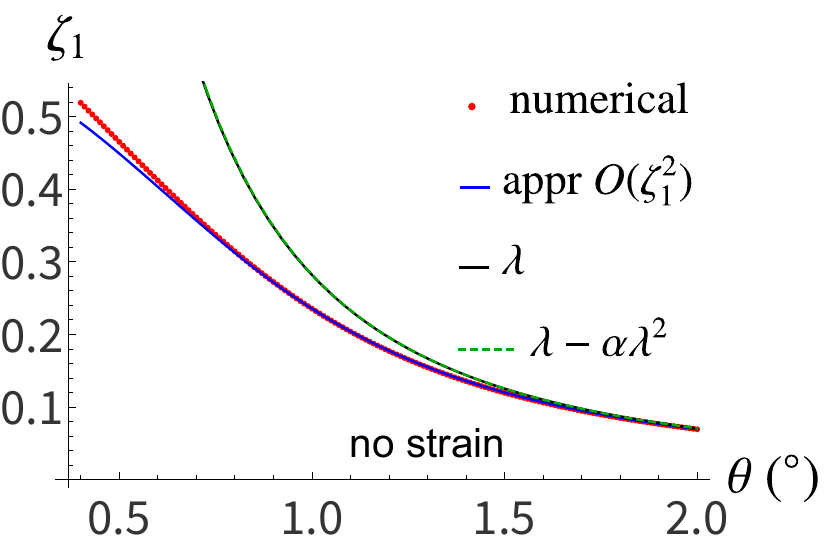}}
        \subfigure[\label{Fig:LatRelaxNoStrain:Outer}]{\includegraphics[width=0.85\columnwidth]{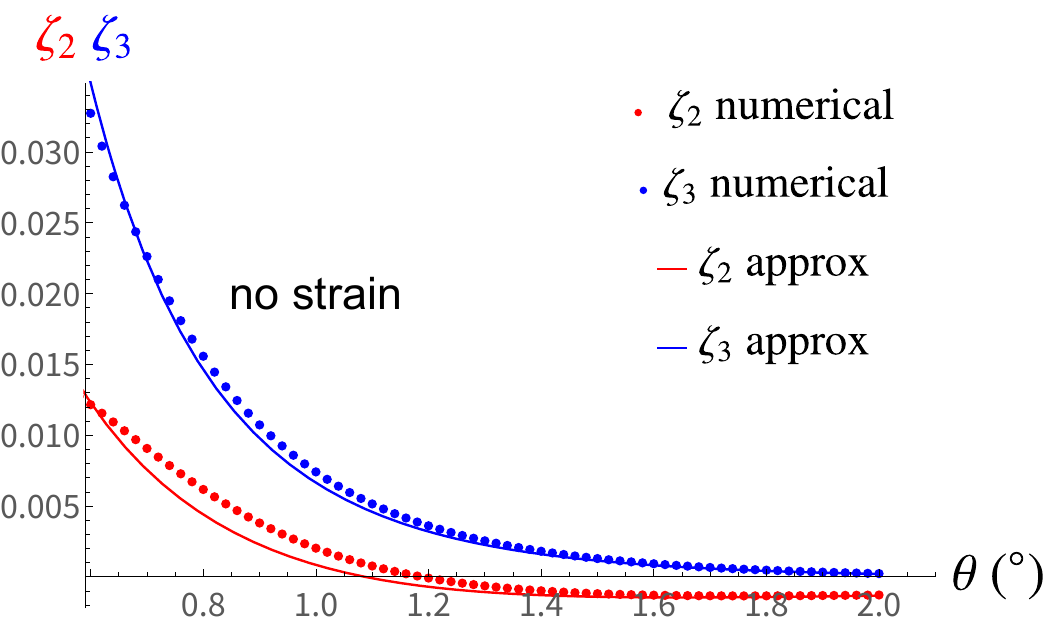}}
	\caption{Numerical solution~\cite{KangLatticeRelax2025} to the Eq.~\ref{Eqn:LatticeUqRelax}, minimizing the sum of the elastic energy $U_E$ and the interlayer adhesion energy $U_B$, is obtained first; it includes many $\fvec g$-shells and achieves convergence. (a) Red dots show the amplitude of this solution at the first shell $\pm \fvec{g}_{1,2,3}$ for the parameters of Ref.~\cite{KaxirasPRB18} in Table~\ref{Tab:RelaxPara},  expressed in terms of the dimensionless variable $\zeta_1$ (see Eq.\ref{Eqn:LatRelaxFormula}). The blue curve is the approximate formula given in the Eq.\ref{Eqn:Zeta1Appr} using our approach, upon which we can  systematically improve as explained in the main text.   The solid black curve and the dashed green curve are the Taylor expansions of the solution to the first and the second order in the dimensionless parameter $\lambda=c_1/(\mathcal{G}\theta^2)$ that quantifies the relative magnitudes of $U_B$ and $U_E$ at a given twist angle $\theta$.
    The Taylor expansions are seen to be accurate above the twist angle $\sim 1.6^{\circ}$, but to increasingly deviate below. In Sec.\ref{Sec:radius of convergence} we explain the poor performance of the Taylor series expansion by finding its radius of convergence which translates to the twist angle $1.14^{\circ}$ and placing the magic angle beyond the reach of the Taylor expansion approach.
    (b) Dots show the analogous numerical results at the second and third shells, demonstrating that in this range of twist angles, the dimensionless amplitudes $\zeta_{2,3}$ are at least an order of magnitude smaller than $\zeta_1$. The solid lines are our approximate formulas; next order improvements can be found in the Fig.~\ref{FigS:LatticeRelax}. 
    }
	\label{Fig:LatRelaxNoStrain}
\end{figure}
We proceed to derive an approximate but accurate and simple analytical formula for $\delta \fvec U$. For this purpose, it is convenient to introduce a dimensionless parameter
\begin{align}
    \lambda = \frac{c_1}{\mathcal{G} \theta^2}   \label{Eq:lambdaDefinition}
\end{align}
that quantifies the ratio between the interlayer adhesion energy and the elastic energy at a given twist angle $\theta$. For smaller $\theta$, $\lambda$ is larger implying relative growth of the interlayer adhesion energy compared to the elastic energy and leading to stronger lattice relaxation. The theoretical model of Ref.~\cite{KaxirasPRB18} gives $\lambda \approx 0.256$ at the first magic angle of $\approx1.05^\circ$. 

In the unstrained case the moire lattice is invariant under the six-fold rotation $C_{6z}$ and the mirror reflections along $xz$ and $yz$ planes. The Helmholtz decomposition of the lattice relaxation field~\cite{JKPRB23} gives
\begin{align}
    \delta \fvec U(\fvec x) = \fvec \nabla \times (\varepsilon(\fvec x) \hat z) + \fvec \nabla \varphi(\fvec x) \ ,  \label{Eqn:HelmholtzDecomp}
\end{align}
where both fields $\varepsilon(\fvec x)$ and $\varphi(\fvec x)$ are periodic, and even under space inversion, i.e.~$\varepsilon(\fvec x) = \varepsilon(- \fvec x)$ and $\varphi(\fvec x) = \varphi(- \fvec x)$. Thus, their Fourier series can be written as
\begin{align}
    & \varepsilon(\fvec x) = \sum_{\fvec g} \tilde{\varepsilon}_{\fvec g} \cos(\fvec g \cdot \fvec x) \quad \mbox{and} \quad 
    \varphi(\fvec x)  = \sum_{\fvec g} \tilde{\varphi}_{\fvec g} \cos(\fvec g \cdot \fvec x) \nonumber \\
    & \Longrightarrow \quad  \fvec W(\fvec g)  = \tilde{\varepsilon}_{\fvec g} \hat z \times \fvec g  - \tilde{\varphi}_{\fvec g}  \fvec g  \ .
\end{align}
As mentioned, in the absence of external strain, the lattice relaxation $\delta \fvec U$ is invariant under $C_{6z}$, and odd under the mirror reflection about $xz$ and $yz$ planes, leading to the symmetry constraints for both $\tilde{\varepsilon}_{\fvec g}$ and $\tilde{\varphi}_{\fvec g}$. As a consequence, for $\fvec g$ on the first three  shells, $\tilde{\varphi}_{\fvec g}$ must vanish~\cite{ShaffiquePRL24}, and $\tilde{\varepsilon}_{\fvec g}$ depends only on  the shell index~\cite{JKPRB23}, 
\begin{align}
    & \tilde{\varepsilon}_{\pm \fvec g_1} = \tilde{\varepsilon}_{\pm \fvec g_2} = \tilde{\varepsilon}_{\pm \fvec g_3} = \zeta_1/(\theta |\fvec G_1|^2) \\
    & \tilde{\varepsilon}_{\pm (\fvec g_1 - \fvec g_2)} = \tilde{\varepsilon}_{\pm (\fvec g_2 - \fvec g_3)} = \tilde{\varepsilon}_{\pm (\fvec g_3 - \fvec g_1)} = \zeta_2/(\theta |\fvec G_1|^2)  \\
    & \tilde{\varepsilon}_{\pm 2\fvec g_1} = \tilde{\varepsilon}_{\pm 2\fvec g_2} = \tilde{\varepsilon}_{\pm 2\fvec g_3} = \zeta_3/(\theta |\fvec G_1|^2)  \  .
\end{align}
In the above we introduced the dimensionless parameters $\zeta_i$ ($i = 1$, $2$, and $3$) which quantify the strength of $\fvec W(\fvec g)$ on the $i$th shell as $|\fvec W(\fvec g)|/a \sim |\fvec g| \tilde{\varepsilon}_{\fvec g}/a \sim  |\fvec g| \zeta_i/(\theta |\fvec G_1|^2 a) \sim \zeta_i$. Explicitly displaying the Fourier amplitudes on the first three shells, the lattice distortion field has the form
\begin{align}
    & \delta \fvec U(\fvec x)  = \frac{2}{|\fvec G_1|^2} \sum_{a = 1}^3 \fvec G_a \left(\zeta_1 \sin(\fvec g_a \cdot \fvec x) + 2\zeta_3 \sin(2\fvec g_a \cdot \fvec x) \right) \nonumber \\
    & \quad + \frac{2 \zeta_2}{|\fvec G_1|^2} \sum_{a = 1}^3 \big( \fvec G_a - \fvec G_{a+1} \big) \sin((\fvec g_a - \fvec g_{a+1} ) \cdot \fvec x)+\ldots  \ , \label{Eqn:LatRelaxFormula}
\end{align}
where for notational simplicity, we define $\fvec G_4 = \fvec G_1$ and $\fvec g_4 = \fvec g_1$, $|\fvec G_1|=4\pi/(\sqrt{3}a)$, and $\ldots$ stand for terms at higher shells than the third. 

The solution to the Eq. \ref{Eqn:LatticeUqRelax} in the absence of the interlayer adhesion term, $U_B$, is simple, namely $\delta \fvec U = 0$. Therefore, when the adhesion term is small compared to the elastic term, we expect the $\fvec W(\fvec g)$ to be small as well. Assume that we have found this solution. We then extract its amplitude on the first shell $\pm \fvec g_{1,2,3}$ and express it in terms of $\zeta_1$. We can then estimate the magnitudes of the solution on the higher $\fvec g$ shells, $\zeta_2$, $\zeta_3$, etc. in terms of $\zeta_1$. We see that for small $\lambda$, and given the hierarchy $| c_1 | \gg |c_{2,3} |$  in the Tab.\ref{Tab:RelaxPara}, the solution for $\zeta_{2,3}$ is down by factors of either $c_{2,3}/c_1$ or $\lambda$ compared to $\zeta_1$. The suppression is even stronger on further shells.
Therefore, $\delta\fvec U$ is almost entirely determined by the Fourier amplitudes on the innermost shell. By solving Eq.~\ref{Eqn:LatticeUqRelax} numerically using the iteration method, the full solution (which we have at each shell) confirms that for $\theta \gtrsim 0.7^{\circ}$, the Fourier series of $\delta \fvec U(\fvec x)$ is indeed dominated by the innermost shell.

Since $\fvec W(\fvec g)$ is small on the outer shells for the twist angles of interest to us, the Fourier components on the outer shells have a negligible impact on the innermost shell. Consequently, $\zeta_1$ can be obtained accurately by setting $\zeta_2 = \zeta_3 = 0$ in Eq.~\ref{Eqn:LatRelaxFormula}, ignoring the higher shells denoted by the ellipsis, and then minimizing $U^{-}_E + U_B$. Explicitly, we start by setting 
\begin{align}
    \delta \fvec U(\fvec x) & \rightarrow \frac{2 \zeta_1}{|\fvec G_1|^2} \sum_{j = 1}^3 \fvec G_j \sin(\fvec g_j \cdot \fvec x), \  \label{Eqn:LatRelaxAppFormula}
\end{align}
substituting the above into Eq.~\ref{Eqn:ElasticEne}, and obtaining
\begin{align}
    U^{-}_E/A_{tot} = 3\mathcal{G} \theta^2 \zeta_1^2 /2 \ ,
\end{align}
where $A_{tot}$ is the total area of the sample. We see that the elastic energy does not depend on the elastic coefficient $\mathcal{K}$ because $\delta \fvec U(\fvec x)$ on the first $\fvec g$-shell is divergence free. The elastic energy cost also grows as a square of the twist angle because the moire period decreases inversely as the twist angle increases and the moire period sets the scale of the spatial variation penalized by the elastic energy. The interlayer adhesion energy, $U_B$ in Eq.~\ref{Eqn:AdhesionEne} can be written as
\begin{align}
    \frac{U_B}{A_{tot}} = A_{tot}^{-1} \sum_{\fvec G} V_{\fvec G} \int\rmd^2\fvec x\ \frac14 \left( e^{i \fvec g_{\fvec G} \cdot \fvec x} e^{i \fvec G \cdot \delta \fvec U(\fvec x)} + c.c. \right). \label{Eqn:UBForm}
\end{align}
To perform the Fourier integral we note that the phase factor associated with the distorted configuration is periodic and can be expanded via Fourier series as~\cite{Guinea23}
\begin{align}
    & e^{i \fvec G \cdot \delta \fvec U(\fvec x)} = \prod_{j = 1}^3 e^{2 i \sin(\fvec g_j \cdot \fvec x) \fvec G \cdot \fvec W(\fvec g_j)} \nonumber \\
    & = \sum_{n_1 = - \infty}^{\infty} \sum_{n_2 = - \infty}^{\infty} \sum_{n_3 = - \infty}^{\infty} \left( \prod_{j = 1}^3 J_{n_j}\left( 2 \zeta_1 \frac{\fvec G \cdot \fvec G_j}{|\fvec G_1|^2} \right) \right) \nonumber \\
    & \times \exp\left(i \big(\sum_{j = 1}^3 n_j \fvec g_j \big) \cdot \fvec x \right) \ ,
\end{align}
where $J_n(x)$ is the Bessel function of the first kind (with integer $n$). Substituting the above into Eq.~\ref{Eqn:UBForm}, we obtain
\begin{align}
    \frac{U_B}{A_{tot}} & = \half \sum_{\fvec G} V_{\fvec G} \sum_{n_1, n_2, n_3 = - \infty}^{\infty} \left( \prod_{j = 1}^3 J_{n_j}\left( 2 \zeta_1 \frac{\fvec G \cdot \fvec G_j}{|\fvec G_1|^2} \right) \right) \nonumber \\
    &  \times  \delta_{\fvec g_{\fvec G} + \sum_{j = 1}^3 n_j \fvec g_j,\ 0} \nonumber \\
    & = 3 c_2 \big( J_1(3\zeta_1) \big)^2 + 3  \sum_{n = - \infty}^{\infty} \left(  c_1 J_n(2\zeta_1) \big( J_{n+1}(\zeta_1) \big)^2 \right. \nonumber \\
    & \quad \left. + c_3 J_n(4 \zeta_1) \left( J_{n+2}(2\zeta_1) \right)^2 \right) \label{Eqn:UBBesselJ}  \ .
\end{align}
The last equality has been derived by noting that $V_{\fvec G} \neq 0$ if and only if $\fvec G$ is located on the first three shells. 

By minimizing $U^-_E+U_B$ with respect to $\zeta_1$ we obtain 
\begin{align}
    0 & = \zeta_1 + \lambda \frac{\rmd}{\rmd \zeta_1} \left( \frac{c_2}{c_1} \big( J_1(3\zeta_1) \big)^2 + \sum_{n = - \infty}^{\infty} \left( J_n(2\zeta_1) \big( J_{n+1}(\zeta_1) \big)^2 \right. \right. \nonumber \\
    & \left. \left. + \frac{c_3}{c_1} J_n(4 \zeta_1) \left( J_{n+2}(2\zeta_1) \right)^2 \right) \right) \ . \label{Eqn:Zeta1ExactEqn}
\end{align}
Using $\rmd J_n(x)/\rmd x = \frac{1}{2}(J_{n-1}(x) - J_{n+1}(x))$ one can obtain a numerical solution of the above equation which will give the optimal $\zeta_1$ in terms of $\lambda$; we did so and because we find a highly accurate closed form solution below which matches this numerical solution, we do not show it in Fig.~\ref{Fig:LatRelaxNoStrain}. 

To proceed analytically we first note that as $\lambda\rightarrow 0$, the above equation is satisfied if $\zeta_1\rightarrow 0$. Physically, this is the limit of short moire period when the elastic energy cost of relaxation dominates the interlayer adhesion energy gain, and the in-plane relaxation vanishes. Therefore, as we increase $\lambda$ and $\zeta_1$ increases from zero, we can approximate the right hand side of Eq.\ref{Eqn:Zeta1ExactEqn} by Taylor expanding it in small $\zeta_1$, noting that the Bessel functions of the first kind, $J_n(z)$, with integer $n$, are entire functions of complex $z$. This is justified as long as the solution for $\zeta_1$ we seek remains small. 
We find that for $0\leq \lambda\lesssim 0.44$, an accurate solution can be obtained by Taylor expanding the right-hand side of Eq.\ref{Eqn:Zeta1ExactEqn} in $\zeta_1$ to second order, giving 
\begin{align}
0\approx
    	3 \lambda \zeta_1^2 + \left(1 + \alpha \lambda \right)\zeta_1 - \lambda,  \label{Eqn:Zeta1Eqn}
\end{align}
where for notational convenience, we introduced the parameter
\begin{align}
    \alpha = \frac{1}{2}+\frac{9}{2}\frac{c_2}{c_1}+4\frac{c_3}{c_1} \ .  \label{Eqn:AlphaFormula}
\end{align}
Solving the quadratic Eq.\ref{Eqn:Zeta1Eqn} gives us an accurate analytical formula for the optimal amplitude on the first $\fvec g$ shell
\begin{align}
    \zeta_1 \approx \frac1{6\lambda} \left( -(1+ \alpha \lambda) + \sqrt{(1 + \alpha \lambda)^2 +12 \lambda^2} \right) \label{Eqn:Zeta1Appr}  \ .
\end{align}

With the parameters listed in Tab.~\ref{Tab:RelaxPara} for Ref.~\cite{KaxirasPRB18}, we can obtain a numerical solution of Eq.~\ref{Eqn:LatticeUqRelax}. We then extract the corresponding $\zeta_j$ for the relaxation on the $j^{th}$ $\fvec g$-shell using
$\zeta_j = \frac{|\fvec g_1|^2}{|\fvec g|^2} \fvec G_{\fvec g} \cdot \fvec W(\fvec g)$,
where $\fvec G_g = \hat z \times \fvec g/\theta$ and $\fvec g$ is on the $j^{th}$ shell. The comparison of such numerical result on the first shell and the approximate formula in Eq.~\ref{Eqn:Zeta1Appr} is presented in Fig.~\ref{Fig:LatRelaxNoStrain}, showing an excellent agreement for the twist angle $\theta$ down to at least $0.7^{\circ}$.

The closed form expressions for the $\zeta_2$ and $\zeta_3$ for the $2$nd and $3$rd shells can be obtained using similar methods. Starting from Eq.~\ref{Eqn:LatRelaxFormula} with $\zeta_1$ from Eq.~\ref{Eqn:Zeta1Appr}, we keep $\zeta_1$ fixed and expand the total energy $U^{-}_E + U_B$ up to the quadratic order of $\zeta_2$ and $\zeta_3$. This approximation for the energy is justified since $\zeta_2$ and $\zeta_3$ are much smaller than $1$ for the twist angle $\theta$ down to $0.2^{\circ}$, notably smaller than the first magic angle.  Integrating Eq.~\ref{Eqn:ElasticEne} over the space, we found the exact formula for the elastic energy:
\begin{align}
    U^{-}_E/A_{tot} = \frac32 \mathcal{G} \theta^2 \left( \zeta_1^2 + 9 \zeta_2^2 + 16 \zeta_3^2 \right) \ .
\end{align}
We also obtain a (complicated) formula for $U_B$ that is presented in the appendix. Minimizing $U^-_E+U_B$, and expanding the Bessel functions to $\mathcal{O}\left(\zeta_1^2\right)$, we obtain the approximate solutions for $\zeta_2$ and $\zeta_3$:
\begin{align}
    \zeta_2 & \approx \frac{\lambda}6 \zeta_1 + \frac{\lambda}3 \frac{c_2}{c_1} - \frac{\lambda}6 \zeta_1 \left( \zeta_1 + 4\frac{c_3}{c_1} \right),   \label{Eqn:Zeta2Appr}  \\
    \zeta_3 & \approx \frac{\lambda}8 \zeta_1 + \frac{\lambda}4 \frac{c_3}{c_1}. \label{Eqn:Zeta3Appr} 
\end{align}
Fig.~\ref{Fig:LatRelaxNoStrain:Outer} shows the plots of $\zeta_2$ and $\zeta_3$ with parameters specified in Tab.~\ref{Tab:RelaxPara} for the model proposed in Ref.~\cite{KaxirasPRB18}. Clearly, the analytical formulas in Eqs.~\ref{Eqn:Zeta2Appr} and \ref{Eqn:Zeta3Appr} match with the full numerical solution very well for $\theta \gtrsim 0.7^{\circ}$. In addition, when the $\lambda \ll 1$ (as is the case for $\theta \gtrsim 1^{\circ}$) then $\zeta_1 \approx \lambda$ as seen from Eq.\ref{Eqn:Zeta1Appr}, and
$\zeta_2/\zeta_1 \sim \zeta_3/\zeta_1 \sim \max(\lambda, |c_2|/c_1) \ll 1$, confirming our a'priori estimates. Therefore, the Fourier components of the distortion on the outer shells are much smaller than on the first shell, as also confirmed in Fig.~\ref{Fig:LatRelaxNoStrain}. This hierarchy in the lattice relaxation not only justifies the assumption when deriving Eq.~\ref{Eqn:Zeta2Appr} and \ref{Eqn:Zeta3Appr}, but also allows us to neglect the impact of outer shell relaxation on the electronic Hamiltonian for $\theta \gtrsim 1^{\circ}$, as shown in an upcoming paper. 

We thus obtain a highly accurate expression for the difference between the in-plane layer displacements in the two layers $\fvec U^-  = \fvec U_t - \fvec U_b$. The accuracy of our formulas for $\zeta_i$ can be further improved, as explicitly worked out in Sec.~\ref{SecS:Improve} in Appendix.  



\subsection{Radius of convergence for the lattice relaxation}
\label{Sec:radius of convergence}
In order to compare with the results of Refs.~\cite{Guinea23,ShaffiquePRL24} we expand our solution in the Eq.~\ref{Eqn:Zeta1Appr} in a power series of $\lambda$, finding $\zeta_1 \approx \lambda - \alpha \lambda^2 + \mathcal{O}(\lambda^3)$. Although we do not rely on such power series expansion, and in fact discuss its shortcomings below, we note that our leading term $\zeta_1 \approx \lambda$ agrees with the $\mathcal{O}(\lambda)$ result obtained in Ref.~\cite{Guinea23} and Ref.\cite{ShaffiquePRL24} provided we identify the latter's $c_1/(2\sin^2\frac{\theta}{2})$ and $2\alpha_1$ with our $\lambda$ and $\zeta_1$ respectively. Expanding to such a low order agrees with the full numerical solution on the first $\fvec g$ shell for $\theta \gtrsim 1.6^{\circ}$, but increasingly overestimates the relaxation at smaller angles; the deviation is $\sim 20\%$ at the (first) magic angle.
Our prefactor of $\lambda^2$ also agrees with the $\mathcal{O}(\lambda^2)$ prefactor in Ref.~\cite{ShaffiquePRL24} provided $c_2=c_3=0$ in which case $\alpha=0.5$. This prefactor is significantly corrected by non-zero $c_2$ and $c_3$ as shown in Eq.~\ref{Eqn:AlphaFormula}, and the resulting prefactor $\alpha \approx -0.0076$ is tiny for the model proposed in Ref.~\cite{KaxirasPRB18}.

\begin{figure}[t] 
	\centering
	\includegraphics[width=0.6\columnwidth]{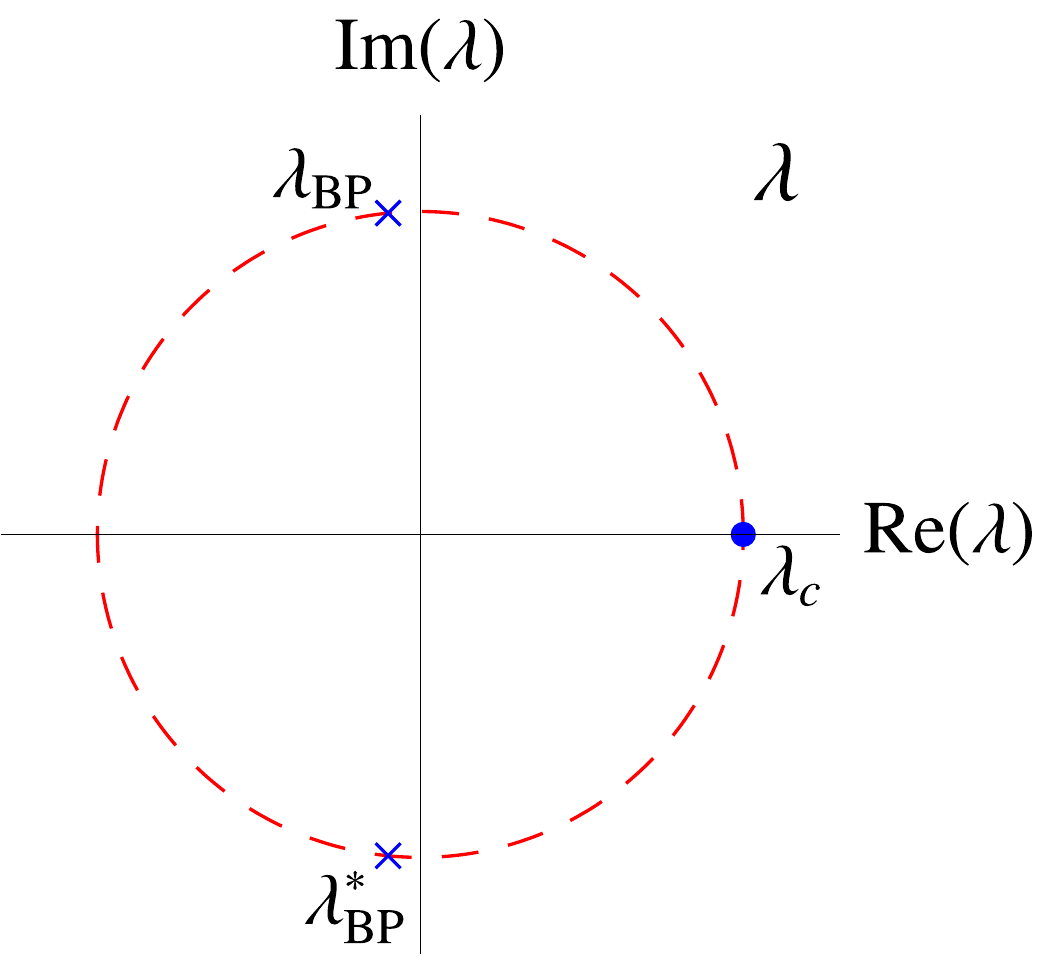}
	\caption{Schematic plot for the branch points of the function $\zeta_1(\lambda)$ in the complex $\lambda$-plane, whose magnitude $|\lambda_{BP}|$ dictates the radius of convergence, $\lambda_c$, of the power series expansion of $\zeta_1(\lambda)$ about $\lambda=0$. Note that the two branch points, marked by the blue crosses, are complex conjugates of each other.}
	\label{Fig:Converge}
\end{figure}

Our closed form solution allows us to understand the convergence of the power series expansion in $\lambda$ by studying the analytic properties in the complex $\lambda$ plane. The solution in Eq.~\ref{Eqn:Zeta1Appr} contains a branch cut with branch points at
$\lambda_{\text{BP}} = (- \alpha + 2\sqrt{3}i)/(\alpha^2 + 12)$ and $\lambda^*_{\text{BP}}$.
A complex circle centered at the origin and passing through the branch points intersects the real $\lambda$ axis at  
$\lambda_c=1/\sqrt{12 + \alpha^2}$ as shown in Fig.~\ref{Fig:Converge}; $\lambda_c$ is the radius of convergence of the power series~\cite{WhittakerBook}. As a consequence, the power expansion of Eq.\ref{Eqn:Zeta1Appr} in $\lambda$  must diverge if $\lambda > \lambda_c$. With the parameters of Ref.~\cite{KaxirasPRB18} specified in Tab.~\ref{Tab:RelaxPara}, we obtain $\lambda_c \simeq 0.289$ with the corresponding angle $\theta_c = 0.99^{\circ}$. Note that our solution in the Eq.\ref{Eqn:Zeta1Appr} is perfectly smooth at $\lambda_c = |\lambda_{\text{BP}}|$ and the singularity occurs only in the power expansion. 

Because the right hand side of the Eq.\ref{Eqn:Zeta1ExactEqn} is an analytic function of $\zeta_1$ and $\lambda$, we can also study its zeros in the complex $\lambda$ plane numerically without approximating the right hand side by a quadratic polynomial in $\zeta_1$. For a fixed complex value of $\lambda$ the zeros of the real part form curves in the complex $\zeta_1$ plane as do the zeros of the imaginary part. Their intersection gives the solutions of the Eq.\ref{Eqn:Zeta1ExactEqn}, the physical one being on the real $\zeta_1$ axis for real $\lambda$ and vanishing with vanishing $\lambda$. For $c_2=c_3=0$ we find that two solutions closest to the origin are $\lambda_{BP}=\lambda_c e^{\pm i\pi\delta}$ with machine precision $\lambda_c=0.2603353755476081$ and $\delta= 0.5568038282560424$ (see Sec.~\ref{SecS:BP} in Appendix for the method to obtain these and for the proof of the branch points' existence). These correspond to the branch points we found approximately using Eq.~\ref{Eqn:Zeta1Appr}.
For the values of $c_2$ and $c_3$ listed in Tab.~\ref{Tab:RelaxPara} for Ref.~\cite{KaxirasPRB18} we find $\lambda_{c}=0.2166638106379033$ and $\delta= 0.4681651233626292$, 
further reducing the radius of convergence of any power series in $\lambda$ solution to the amplitude of the first $\fvec g$ shell of $\delta \fvec U(\fvec x)$. This value of $\lambda_c$ translates to the twist angle $\theta_c = 1.14^\circ$ placing the first magic angle just outside the radius of convergence and making the power series approach impractical for obtaining an increasingly accurate solution near the first magic angle. As mentioned, our closed form solution does not suffer from this problem. We can further improve upon it by expanding the right hand side of the Eq.~\ref{Eqn:Zeta1ExactEqn} to the third order in $\zeta_1$ and solving the corresponding cubic equation for $\zeta_1$ in terms of $\lambda$. We have done so, but because the expressions are somewhat unwieldy and because the Eq.~\ref{Eqn:Zeta1Appr} is sufficiently accurate for our purposes down to and below the first magic angle, we do not write out the solution of the cubic here.

\section{Lattice relaxation in the presence of an external Heterostrain}
\label{Sec:relaxation with strain}

\begin{figure}[t]
	\centering
	\subfigure[\label{Fig:LatRelaxStrain:Inverse}]{\includegraphics[width=0.99\columnwidth]{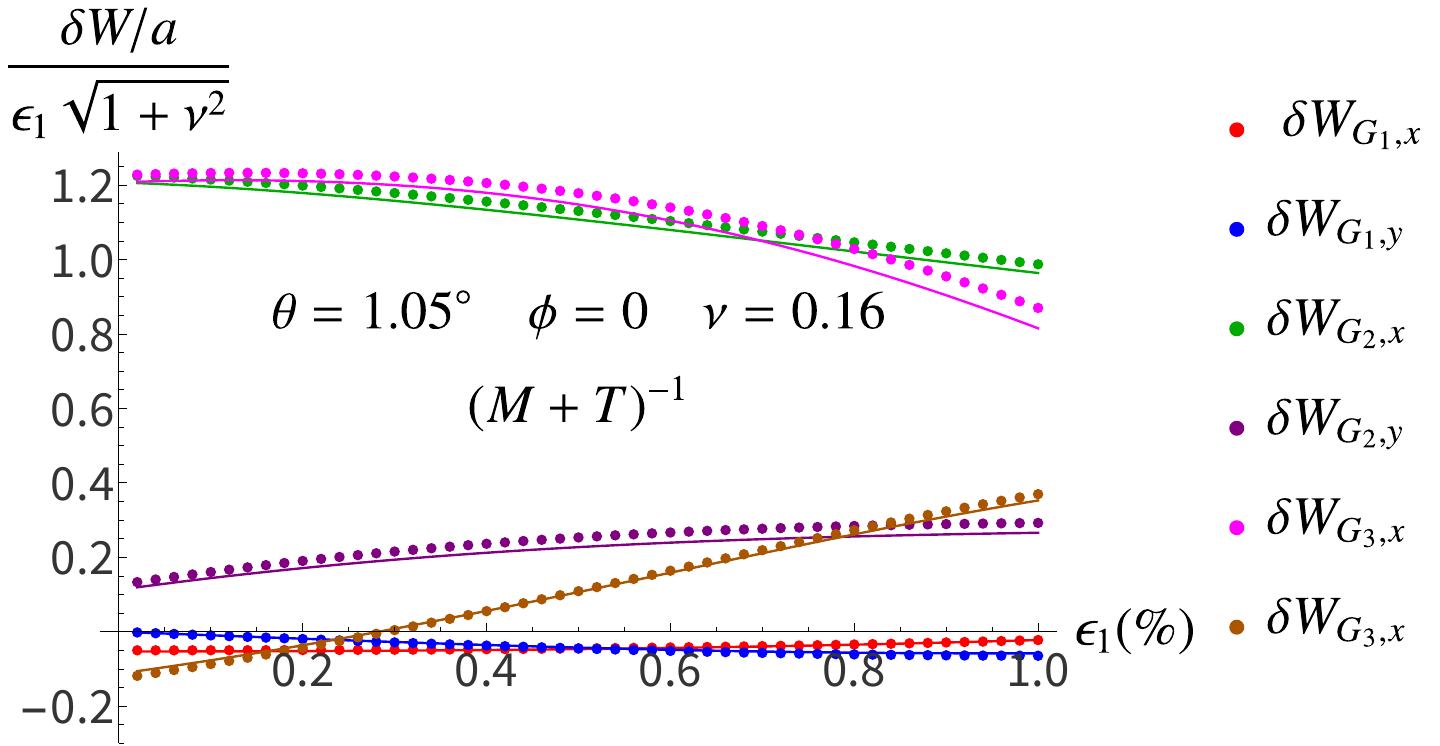}}
        \subfigure[\label{Fig:LatRelaxStrain:Appr}]{\includegraphics[width=0.99\columnwidth]{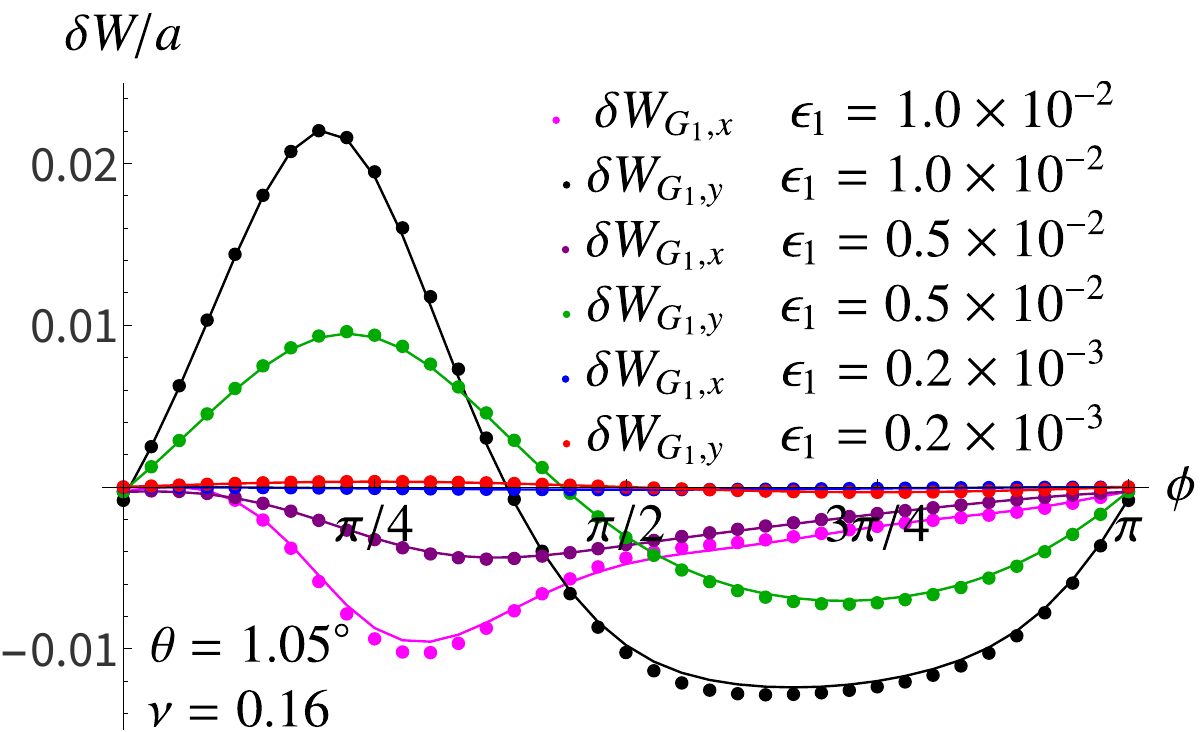}}
	\caption{The heterostrain-induced change of the Fourier amplitudes of the relaxation field, labeled by $\delta \fvec W_{\fvec G}$ and defined in the Eq.~\ref{Eqn:WGDecomp} (see also Eqs. \ref{Eqn:LatRelaxFTStrain}, \ref{Eq:Udef}, \ref{Eq:UminusDef}, and \ref{Eqn:LatDis}) is divided by the graphene lattice constant $a\approx0.246$nm and plotted as a function of (a) $\epsilon_1$ and (b) $\phi$, the two parameters entering $S^{\epsilon}$ introduced in Eq.~\ref{Eqn:StrainMat}. The plots were obtained at the magic angle with $\theta = 1.05^{\circ}$. The remaining strain parameter is set to $\epsilon_2 = -\nu \epsilon_1$ using the Poisson ratio $\nu = 0.16$. Only $\delta \fvec W_{\fvec G}$ with $\fvec G$ at the first shell is sizable, other $\delta \fvec W_{\fvec G}$'s are negligible for realistic samples. (a) Colored dots illustrate the direct numerical solution~\cite{KangLatticeRelax2025} of the Eq.~\ref{Eqn:LatticeUqRelax} resulting in $\delta W$ vs $\epsilon_1$ (with $\phi = 0$), with each component marked by different colors. Because $\delta \fvec W$ grows with increasing heterostrain, we rescaled it by the magnitude of the strain matrix, defined as $\sqrt{\Tr((S^{\epsilon})^2)} = |\epsilon_1| \sqrt{1 + \nu^2}$. The solutions obtained using the simple set of six linear Eqs.~\ref{Eqn:W1FiniteMatrixEqn} are also presented as colored curves for comparison, showing an excellent agreement with the full numerical calculation even when $\epsilon_1$ is as large as $1\%$. (b) The two components of $\delta \fvec W_{\fvec G_1}$ vs $\phi$ (at different $\epsilon_1$) as obtained by (colored dots) numerically solving Eqn.~\ref{Eqn:LatticeUqRelax} and (colored curves) by solving the Eqn.~\ref{Eqn:W1FiniteMatrixEqn}. Again, these two approaches produce solutions in excellent agreement. Although only $\delta \fvec W_{\fvec G_1}$ is plotted here, $\delta \fvec W_{\fvec G_2}$ and $\delta \fvec W_{\fvec G_3}$ can  also be obtained from $\delta \fvec W_{G_1}$ from the symmetry constrains as listed in Eq.~\ref{EqnS:C6WG2} and \ref{EqnS:C6WG3}.}
	\label{Fig:LatRelaxStrain}
\end{figure}

Having discussed the lattice relaxation in the unstrained twisted bilayer, we study the impact of the external heterostrain that is described by a $2\times 2$ symmetric matrix shown in Eq.~\ref{Eqn:StrainMat}. Although the heterostrain is ubiquitous in realistic twisted samples, the two strain parameters $\epsilon_1$ and $\epsilon_2$ are always much smaller than the twist angle $\theta$ in which case the matrix $S^{\epsilon} + i \theta \sigma^2$ introduced in Eq.~\ref{Eqn:gGMap} is non-singular. As a consequence, the lattice relaxation $\delta \fvec U(\fvec x)$ is still periodic and its periods are given by two non-collinear moire lattice vectors $\fvec L_1$ and $\fvec L_2$  defined in the Eq.~\ref{Eqn:LiVec}. The strained lattice relaxation is obtained by solving Eq.~\ref{Eqn:LatticeUqRelax} with the corresponding moire reciprocal lattice vectors $\fvec g$ given in Eq.~\ref{Eqn:gdef}. 

Even with the approximation scheme used in the previous section for the unstrained case, it is hopelessly challenging to obtain a closed form solution of the Eq.~\ref{Eqn:LatticeUqRelax} without making any further assumptions about the magnitude of the heterostrain.
Nevertheless, we can make progress by 
using the fact that in realistic devices the external heterostrain is small i.e.~$|\epsilon_{1,2}| \ll \theta$.
This can be used to derive an approximate but closed-form solution for the change of lattice relaxation due to external heterostrain.  
To this end, we express the lattice relaxation as 
\begin{align}
    \delta \fvec U(\fvec x)  = 2 \sum_{\fvec G \in \mathbb{H}} \fvec W_{\fvec G} \sin(\fvec g_{\fvec G} \cdot \fvec x)      \ , \label{Eqn:LatRelaxFTStrain}
\end{align}
where $\fvec g_{\fvec G}$'s are the heterostrain distorted moire reciprocal lattice vectors defined in Eq.~\ref{Eqn:gGMap} and where we introduced $\mathbb{H}$ to be half of the set of nonzero $\fvec G$ vectors, that includes one and only one vector for each nonzero vector pair $(\fvec G,\ - \fvec G)$. This can be done because $\fvec g_{\fvec G}$ is linear in $\fvec G$ and $\fvec{W}_{\fvec G}$ is odd under $\fvec G\rightarrow -\fvec G$, as discussed below Eq.\ref{Eqn:WgDef}. 
The geometric effect of heterostrain on the reciprocal lattice vectors in the Eq.~\ref{Eqn:gGMap} can be isolated as 
\begin{align}
    & \fvec g_{\fvec G} = \fvec g^{(0)}_{\fvec G} + \delta \fvec g_{\fvec G}   \qquad  \mbox{with}    \label{Eqn:gGDecomp} \\
 & g^{(0)}_{\fvec G, \mu} = \theta i \sigma^{2}_{\mu\nu} G_{\nu} \quad \mbox{and} \quad \delta g_{\fvec G, \mu} = S^{\epsilon}_{\mu\nu} G_{\nu}.  \label{Eqn:gGDef}
\end{align}
In addition, we decompose $\fvec W_{\fvec G}$ into two parts,
\begin{align}
    \fvec W_{\fvec G} = \fvec W^{(0)}_{\fvec G} + \delta \fvec W_{\fvec G},\label{Eqn:WGDecomp} 
\end{align}
where $\fvec W^{(0)}_{\fvec G}$ is the solution of the Eq.~\ref{Eqn:LatticeUqRelax} in the absence of external heterostrain. It has already been determined in the previous section: for $\fvec G$ on the first three shells 
\begin{equation}
\fvec W^{(0)}_{\fvec G\in j^\text{th} \text{shell}} = \zeta_j \frac{\fvec G}{|\fvec G_1|^2},\;\;\;\; \label{Eqn:W0}
j=1,2,3,
\end{equation}
while for $j > 3$, $\fvec W^{(0)}_{\fvec G}$ is too small and thus can be neglected. $\delta \fvec W_{\fvec G}$ gives the heterostrain induced change of the Fourier coefficients of the lattice relaxation, $\delta\fvec U(\fvec x)$, at the reciprocal lattice vectors $\fvec g_{\fvec G}$. Note that these reciprocal lattice vectors are themselves modified by the heterostrain as shown in the Eq.\ref{Eqn:gGDecomp}. 

The equation for $\delta \fvec W_{\fvec G}$ is obtained by substituting Eq.~\ref{Eqn:gGDecomp} and \ref{Eqn:WGDecomp} into Eq.~\ref{Eqn:LatticeUqRelax}. To present its explicit form, we first decompose the matrix introduced in Eq.~\ref{Eqn:MMatrix} into three parts, organized by the powers of $\delta \fvec g$,  
\begin{align}
    M(\fvec g_{\fvec G}) & = M^{(0)}_{\fvec G} + M^{(1)}_{\fvec G} + M^{(2)}_{\fvec G}, \label{Eqn:MMat} \\
    M^{(0)}_{\fvec G, \mu\nu} & =\mathcal{G} |\fvec g_{\fvec G}^{(0)}|^2 \delta_{\mu\nu} + \mathcal{K} g_{\fvec G, \mu}^{(0)} g_{\fvec G, \nu}^{(0)}, \label{Eqn:M0} \\
    M^{(1)}_{\fvec G, \mu\nu} & = 2 \mathcal{G} (\fvec g_{\fvec G}^{(0)} \cdot \delta \fvec g_{\fvec G} ) \delta_{\mu\nu} + \nonumber \\
    & \quad \mathcal{K} \left( g_{\fvec G, \mu}^{(0)} \delta g_{\fvec G, \nu} + \delta g_{\fvec G, \mu} g_{\fvec G, \nu}^{(0)}  \right), \label{Eqn:M1} \\
    M^{(2)}_{\fvec G, \mu\nu} & = \mathcal{G} |\delta \fvec g_{\fvec G}|^2 \delta_{\mu\nu} + \mathcal{K} \delta g_{\fvec G, \mu} \delta g_{\fvec G, \nu}   \ .  \label{Eqn:M2}
\end{align}
Then, as explicitly derived in the Appendix, the Eq.~\ref{Eqn:LatticeUqRelax} results in the following equation for $\delta \fvec W$:
\begin{align}
    0 & =\left( M_{\fvec G}^{(0)} + M_{\fvec G}^{(1)} + M_{\fvec G}^{(2)} \right)_{\mu\nu} \left( \fvec W^{(0)}_{\fvec G} + \delta \fvec W_{\fvec G} \right)_{\nu} +\nonumber \\
    & \left. \frac{\partial}{\partial W_{\fvec G, \mu}} \half \sum_{\fvec G'} V_{\fvec G'} B(\fvec G',\  \{ \fvec W \}) \right|_{\fvec W = \fvec W^{(0)} + \delta \fvec W }   \ .  \label{Eqn:WEqn}
\end{align}
In the above we introduced the function $B(\fvec G, \{\fvec W\})$ which is most easily defined via a Fourier integral as 
\begin{align}
    & B(\fvec G, \{ W \})  =  \frac{1}{A_{tot}} \int\rmd^2 \fvec x\ e^{i\fvec g_{\fvec G} \cdot \fvec x} \times \nonumber \\
    & \prod_{\fvec G' \in \mathbb{H}} \left(  \sum_{ n = - \infty }^{\infty}  J_{n} \left( 2 \fvec G \cdot \fvec W_{\fvec G'} \right) e^{i n \fvec g_{\fvec G'} \cdot \fvec x } \right) \ .
\label{Eqn:BDef}
\end{align}
Here $J_n(x)$ is the Bessel function of the first kind, the set $\mathbb{H}$ is introduced in Eq.~\ref{Eqn:LatRelaxFTStrain}.
Of course, the above integral can be performed analytically, but we find it more transparent to present it this way.

So far, no approximations have been made and the obtained nonlinear equation for $\delta \fvec  W$ is still difficult to solve. However, $\delta \fvec W$ must vanish in the absence of strain. Therefore, for $|\epsilon_{1,2}| \ll \theta\ll1$, we expect $|\delta \fvec W| \ll |\fvec W^{(0)}|$, leading to
$ |\fvec G \cdot \delta \fvec W_{\fvec G'}| \ll |\fvec G \cdot \fvec W^{(0)}_{\fvec G'}| \lesssim \zeta_1 \sim \lambda \lesssim 1.$
As a consequence, the last term in Eq.~\ref{Eqn:WEqn} needs to be known only near the solution, $\delta \fvec W$, which we argue is small for small heterostrain. Therefore, the complicated nonlinear term can be approximated by Taylor series in powers of $\delta \fvec W$. Expanding it up to the linear order in $\fvec G \cdot \delta \fvec W$ leads to a set of coupled linear equations for $\delta \fvec W$. 
Since $\fvec W^{(0)}$ is the solution for the unstrained case, the Eq.\ref{Eqn:WEqn} holds when $\delta \fvec g = \delta \fvec W = 0$. The coupled linear equations can be expressed as
\begin{align}
    0 & = \sum_{\nu} \left(  M_{\fvec G}^{(1)} + M_{\fvec G}^{(2)} \right)_{\mu\nu} W^{(0)}_{\fvec G, \nu}
     \nonumber \\
    & + \sum_{\nu} \left( M_{\fvec G}^{(0)} + M_{\fvec G}^{(1)} + M_{\fvec G}^{(2)} \right)_{\mu\nu} \delta W_{\fvec G, \nu} \nonumber \\
    & + \sum_{\fvec G', \nu} \big( T(\{ \fvec W^{(0)} \}) \big)_{\fvec G,\mu, \fvec G',\nu} \delta W_{\fvec G', \nu},   \label{Eqn:DWMatrixEqn}
\end{align}
where, for notational convenience, we introduced $T(\{ \fvec W^{(0)} \})$ defined as
\begin{equation}
   \big( T(\{ \fvec W^{(0)} \}) \big)_{\fvec G,\mu, \fvec G',\nu}  = \sum_{\fvec G''} \frac{V_{\fvec G''}}{2} \left. \frac{\partial^2 B(\fvec G'',\  \{ \fvec W \}) }{\partial W_{\fvec G, \mu} \partial W_{\fvec G', \nu}}  \right|_{\fvec W = \fvec W^{(0)}}.  \label{Eqn:TMatrix}
\end{equation}
Note that the first term in the Eq.(\ref{Eqn:DWMatrixEqn}) is a constant, independent of $\delta \fvec W$, and the last two terms are linearly proportional to $\delta \fvec W$. Therefore, we have successfully transformed the nonlinear equation (\ref{Eqn:WEqn}) into a linear equation for $\delta \fvec W$. However, due to the infinite number of $\fvec G$ vectors, the $T$ matrix is also of infinite size, and thus needs to be truncated to be practical and to allow obtaining $\delta \fvec W$. In the appendix, we present a detailed argument showing that $\delta \fvec W_{\fvec G}$ is much larger for $\fvec G$ on the first shell than on further shells, allowing us to neglect all $\delta \fvec W_{\fvec G}$'s except $\delta \fvec W_{\fvec G_a}$ ($a = 1$, $2$, and $3$) in Eq.~\ref{Eqn:DWMatrixEqn}. Thus, we obtain the reduced form of the Eq.~\ref{Eqn:DWMatrixEqn}
\begin{align}
     & \sum_{b = 1}^3 \sum_{\nu = 1}^2 (\mathcal{M} + \mathcal{T})_{a \mu, b \nu} \delta W_{\fvec G_b \nu} = - u_{a \mu}, \label{Eqn:W1FiniteMatrixEqn} 
\end{align}
where the reduced form of the first term in the Eq.~\ref{Eqn:DWMatrixEqn}, entering the right hand side of the above equation, can be expressed via the heterostrain distorted reciprocal moire lattice vectors $\fvec g_a$ in the Eq.\ref{Eqn:gdef}, the reciprocal moire vectors in the absence of the heterostrain $\fvec g^{(0)}_a$, as well as their difference $\delta\fvec g_a = \fvec g_a-\fvec g_a^{(0)}$, as
%
\begin{align}
     & u_{a \mu} =   \frac{\zeta_1}{|\fvec G_1|^2} \left[ \mG \left(|\fvec g_a|^2 - |\fvec g^{(0)}_a|^2 \right) G_{a, \mu}   + \mK (\delta \fvec g_a \cdot \fvec G_a) g_{a,\mu} \right].                \label{Eqn:uForm}
     \end{align}     
The block diagonal $6\times 6$ matrix $\mathcal{M}$ in the Eq.~\ref{Eqn:W1FiniteMatrixEqn} can also be conveniently expressed in terms of $\fvec g_a$'s and the elastic coefficients $\mG$ and $\mK$ as 
     \begin{align}
     & \mathcal{M}_{a\mu, b\nu}  = \delta_{ab} \left( \mathcal{G} |\fvec g_a|^2 \delta_{\mu\nu} + \mathcal{K} g_{a, \mu} g_{a, \nu} \right) \ 
 . \label{Eqn:MForm} 
\end{align}
Although the matrix $\mT$ with its elements defined as $\mathcal{T}_{a\mu, b\nu} = \big( T(\{ \fvec W^{(0)} \}) \big)_{\fvec G_a,\mu, \fvec G_b,\nu}$ is independent of the external heterostrain, its expression is still quite complicated because of the formally infinite products of infinite sums of Bessel functions. Therefore, further approximations need to be made to the $\mathcal{T}$ matrix elements in order to obtain the closed-form expression for $\delta \fvec W$. 
Because the unstrained solution is dominated by the first shell, we keep only $\fvec W^{(0)}_{\fvec G_a} = \zeta_1 \fvec G_a/|\fvec G_1|^2$ and set all other $\fvec W^{(0)}_{\fvec G}$'s to zero. 
This turns every element of the six-by-six matrix $\mathcal{T}$  into a nonlinear function of $\zeta_1$ alone. Next, we expand these nonlinear functions up to the order $\zeta_1^2$. This effectively truncates the infinite product onto only the first shell of $\mathbb{H}$, as well as turns the infinite sum into a finite sum over a manageable number of terms, leading to closed form expressions for the diagonal blocks $\mathcal{T}_{a\mu, a\nu}$ and the off-diagonal blocks $\mathcal{T}_{a\mu, b\nu}$ ($a\neq b$):
\begin{widetext}
    \begin{align}
        \mathcal{T}_{a\mu, a\nu} & = c_1 \left[\left(3 \zeta_1 - \frac98 \zeta_1^2\right) |\fvec G_1|^2 \delta_{\mu\nu} + \left(\zeta_1 + \frac{\zeta_1^2}2\right) G_{a,\mu} G_{a, \nu}  \right]  + 4 c_3 G_{a, \mu} G_{a,\nu},  \label{Eqn:DiagTAppr} \\
        \mathcal{T}_{a\mu, b\nu} & = c_1 \left[ \left(1 + \zeta_1 - \frac74 \zeta_1^2\right) G_{c, \mu}  G_{c,\nu} - \frac32 \zeta_1 \left(1 - \frac{\zeta_1}2 \right)|\fvec G_1|^2 \delta_{\mu\nu}   \right] + 6 c_3 \zeta_1 \left(  |\fvec G_1|^2 \delta_{\mu\nu} - 2 G_{c, \mu} G_{c, \nu} \right)  \nonumber \\
    & \quad - c_2 \left[ (\fvec G_a - \fvec G_b)_{\mu} (\fvec G_a - \fvec G_b)_{\nu} + \frac32 \zeta_1 \left( \frac92 |\fvec G_1|^2 \delta_{\mu\nu} - (\fvec G_a - \fvec G_b)_{\mu} (\fvec G_a - \fvec G_b)_{\nu} \right)   \right]   \qquad \mbox{with} \quad a\neq b \ .  \label{Eqn:OffTAppr}
    \end{align}
\end{widetext}
For convenience, here we introduced the notation $\fvec G_c = - (\fvec G_a + \fvec G_b)$ when $a\neq b$, with no sum over repeated indices. 
As a result, $\delta \fvec W_{\fvec G_a}$ can be solved simply by inverting the $6\times6$ matrix $\mathcal{M}+\mathcal{T}$, which can always be done analytically using the cofactor method. 
Thus,
\begin{align}
    \delta W_{\fvec G_a, \mu} = - \sum_{b = 1}^3 \sum_{\nu = 1}^2 (\mathcal{M} + \mathcal{T})^{-1}_{a\mu, b\nu} u_{b\nu} \ ,  \label{Eqn:DWAppr}
\end{align}
and the heterostrained lattice relaxation can be accurately approximated as
%
    \begin{eqnarray}\label{Eqn:main formula}
     \delta \fvec U(\fvec x)  \approx 2 \sum_{a = 1}^3  &&\left[ \left(\frac{\zeta_1 \fvec G_a}{|\fvec G_1|^2}+\delta \fvec W_{\fvec G_a}\right) \sin(\fvec g_a \cdot \fvec x)\right.\nonumber\\
    &&\left. +  \frac{\zeta_2 \big(\fvec G_a - \fvec G_{a+1} \big)}{|\fvec G_1|^2} \sin((\fvec g_a - \fvec g_{a+1} ) \cdot \fvec x)\right.\nonumber\\
    &&\left.+ \frac{2\zeta_3\fvec G_a}{|\fvec G_1|^2} \sin(2\fvec g_a \cdot \fvec x) \right],  
\end{eqnarray}
where $\fvec G_4 = \fvec G_1$ and $\fvec g_4 = \fvec g_1$, and $|\fvec G_1|=4\pi/(\sqrt{3}a)$. Recall that $\fvec g_a$'s
are the heterostrain distorted reciprocal moire lattice vectors in the Eq.\ref{Eqn:gdef}, while $\fvec G_a$'s are unaffected by the heterostrain (see Eqs. \ref{Eqn:GVectors},\ref{Eqn:G3Vector}).
If we only keep the first shell, i.e. set $\zeta_2=\zeta_3=0$,  and use $\zeta_1$ from the Eq.~\ref{Eqn:Zeta1Appr}, then the above result is accurate to within $4\%$ for $\theta \gtrsim 1^{\circ}$, and as we shown in an upcoming paper, can be safely used to obtain the electronic spectrum (See also Ref.~\cite{JonahHFarXiv24, JonahIKSarXiv25}). Higher accuracy can be achieved by including $\zeta_{2}$ and $\zeta_{3}$ from the Eq.~\ref{Eqn:Zeta2Appr} and Eq.~\ref{Eqn:Zeta3Appr}, respectively, allowing us to go below the first magic angle with high accuracy even at $\sim 0.7^\circ$.
Fig.~\ref{Fig:LatRelaxStrain:Inverse} compares the full numerical solution for $\delta \fvec W$ and our analytical solution in the Eq.~\ref{Eqn:DWAppr} for a particular form of the strain matrix $S^{\epsilon}$ with $\phi = 0$ and $\epsilon_2 = -0.16 \epsilon_1$. As can be seen, the agreement between the two is very good all the way up to $\epsilon_1\approx 0.01=1\%$, or $\epsilon_1/\theta \lesssim 0.55$.  

Although the Eq.~\ref{Eqn:DWAppr} provides a closed form expression for the heterostrain induced correction to the relaxed atomic configuration, it involves an inverse of the $6 \times 6$ matrix which is cumbersome to express in a simple analytical form. We can further simplify it by noting that there is a hierarchy between the sizes of the matrix $\mathcal{M}$ -- which is simple to invert because it is block diagonal and effectively involves only $2 \times 2$ matrices -- and the matrix $\mathcal{T}$. Namely, 
$  |\mathcal{T}_{a\mu, b\nu}|/|\mathcal{M}_{c\rho, c\sigma}| \lesssim c_1 |\fvec G_1|^2 / \mathcal{G} |\fvec g_1|^2 \sim \lambda.$
Therefore, when $\lambda \ll 1$, the inverse of the matrix $\mathcal{M} + \mathcal{T}$ can be Taylor expanded as
\begin{align}
    (\mathcal{M} + \mathcal{T})^{-1} & = \mathcal{M}^{-1} - \mathcal{M}^{-1} \mathcal{T} \mathcal{M}^{-1} \nonumber \\
    & + \mathcal{M}^{-1} \mathcal{T} \mathcal{M}^{-1}  \mathcal{T} \mathcal{M}^{-1} + \ldots   \label{Eqn:InverseMpT}
\end{align}
where $\mathcal{M}^{-1}$ can be explicitly expressed as
\begin{align}
    \big( \mathcal{M}^{-1} \big)_{a\mu, b\nu} = \delta_{ab} \left( \frac{\delta_{\mu\nu}}{\mathcal{G}|\fvec g_a|^2}  - \frac{\mathcal{K} g_{a,\mu} g_{a,\nu}}{\mathcal{G} (\mathcal{G} + \mathcal{K}) |\fvec g_a|^4}   \right) \ .
\end{align}

\begin{widetext}
    \begin{figure*}[t] 
	\centering
	\subfigure[\label{FigS:DWSeries:G1xEps02}]{\includegraphics[width=0.98\columnwidth]{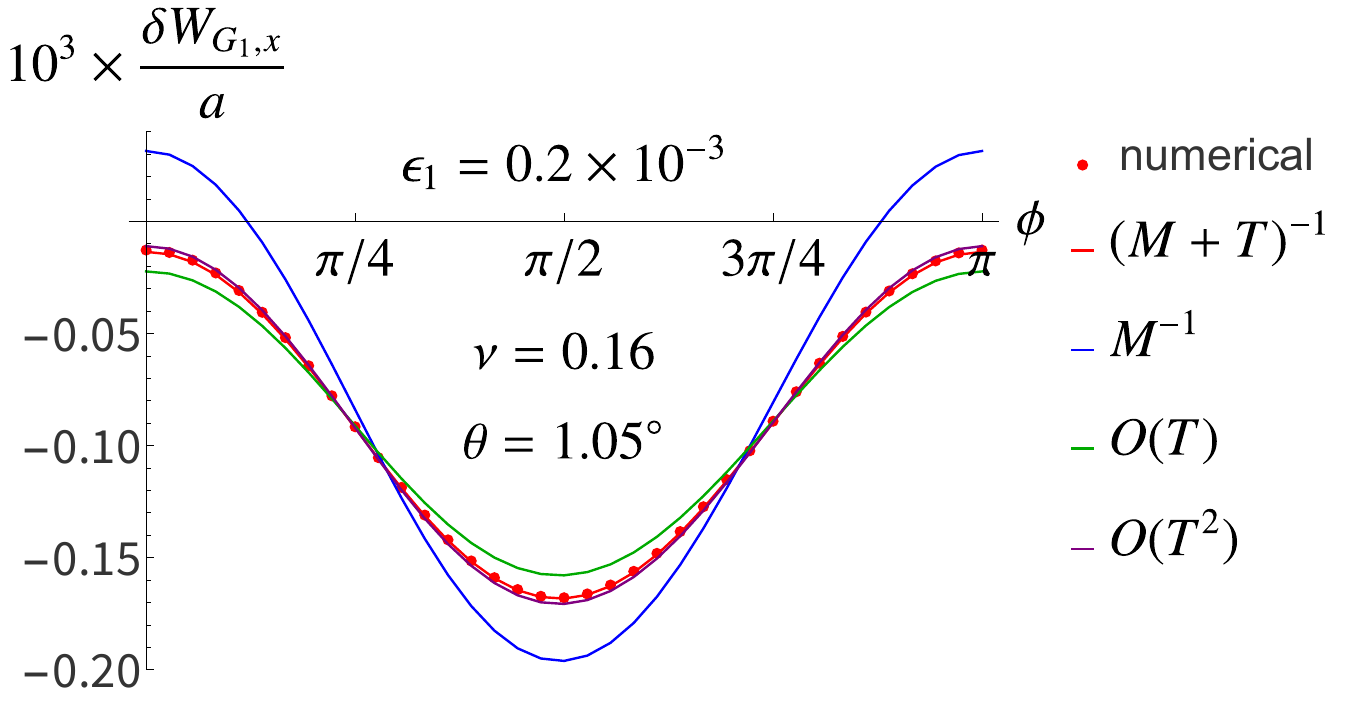}}
    \subfigure[\label{FigS:DWSeries:G1yEps02}]{\includegraphics[width=0.98\columnwidth]{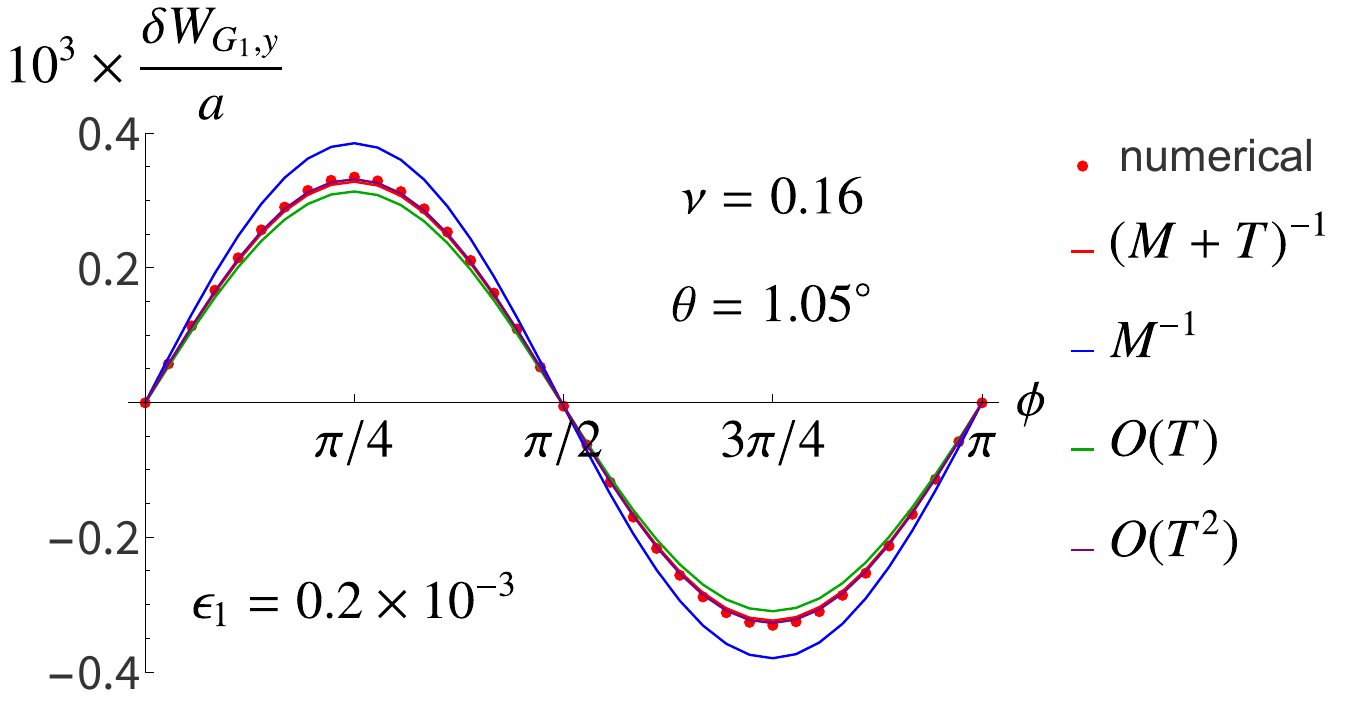}}
    \subfigure[\label{FigS:DWSeries:G1xEps50}]{\includegraphics[width=0.98\columnwidth]{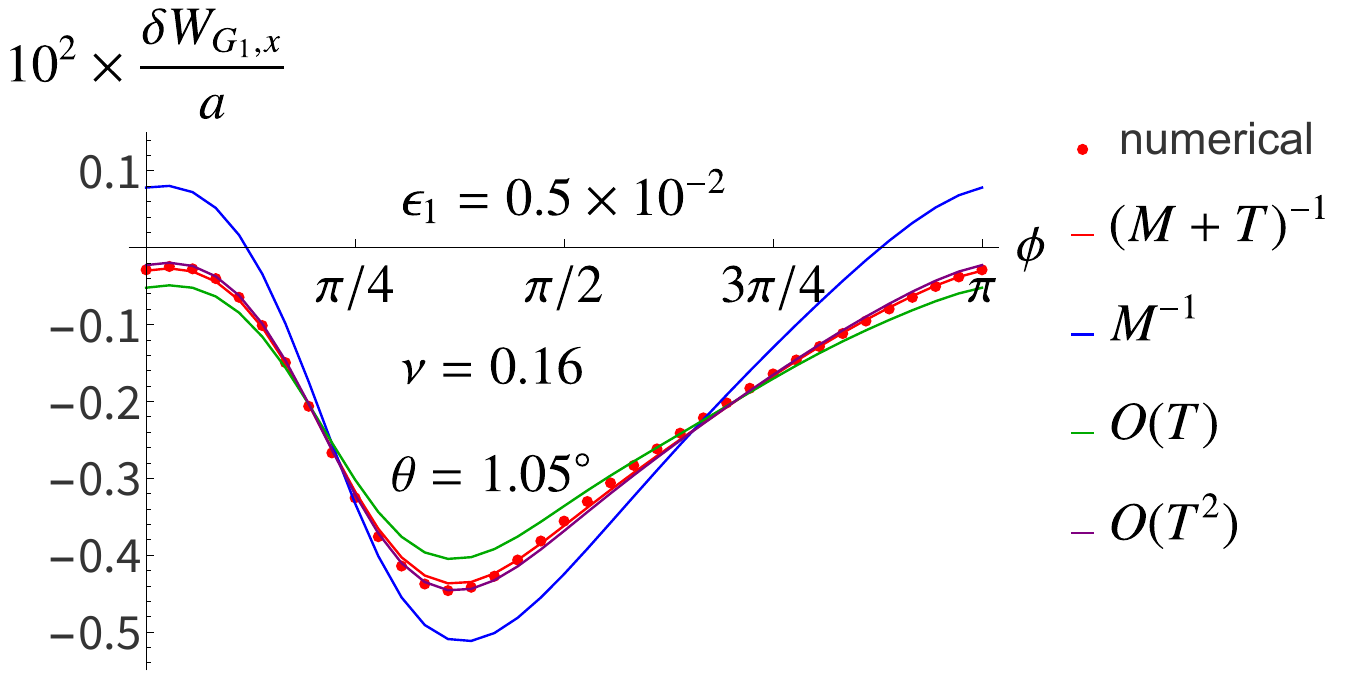}}
    \subfigure[\label{FigS:DWSeries:G1yEps50}]{\includegraphics[width=0.98\columnwidth]{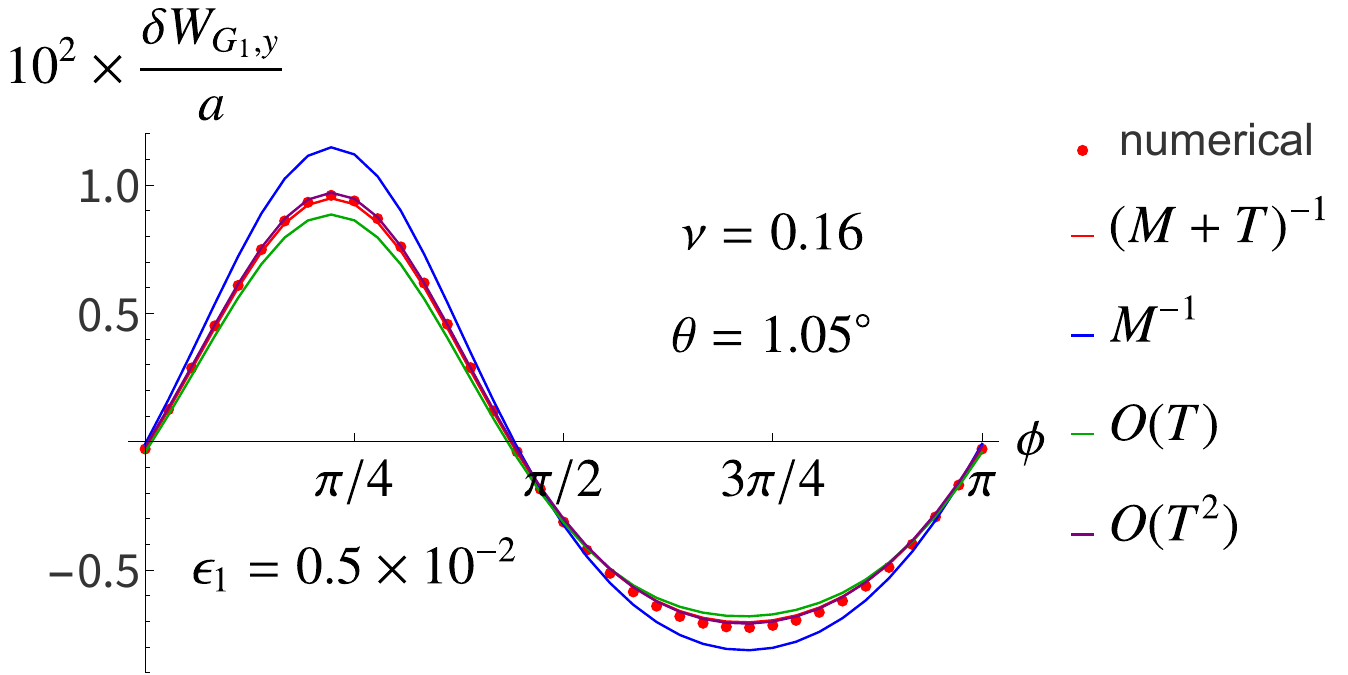}}
    \subfigure[\label{FigS:DWSeries:G1xEps100}]{\includegraphics[width=0.98\columnwidth]{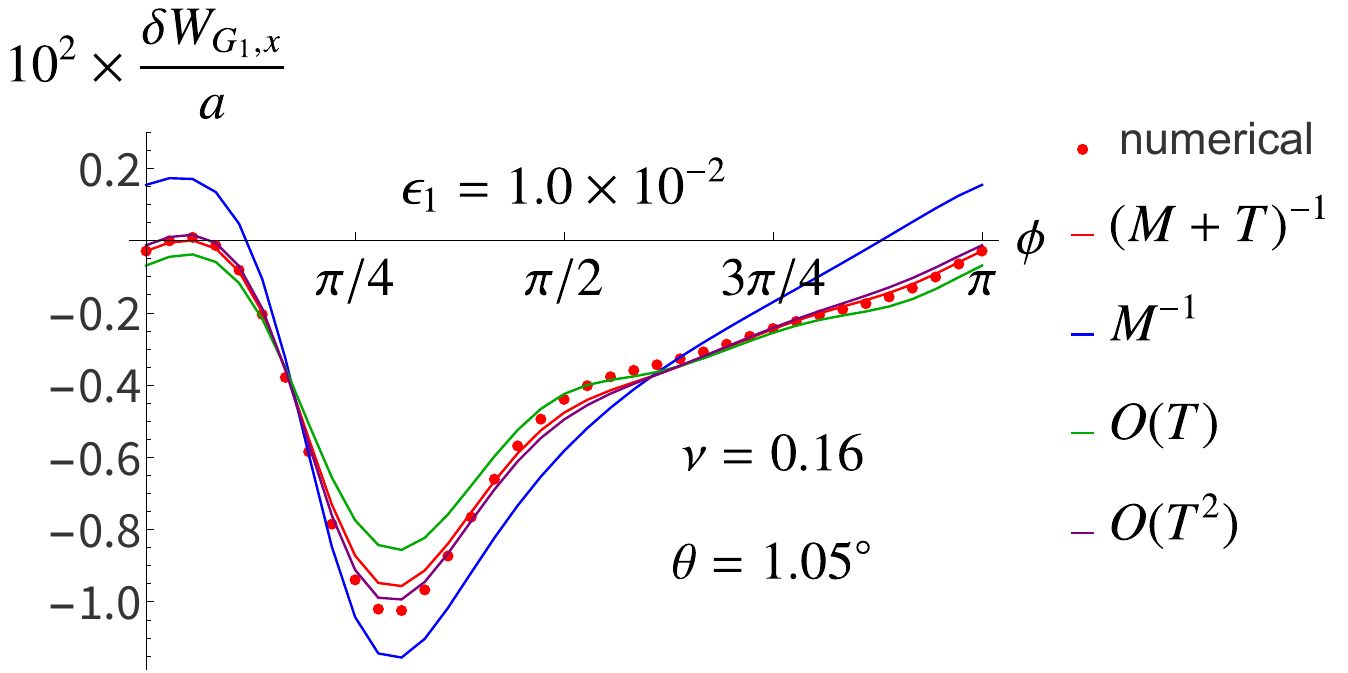}}
    \subfigure[\label{FigS:DWSeries:G1yEps100}]{\includegraphics[width=0.98\columnwidth]{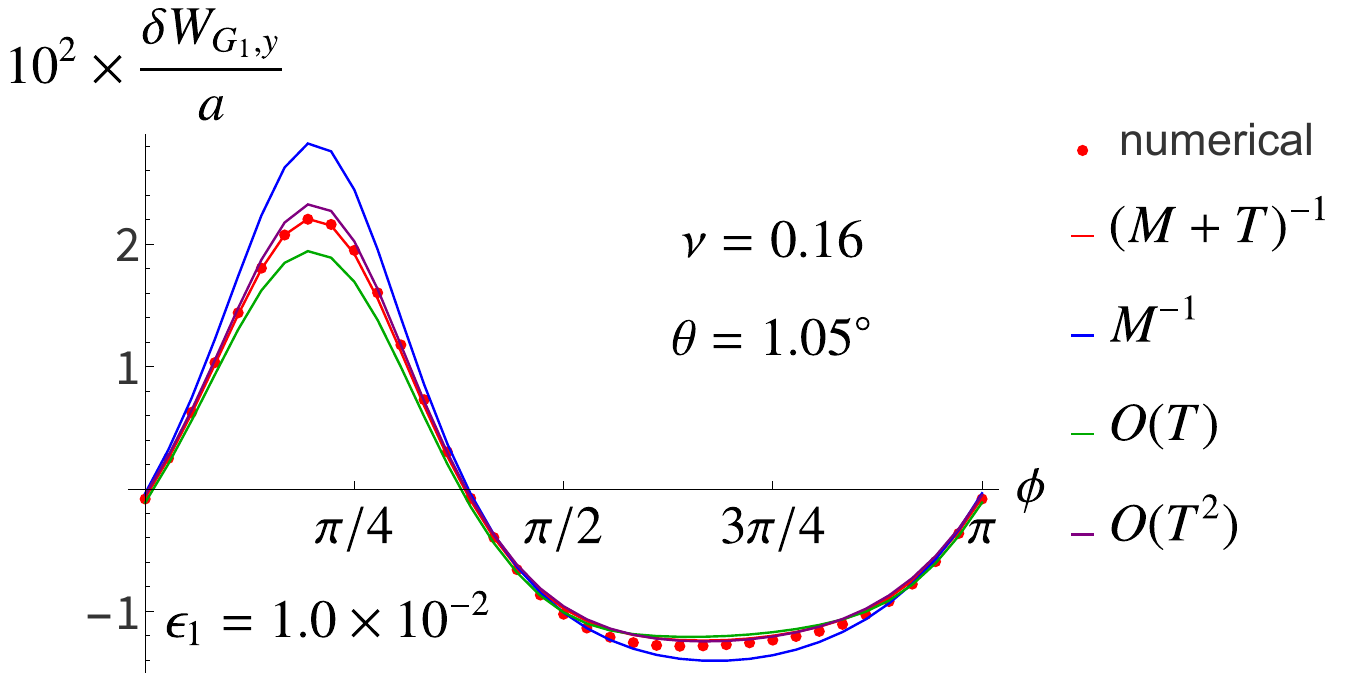}}
	\caption{Similar to Fig.~\ref{Fig:LatRelaxStrain:Appr}, the $x$ component (left) and the $y$ components (right) of $\delta \fvec W_{\fvec G_1}$ vs $\phi$, at the first magic angle $\theta = 1.05^{\circ}$, with $\epsilon_1 = 0.02\%$ (1st row), $\epsilon_1 = 0.5\%$ (2nd row), and $\epsilon_1 = 1\%$  (3rd row). The remaining strain parameter is set to  $\epsilon_2 = -\nu \epsilon_1$ using the Poisson ratio $\nu = 0.16$. Red dots illustrate $\delta \fvec W$ obtained by numerically solving the Eq.~\ref{Eqn:LatticeUqRelax}~\cite{KangLatticeRelax2025}. The solid red curves are the solutions obtained from the Eq.~\ref{Eqn:DWAppr}. Other colored curves give $\delta \fvec W$ obtained by truncating $(\mM + \mT)^{-1}$ in Eq.~\ref{Eqn:DWAppr} at different orders of $\mT$ in the power series expansion (see Eq.~\ref{Eqn:InverseMpT}), showing that the truncation at $\mathcal{O}(\mT^2)$ is almost identical to $(\mM + \mT)^{-1}$. }
	\label{Fig:DWSeries}
\end{figure*}
\end{widetext}

The power series in Eq.~\ref{Eqn:InverseMpT} can be truncated at any order to produce an approximate closed-form expression for $(\mathcal{M} + \mathcal{T})^{-1}$. The accuracy of the truncation order depends on the value of $\lambda$. For $\theta \gtrsim 1.4^{\circ}$, or equivalently $\lambda \lesssim 0.1$, the truncation at $0$th order is highly accurate as shown in the Fig.~\ref{FigS:DWSeriesTwist14}, leading  to 
\begin{align}
   & \delta W_{\fvec G_a,\mu}  \approx - \sum_{\nu} \big( \mM \big)^{-1}_{a\mu, a\nu} u_{a,\nu}= \nonumber \\
    & \frac{\zeta_1}{|\fvec G_1|^2} \left[  \left( \frac{|\fvec g_1^{(0)}|^2}{|\fvec g_a|^2}-1 \right) G_{a, \mu} - \frac{\mK}{\mG + \mK} \frac{|\fvec g_a^{(0)}|^2}{|\fvec g_a|^4} (\delta \fvec g_a \cdot \fvec G_a) g_{a,\mu} \right].
\end{align}
At the first magic angle, the accuracy of different higher order in $\mathcal{T}$ truncation orders is illustrated in the Fig.~\ref{Fig:DWSeries}, where they are compared with the full numerical solution of the Eq.~\ref{Eqn:LatticeUqRelax} as a function of the strain orientation angle $\phi$ for different values of $\epsilon_1$. 


\subsection{Lattice Corrugation}
\label{Sec:LatCorrugation}
In addition to the in-plane relaxation, the atoms of the twisted bilayer graphene also deform along the out-of-plane direction. Such deformation is described by the lattice corrugation field, defined as 
\begin{align}
  h(\fvec x) \hat z= \fvec U^{\perp}_t(\fvec x) - \fvec U^{\perp}_b(\fvec x)   \ ,   \label{Eqn:hCorrug}
\end{align}
where $\fvec U^{\perp}_j$ is the lattice distortion on the layer $j$ along the out-of-plane direction. $h(\fvec x)$ is determined by the local interlayer stacking, achieving the maximum $h_{AA}$ at AA stacking and the minimum $h_{AB}$ at AB stacking~\cite{OshiyamaPRB14}. 

The interlayer stacking is given by the relative in-plane distortion $\fvec U^-$. For AA stacking, $\fvec U^- = n_1 \fvec a_1 + n_2 \fvec a_2$ where $n_{1,2}$ are two arbitrary integers and $\fvec a_{1,2}$, as defined in Eq.~\ref{Eqn:MLGLatVecs}, are the monolayer graphene lattice vectors. For AB/BA stacking, $\fvec U^- =  (n_1 \pm \frac13) \fvec a_1 + ( n_2 \pm \frac13) \fvec a_2$. It is straightforward to write the corrugation field $h$ as a function of the position $\fvec x$ as has been done in Eq.~\ref{Eqn:hCorrug}. However, it is more convenient to express $h$ as a function of $\fvec U^-$, i.e.~$h = h(\fvec U^-)$ because of its dependence on the interlayer stacking. In the rest of this section, we consider various symmetries which constrain the form of $h(\fvec U^-)$ and obtain its approximate formula.  

The corrugation field $h$ depends on the local interlayer stacking that doesn't change under the shift $\fvec U^-(\fvec x) \rightarrow \fvec U^-(\fvec x) + \fvec a_j$ (with $j = 1,2$).  Therefore, $h(\fvec U^-)$ is a periodic function $h(\fvec U^-) = h(\fvec U^- + \fvec a_j)$, that can be expressed via Fourier series as 
\begin{align}
    h(\fvec U^-) = \sum_{\fvec G} h_{\fvec G} e^{i \fvec G \cdot \fvec U^-} \ .   \label{Eqn:hG}
\end{align}
If the interlayer stacking is also invariant under $C_{6z}$ and mirror reflections along $xz$ and $yz$ planes, then the Fourier components $h_{\fvec G}$ are also unchanged under these transformations. Specifically, $h_{\fvec G}$ is identical at each of the six $\fvec G$s on the first shell. Note that $h = h_{AA}$ for AA stacking and $h = h_{AB}$ for AB stacking. Approximating $h(\fvec U^-)$ by keeping $h_{\fvec G}$ in the Eq.~\ref{Eqn:hG} only at $\fvec G=0$ and the first shell, i.e. at $|\fvec G| \leq |\fvec G_1|$, we obtain~\cite{LiangPRX}
\begin{align}
     h(\fvec x) & = \frac13 (h_{AA} + 2 h_{AB})  \nonumber \\
    & + \frac29 (h_{AA} - h_{AB}) \sum_{a  = 1}^3 \cos\big( \fvec G_a \cdot \fvec U^-(\fvec x) \big)
\end{align}
where the layer antisymmetric in-plane distortion $\fvec U^-$ is given by the Eq.~\ref{Eq:Udef}.

\section{summary}
In this work, we have compared the lattice relaxation from two models in Ref.~\cite{KoshinoPRB17,*KoshinoPRB17Erratum} and \cite{KaxirasPRB18} with the experimental results obtained in Ref.~\cite{BediakoNM21}, and found that the latter model (Ref.~\cite{KaxirasPRB18}) follows the experimental data much more closely. Using this model we then derived a closed form expression for the lattice relaxation. In the absence of external heterostrain, we found our analytical expression for the lattice relaxation which is highly accurate for the twist angles down to $\sim 0.7^{\circ}$. 
Introducing the dimensionless parameter $\lambda$ that quantifies the ratio between the interlayer adhesion energy and the elastic energy at a given twist angle, we also found that when the lattice relaxation is expanded in the Taylor series in $\lambda$, it becomes divergent at a value of $\lambda$ which translates to the twist angle slightly above the magic value for realistic models of the elastic energy and the interlayer adhesion energy. We related this divergence to the existence of the branch points of the solution in the complex $\lambda$ plane, and numerically determined the radius of convergence of the Taylor series with high precision.

We also derived a closed form expression for the strain-induced change of the lattice relaxation. This formula is accurate at least up to $1\%$ heterostrain. It can be used to understand the electronic spectrum of the twisted bilayer graphene in the presence of realistic heterostrain. Incorporating these results into the electronic continuum model will be a subject of an upcoming paper.

\acknowledgments
J.~K.~acknowledges the support from the NSFC Grant No.~12074276, the Double First-Class Initiative Fund of ShanghaiTech University, and the start-up grant of ShanghaiTech University. O.V. was funded in part by the Gordon and Betty Moore Foundation's EPiQS Initiative Grant GBMF11070 and acknowledges support from the National High Magnetic Field Laboratory funded by the National Science Foundation (Grant No. DMR-2128556) and the State of Florida.

\bibliography{LatticeRelaxation} 

\begin{thebibliography}{36}%
\makeatletter
\providecommand \@ifxundefined [1]{%
 \@ifx{#1\undefined}
}%
\providecommand \@ifnum [1]{%
 \ifnum #1\expandafter \@firstoftwo
 \else \expandafter \@secondoftwo
 \fi
}%
\providecommand \@ifx [1]{%
 \ifx #1\expandafter \@firstoftwo
 \else \expandafter \@secondoftwo
 \fi
}%
\providecommand \natexlab [1]{#1}%
\providecommand \enquote  [1]{``#1''}%
\providecommand \bibnamefont  [1]{#1}%
\providecommand \bibfnamefont [1]{#1}%
\providecommand \citenamefont [1]{#1}%
\providecommand \href@noop [0]{\@secondoftwo}%
\providecommand \href [0]{\begingroup \@sanitize@url \@href}%
\providecommand \@href[1]{\@@startlink{#1}\@@href}%
\providecommand \@@href[1]{\endgroup#1\@@endlink}%
\providecommand \@sanitize@url [0]{\catcode `\\12\catcode `\$12\catcode
  `\&12\catcode `\#12\catcode `\^12\catcode `\_12\catcode `\%12\relax}%
\providecommand \@@startlink[1]{}%
\providecommand \@@endlink[0]{}%
\providecommand \url  [0]{\begingroup\@sanitize@url \@url }%
\providecommand \@url [1]{\endgroup\@href {#1}{\urlprefix }}%
\providecommand \urlprefix  [0]{URL }%
\providecommand \Eprint [0]{\href }%
\providecommand \doibase [0]{https://doi.org/}%
\providecommand \selectlanguage [0]{\@gobble}%
\providecommand \bibinfo  [0]{\@secondoftwo}%
\providecommand \bibfield  [0]{\@secondoftwo}%
\providecommand \translation [1]{[#1]}%
\providecommand \BibitemOpen [0]{}%
\providecommand \bibitemStop [0]{}%
\providecommand \bibitemNoStop [0]{.\EOS\space}%
\providecommand \EOS [0]{\spacefactor3000\relax}%
\providecommand \BibitemShut  [1]{\csname bibitem#1\endcsname}%
\let\auto@bib@innerbib\@empty
\bibitem [{\citenamefont {Cao}\ \emph {et~al.}(2018{\natexlab{a}})\citenamefont
  {Cao}, \citenamefont {Fatemi}, \citenamefont {Demir}, \citenamefont {Fang},
  \citenamefont {Tomarken}, \citenamefont {Luo}, \citenamefont
  {Sanchez-Yamagishi}, \citenamefont {Watanabe}, \citenamefont {Taniguchi},
  \citenamefont {Kaxiras}, \citenamefont {Ashoori},\ and\ \citenamefont
  {Jarillo-Herrero}}]{Cao2018Insulator}%
  \BibitemOpen
  \bibfield  {author} {\bibinfo {author} {\bibfnamefont {Y.}~\bibnamefont
  {Cao}}, \bibinfo {author} {\bibfnamefont {V.}~\bibnamefont {Fatemi}},
  \bibinfo {author} {\bibfnamefont {A.}~\bibnamefont {Demir}}, \bibinfo
  {author} {\bibfnamefont {S.}~\bibnamefont {Fang}}, \bibinfo {author}
  {\bibfnamefont {S.~L.}\ \bibnamefont {Tomarken}}, \bibinfo {author}
  {\bibfnamefont {J.~Y.}\ \bibnamefont {Luo}}, \bibinfo {author} {\bibfnamefont
  {J.~D.}\ \bibnamefont {Sanchez-Yamagishi}}, \bibinfo {author} {\bibfnamefont
  {K.}~\bibnamefont {Watanabe}}, \bibinfo {author} {\bibfnamefont
  {T.}~\bibnamefont {Taniguchi}}, \bibinfo {author} {\bibfnamefont
  {E.}~\bibnamefont {Kaxiras}}, \bibinfo {author} {\bibfnamefont {R.~C.}\
  \bibnamefont {Ashoori}},\ and\ \bibinfo {author} {\bibfnamefont
  {P.}~\bibnamefont {Jarillo-Herrero}},\ }\bibfield  {title} {\bibinfo {title}
  {Correlated insulator behaviour at half-filling in magic-angle graphene
  superlattices},\ }\href {https://doi.org/10.1038/nature26154} {\bibfield
  {journal} {\bibinfo  {journal} {Nature}\ }\textbf {\bibinfo {volume} {556}},\
  \bibinfo {pages} {80} (\bibinfo {year} {2018}{\natexlab{a}})}\BibitemShut
  {NoStop}%
\bibitem [{\citenamefont {Cao}\ \emph {et~al.}(2018{\natexlab{b}})\citenamefont
  {Cao}, \citenamefont {Fatemi}, \citenamefont {Fang}, \citenamefont
  {Watanabe}, \citenamefont {Taniguchi}, \citenamefont {Kaxiras},\ and\
  \citenamefont {Jarillo-Herrero}}]{Cao2018SC}%
  \BibitemOpen
  \bibfield  {author} {\bibinfo {author} {\bibfnamefont {Y.}~\bibnamefont
  {Cao}}, \bibinfo {author} {\bibfnamefont {V.}~\bibnamefont {Fatemi}},
  \bibinfo {author} {\bibfnamefont {S.}~\bibnamefont {Fang}}, \bibinfo {author}
  {\bibfnamefont {K.}~\bibnamefont {Watanabe}}, \bibinfo {author}
  {\bibfnamefont {T.}~\bibnamefont {Taniguchi}}, \bibinfo {author}
  {\bibfnamefont {E.}~\bibnamefont {Kaxiras}},\ and\ \bibinfo {author}
  {\bibfnamefont {P.}~\bibnamefont {Jarillo-Herrero}},\ }\bibfield  {title}
  {\bibinfo {title} {Unconventional superconductivity in magic-angle graphene
  superlattices},\ }\href {https://doi.org/10.1038/nature26160} {\bibfield
  {journal} {\bibinfo  {journal} {Nature}\ }\textbf {\bibinfo {volume} {556}},\
  \bibinfo {pages} {43} (\bibinfo {year} {2018}{\natexlab{b}})}\BibitemShut
  {NoStop}%
\bibitem [{\citenamefont {Bistritzer}\ and\ \citenamefont
  {MacDonald}(2011)}]{BMModel}%
  \BibitemOpen
  \bibfield  {author} {\bibinfo {author} {\bibfnamefont {R.}~\bibnamefont
  {Bistritzer}}\ and\ \bibinfo {author} {\bibfnamefont {A.~H.}\ \bibnamefont
  {MacDonald}},\ }\bibfield  {title} {\bibinfo {title} {Moire bands in twisted
  double-layer graphene},\ }\href {https://doi.org/10.1073/pnas.1108174108}
  {\bibfield  {journal} {\bibinfo  {journal} {Proc. Natl. Acad. Sci. USA}\
  }\textbf {\bibinfo {volume} {108}},\ \bibinfo {pages} {12233} (\bibinfo
  {year} {2011})},\ \Eprint
  {https://arxiv.org/abs/https://www.pnas.org/doi/pdf/10.1073/pnas.1108174108}
  {https://www.pnas.org/doi/pdf/10.1073/pnas.1108174108} \BibitemShut {NoStop}%
\bibitem [{\citenamefont {Fang}\ and\ \citenamefont
  {Kaxiras}(2016)}]{KaxirasTBM16}%
  \BibitemOpen
  \bibfield  {author} {\bibinfo {author} {\bibfnamefont {S.}~\bibnamefont
  {Fang}}\ and\ \bibinfo {author} {\bibfnamefont {E.}~\bibnamefont {Kaxiras}},\
  }\bibfield  {title} {\bibinfo {title} {Electronic structure theory of weakly
  interacting bilayers},\ }\href {https://doi.org/10.1103/PhysRevB.93.235153}
  {\bibfield  {journal} {\bibinfo  {journal} {Phys. Rev. B}\ }\textbf {\bibinfo
  {volume} {93}},\ \bibinfo {pages} {235153} (\bibinfo {year}
  {2016})}\BibitemShut {NoStop}%
\bibitem [{\citenamefont {Carr}\ \emph
  {et~al.}(2018{\natexlab{a}})\citenamefont {Carr}, \citenamefont {Massatt},
  \citenamefont {Torrisi}, \citenamefont {Cazeaux}, \citenamefont {Luskin},\
  and\ \citenamefont {Kaxiras}}]{KaxirasPRB18}%
  \BibitemOpen
  \bibfield  {author} {\bibinfo {author} {\bibfnamefont {S.}~\bibnamefont
  {Carr}}, \bibinfo {author} {\bibfnamefont {D.}~\bibnamefont {Massatt}},
  \bibinfo {author} {\bibfnamefont {S.~B.}\ \bibnamefont {Torrisi}}, \bibinfo
  {author} {\bibfnamefont {P.}~\bibnamefont {Cazeaux}}, \bibinfo {author}
  {\bibfnamefont {M.}~\bibnamefont {Luskin}},\ and\ \bibinfo {author}
  {\bibfnamefont {E.}~\bibnamefont {Kaxiras}},\ }\bibfield  {title} {\bibinfo
  {title} {Relaxation and domain formation in incommensurate two-dimensional
  heterostructures},\ }\href {https://doi.org/10.1103/PhysRevB.98.224102}
  {\bibfield  {journal} {\bibinfo  {journal} {Phys. Rev. B}\ }\textbf {\bibinfo
  {volume} {98}},\ \bibinfo {pages} {224102} (\bibinfo {year}
  {2018}{\natexlab{a}})}\BibitemShut {NoStop}%
\bibitem [{\citenamefont {Koshino}\ \emph {et~al.}(2018)\citenamefont
  {Koshino}, \citenamefont {Yuan}, \citenamefont {Koretsune}, \citenamefont
  {Ochi}, \citenamefont {Kuroki},\ and\ \citenamefont {Fu}}]{LiangPRX}%
  \BibitemOpen
  \bibfield  {author} {\bibinfo {author} {\bibfnamefont {M.}~\bibnamefont
  {Koshino}}, \bibinfo {author} {\bibfnamefont {N.~F.~Q.}\ \bibnamefont
  {Yuan}}, \bibinfo {author} {\bibfnamefont {T.}~\bibnamefont {Koretsune}},
  \bibinfo {author} {\bibfnamefont {M.}~\bibnamefont {Ochi}}, \bibinfo {author}
  {\bibfnamefont {K.}~\bibnamefont {Kuroki}},\ and\ \bibinfo {author}
  {\bibfnamefont {L.}~\bibnamefont {Fu}},\ }\bibfield  {title} {\bibinfo
  {title} {Maximally localized wannier orbitals and the extended hubbard model
  for twisted bilayer graphene},\ }\href
  {https://doi.org/10.1103/PhysRevX.8.031087} {\bibfield  {journal} {\bibinfo
  {journal} {Phys. Rev. X}\ }\textbf {\bibinfo {volume} {8}},\ \bibinfo {pages}
  {031087} (\bibinfo {year} {2018})}\BibitemShut {NoStop}%
\bibitem [{\citenamefont {Bi}\ \emph {et~al.}(2019)\citenamefont {Bi},
  \citenamefont {Yuan},\ and\ \citenamefont {Fu}}]{LiangStrain}%
  \BibitemOpen
  \bibfield  {author} {\bibinfo {author} {\bibfnamefont {Z.}~\bibnamefont
  {Bi}}, \bibinfo {author} {\bibfnamefont {N.~F.~Q.}\ \bibnamefont {Yuan}},\
  and\ \bibinfo {author} {\bibfnamefont {L.}~\bibnamefont {Fu}},\ }\bibfield
  {title} {\bibinfo {title} {Designing flat bands by strain},\ }\href
  {https://doi.org/10.1103/PhysRevB.100.035448} {\bibfield  {journal} {\bibinfo
   {journal} {Phys. Rev. B}\ }\textbf {\bibinfo {volume} {100}},\ \bibinfo
  {pages} {035448} (\bibinfo {year} {2019})}\BibitemShut {NoStop}%
\bibitem [{\citenamefont {Kang}\ and\ \citenamefont {Vafek}(2018)}]{KangPRX18}%
  \BibitemOpen
  \bibfield  {author} {\bibinfo {author} {\bibfnamefont {J.}~\bibnamefont
  {Kang}}\ and\ \bibinfo {author} {\bibfnamefont {O.}~\bibnamefont {Vafek}},\
  }\bibfield  {title} {\bibinfo {title} {Symmetry, maximally localized wannier
  states, and a low-energy model for twisted bilayer graphene narrow bands},\
  }\href {https://doi.org/10.1103/PhysRevX.8.031088} {\bibfield  {journal}
  {\bibinfo  {journal} {Phys. Rev. X}\ }\textbf {\bibinfo {volume} {8}},\
  \bibinfo {pages} {031088} (\bibinfo {year} {2018})}\BibitemShut {NoStop}%
\bibitem [{\citenamefont {Kang}\ and\ \citenamefont {Vafek}(2023)}]{JKPRB23}%
  \BibitemOpen
  \bibfield  {author} {\bibinfo {author} {\bibfnamefont {J.}~\bibnamefont
  {Kang}}\ and\ \bibinfo {author} {\bibfnamefont {O.}~\bibnamefont {Vafek}},\
  }\bibfield  {title} {\bibinfo {title} {Pseudomagnetic fields, particle-hole
  asymmetry, and microscopic effective continuum hamiltonians of twisted
  bilayer graphene},\ }\href {https://doi.org/10.1103/PhysRevB.107.075408}
  {\bibfield  {journal} {\bibinfo  {journal} {Phys. Rev. B}\ }\textbf {\bibinfo
  {volume} {107}},\ \bibinfo {pages} {075408} (\bibinfo {year}
  {2023})}\BibitemShut {NoStop}%
\bibitem [{\citenamefont {Carr}\ \emph
  {et~al.}(2018{\natexlab{b}})\citenamefont {Carr}, \citenamefont {Massatt},
  \citenamefont {Torrisi}, \citenamefont {Cazeaux}, \citenamefont {Luskin},\
  and\ \citenamefont {Kaxiras}}]{LuskinPRB22}%
  \BibitemOpen
  \bibfield  {author} {\bibinfo {author} {\bibfnamefont {S.}~\bibnamefont
  {Carr}}, \bibinfo {author} {\bibfnamefont {D.}~\bibnamefont {Massatt}},
  \bibinfo {author} {\bibfnamefont {S.~B.}\ \bibnamefont {Torrisi}}, \bibinfo
  {author} {\bibfnamefont {P.}~\bibnamefont {Cazeaux}}, \bibinfo {author}
  {\bibfnamefont {M.}~\bibnamefont {Luskin}},\ and\ \bibinfo {author}
  {\bibfnamefont {E.}~\bibnamefont {Kaxiras}},\ }\bibfield  {title} {\bibinfo
  {title} {Relaxation and domain formation in incommensurate two-dimensional
  heterostructures},\ }\href {https://doi.org/10.1103/PhysRevB.98.224102}
  {\bibfield  {journal} {\bibinfo  {journal} {Phys. Rev. B}\ }\textbf {\bibinfo
  {volume} {98}},\ \bibinfo {pages} {224102} (\bibinfo {year}
  {2018}{\natexlab{b}})}\BibitemShut {NoStop}%
\bibitem [{\citenamefont {Nam}\ and\ \citenamefont
  {Koshino}(2017)}]{KoshinoPRB17}%
  \BibitemOpen
  \bibfield  {author} {\bibinfo {author} {\bibfnamefont {N.~N.~T.}\
  \bibnamefont {Nam}}\ and\ \bibinfo {author} {\bibfnamefont {M.}~\bibnamefont
  {Koshino}},\ }\bibfield  {title} {\bibinfo {title} {Lattice relaxation and
  energy band modulation in twisted bilayer graphene},\ }\href
  {https://doi.org/10.1103/PhysRevB.96.075311} {\bibfield  {journal} {\bibinfo
  {journal} {Phys. Rev. B}\ }\textbf {\bibinfo {volume} {96}},\ \bibinfo
  {pages} {075311} (\bibinfo {year} {2017})}\BibitemShut {NoStop}%
\bibitem [{\citenamefont {Nam}\ and\ \citenamefont
  {Koshino}(2020)}]{KoshinoPRB17Erratum}%
  \BibitemOpen
  \bibfield  {author} {\bibinfo {author} {\bibfnamefont {N.~N.~T.}\
  \bibnamefont {Nam}}\ and\ \bibinfo {author} {\bibfnamefont {M.}~\bibnamefont
  {Koshino}},\ }\bibfield  {title} {\bibinfo {title} {Erratum: Lattice
  relaxation and energy band modulation in twisted bilayer graphene [phys. rev.
  b 96, 075311 (2017)]},\ }\href {https://doi.org/10.1103/PhysRevB.101.099901}
  {\bibfield  {journal} {\bibinfo  {journal} {Phys. Rev. B}\ }\textbf {\bibinfo
  {volume} {101}},\ \bibinfo {pages} {099901} (\bibinfo {year}
  {2020})}\BibitemShut {NoStop}%
\bibitem [{\citenamefont {Kwan}\ \emph {et~al.}(2021)\citenamefont {Kwan},
  \citenamefont {Wagner}, \citenamefont {Soejima}, \citenamefont {Zaletel},
  \citenamefont {Simon}, \citenamefont {Parameswaran},\ and\ \citenamefont
  {Bultinck}}]{NickIKSPRX}%
  \BibitemOpen
  \bibfield  {author} {\bibinfo {author} {\bibfnamefont {Y.~H.}\ \bibnamefont
  {Kwan}}, \bibinfo {author} {\bibfnamefont {G.}~\bibnamefont {Wagner}},
  \bibinfo {author} {\bibfnamefont {T.}~\bibnamefont {Soejima}}, \bibinfo
  {author} {\bibfnamefont {M.~P.}\ \bibnamefont {Zaletel}}, \bibinfo {author}
  {\bibfnamefont {S.~H.}\ \bibnamefont {Simon}}, \bibinfo {author}
  {\bibfnamefont {S.~A.}\ \bibnamefont {Parameswaran}},\ and\ \bibinfo {author}
  {\bibfnamefont {N.}~\bibnamefont {Bultinck}},\ }\bibfield  {title} {\bibinfo
  {title} {Kekul\'e spiral order at all nonzero integer fillings in twisted
  bilayer graphene},\ }\href {https://doi.org/10.1103/PhysRevX.11.041063}
  {\bibfield  {journal} {\bibinfo  {journal} {Phys. Rev. X}\ }\textbf {\bibinfo
  {volume} {11}},\ \bibinfo {pages} {041063} (\bibinfo {year}
  {2021})}\BibitemShut {NoStop}%
\bibitem [{\citenamefont {Parker}\ \emph {et~al.}(2021)\citenamefont {Parker},
  \citenamefont {Soejima}, \citenamefont {Hauschild}, \citenamefont {Zaletel},\
  and\ \citenamefont {Bultinck}}]{NickStrainCNPPRL}%
  \BibitemOpen
  \bibfield  {author} {\bibinfo {author} {\bibfnamefont {D.~E.}\ \bibnamefont
  {Parker}}, \bibinfo {author} {\bibfnamefont {T.}~\bibnamefont {Soejima}},
  \bibinfo {author} {\bibfnamefont {J.}~\bibnamefont {Hauschild}}, \bibinfo
  {author} {\bibfnamefont {M.~P.}\ \bibnamefont {Zaletel}},\ and\ \bibinfo
  {author} {\bibfnamefont {N.}~\bibnamefont {Bultinck}},\ }\bibfield  {title}
  {\bibinfo {title} {Strain-induced quantum phase transitions in magic-angle
  graphene},\ }\href {https://doi.org/10.1103/PhysRevLett.127.027601}
  {\bibfield  {journal} {\bibinfo  {journal} {Phys. Rev. Lett.}\ }\textbf
  {\bibinfo {volume} {127}},\ \bibinfo {pages} {027601} (\bibinfo {year}
  {2021})}\BibitemShut {NoStop}%
\bibitem [{\citenamefont {Wang}\ \emph
  {et~al.}(2023{\natexlab{a}})\citenamefont {Wang}, \citenamefont {Finney},
  \citenamefont {Sharpe}, \citenamefont {Rodenbach}, \citenamefont {Hsueh},
  \citenamefont {Watanabe}, \citenamefont {Taniguchi}, \citenamefont {Kastner},
  \citenamefont {Vafek},\ and\ \citenamefont {Goldhaber-Gordon}}]{XiaoyuPNAS}%
  \BibitemOpen
  \bibfield  {author} {\bibinfo {author} {\bibfnamefont {X.}~\bibnamefont
  {Wang}}, \bibinfo {author} {\bibfnamefont {J.}~\bibnamefont {Finney}},
  \bibinfo {author} {\bibfnamefont {A.~L.}\ \bibnamefont {Sharpe}}, \bibinfo
  {author} {\bibfnamefont {L.~K.}\ \bibnamefont {Rodenbach}}, \bibinfo {author}
  {\bibfnamefont {C.~L.}\ \bibnamefont {Hsueh}}, \bibinfo {author}
  {\bibfnamefont {K.}~\bibnamefont {Watanabe}}, \bibinfo {author}
  {\bibfnamefont {T.}~\bibnamefont {Taniguchi}}, \bibinfo {author}
  {\bibfnamefont {M.~A.}\ \bibnamefont {Kastner}}, \bibinfo {author}
  {\bibfnamefont {O.}~\bibnamefont {Vafek}},\ and\ \bibinfo {author}
  {\bibfnamefont {D.}~\bibnamefont {Goldhaber-Gordon}},\ }\bibfield  {title}
  {\bibinfo {title} {Unusual magnetotransport in twisted bilayer graphene from
  strain-induced open fermi surfaces},\ }\href
  {https://doi.org/10.1073/pnas.2307151120} {\bibfield  {journal} {\bibinfo
  {journal} {Proc. Natl. Acad. Sci. USA}\ }\textbf {\bibinfo {volume} {120}},\
  \bibinfo {pages} {e2307151120} (\bibinfo {year} {2023}{\natexlab{a}})},\
  \Eprint
  {https://arxiv.org/abs/https://www.pnas.org/doi/pdf/10.1073/pnas.2307151120}
  {https://www.pnas.org/doi/pdf/10.1073/pnas.2307151120} \BibitemShut {NoStop}%
\bibitem [{\citenamefont {Huder}\ \emph {et~al.}(2018)\citenamefont {Huder},
  \citenamefont {Artaud}, \citenamefont {Le~Quang}, \citenamefont
  {de~Laissardi\`ere}, \citenamefont {Jansen}, \citenamefont {Lapertot},
  \citenamefont {Chapelier},\ and\ \citenamefont {Renard}}]{RenardPRL18}%
  \BibitemOpen
  \bibfield  {author} {\bibinfo {author} {\bibfnamefont {L.}~\bibnamefont
  {Huder}}, \bibinfo {author} {\bibfnamefont {A.}~\bibnamefont {Artaud}},
  \bibinfo {author} {\bibfnamefont {T.}~\bibnamefont {Le~Quang}}, \bibinfo
  {author} {\bibfnamefont {G.~T.}\ \bibnamefont {de~Laissardi\`ere}}, \bibinfo
  {author} {\bibfnamefont {A.~G.~M.}\ \bibnamefont {Jansen}}, \bibinfo {author}
  {\bibfnamefont {G.}~\bibnamefont {Lapertot}}, \bibinfo {author}
  {\bibfnamefont {C.}~\bibnamefont {Chapelier}},\ and\ \bibinfo {author}
  {\bibfnamefont {V.~T.}\ \bibnamefont {Renard}},\ }\bibfield  {title}
  {\bibinfo {title} {Electronic spectrum of twisted graphene layers under
  heterostrain},\ }\href {https://doi.org/10.1103/PhysRevLett.120.156405}
  {\bibfield  {journal} {\bibinfo  {journal} {Phys. Rev. Lett.}\ }\textbf
  {\bibinfo {volume} {120}},\ \bibinfo {pages} {156405} (\bibinfo {year}
  {2018})}\BibitemShut {NoStop}%
\bibitem [{\citenamefont {Kerelsky}\ \emph {et~al.}(2019)\citenamefont
  {Kerelsky}, \citenamefont {McGilly}, \citenamefont {Kennes}, \citenamefont
  {Xian}, \citenamefont {Yankowitz}, \citenamefont {Chen}, \citenamefont
  {Watanabe}, \citenamefont {Taniguchi}, \citenamefont {Hone}, \citenamefont
  {Dean}, \citenamefont {Rubio},\ and\ \citenamefont
  {Pasupathy}}]{Pasupathy19}%
  \BibitemOpen
  \bibfield  {author} {\bibinfo {author} {\bibfnamefont {A.}~\bibnamefont
  {Kerelsky}}, \bibinfo {author} {\bibfnamefont {L.~J.}\ \bibnamefont
  {McGilly}}, \bibinfo {author} {\bibfnamefont {D.~M.}\ \bibnamefont {Kennes}},
  \bibinfo {author} {\bibfnamefont {L.}~\bibnamefont {Xian}}, \bibinfo {author}
  {\bibfnamefont {M.}~\bibnamefont {Yankowitz}}, \bibinfo {author}
  {\bibfnamefont {S.}~\bibnamefont {Chen}}, \bibinfo {author} {\bibfnamefont
  {K.}~\bibnamefont {Watanabe}}, \bibinfo {author} {\bibfnamefont
  {T.}~\bibnamefont {Taniguchi}}, \bibinfo {author} {\bibfnamefont
  {J.}~\bibnamefont {Hone}}, \bibinfo {author} {\bibfnamefont {C.}~\bibnamefont
  {Dean}}, \bibinfo {author} {\bibfnamefont {A.}~\bibnamefont {Rubio}},\ and\
  \bibinfo {author} {\bibfnamefont {A.~N.}\ \bibnamefont {Pasupathy}},\
  }\bibfield  {title} {\bibinfo {title} {Maximized electron interactions at the
  magic angle in twisted bilayer graphene},\ }\href
  {https://doi.org/10.1038/s41586-019-1431-9} {\bibfield  {journal} {\bibinfo
  {journal} {Nature}\ }\textbf {\bibinfo {volume} {572}},\ \bibinfo {pages}
  {95} (\bibinfo {year} {2019})}\BibitemShut {NoStop}%
\bibitem [{\citenamefont {Xie}\ \emph {et~al.}(2019)\citenamefont {Xie},
  \citenamefont {Lian}, \citenamefont {J{\"a}ck}, \citenamefont {Liu},
  \citenamefont {Chiu}, \citenamefont {Watanabe}, \citenamefont {Taniguchi},
  \citenamefont {Bernevig},\ and\ \citenamefont {Yazdani}}]{XieNature2019}%
  \BibitemOpen
  \bibfield  {author} {\bibinfo {author} {\bibfnamefont {Y.}~\bibnamefont
  {Xie}}, \bibinfo {author} {\bibfnamefont {B.}~\bibnamefont {Lian}}, \bibinfo
  {author} {\bibfnamefont {B.}~\bibnamefont {J{\"a}ck}}, \bibinfo {author}
  {\bibfnamefont {X.}~\bibnamefont {Liu}}, \bibinfo {author} {\bibfnamefont
  {C.-L.}\ \bibnamefont {Chiu}}, \bibinfo {author} {\bibfnamefont
  {K.}~\bibnamefont {Watanabe}}, \bibinfo {author} {\bibfnamefont
  {T.}~\bibnamefont {Taniguchi}}, \bibinfo {author} {\bibfnamefont {B.~A.}\
  \bibnamefont {Bernevig}},\ and\ \bibinfo {author} {\bibfnamefont
  {A.}~\bibnamefont {Yazdani}},\ }\bibfield  {title} {\bibinfo {title}
  {Spectroscopic signatures of many-body correlations in magic-angle twisted
  bilayer graphene},\ }\href {https://doi.org/10.1038/s41586-019-1422-x}
  {\bibfield  {journal} {\bibinfo  {journal} {Nature}\ }\textbf {\bibinfo
  {volume} {572}},\ \bibinfo {pages} {101} (\bibinfo {year}
  {2019})}\BibitemShut {NoStop}%
\bibitem [{\citenamefont {Wong}\ \emph {et~al.}(2020)\citenamefont {Wong},
  \citenamefont {Nuckolls}, \citenamefont {Oh}, \citenamefont {Lian},
  \citenamefont {Xie}, \citenamefont {Jeon}, \citenamefont {Watanabe},
  \citenamefont {Taniguchi}, \citenamefont {Bernevig},\ and\ \citenamefont
  {Yazdani}}]{WongNature2020}%
  \BibitemOpen
  \bibfield  {author} {\bibinfo {author} {\bibfnamefont {D.}~\bibnamefont
  {Wong}}, \bibinfo {author} {\bibfnamefont {K.~P.}\ \bibnamefont {Nuckolls}},
  \bibinfo {author} {\bibfnamefont {M.}~\bibnamefont {Oh}}, \bibinfo {author}
  {\bibfnamefont {B.}~\bibnamefont {Lian}}, \bibinfo {author} {\bibfnamefont
  {Y.}~\bibnamefont {Xie}}, \bibinfo {author} {\bibfnamefont {S.}~\bibnamefont
  {Jeon}}, \bibinfo {author} {\bibfnamefont {K.}~\bibnamefont {Watanabe}},
  \bibinfo {author} {\bibfnamefont {T.}~\bibnamefont {Taniguchi}}, \bibinfo
  {author} {\bibfnamefont {B.~A.}\ \bibnamefont {Bernevig}},\ and\ \bibinfo
  {author} {\bibfnamefont {A.}~\bibnamefont {Yazdani}},\ }\bibfield  {title}
  {\bibinfo {title} {Cascade of electronic transitions in magic-angle twisted
  bilayer graphene},\ }\href {https://doi.org/10.1038/s41586-020-2339-0}
  {\bibfield  {journal} {\bibinfo  {journal} {Nature}\ }\textbf {\bibinfo
  {volume} {582}},\ \bibinfo {pages} {198} (\bibinfo {year}
  {2020})}\BibitemShut {NoStop}%
\bibitem [{\citenamefont {Kazmierczak}\ \emph {et~al.}(2021)\citenamefont
  {Kazmierczak}, \citenamefont {Van~Winkle}, \citenamefont {Ophus},
  \citenamefont {Bustillo}, \citenamefont {Carr}, \citenamefont {Brown},
  \citenamefont {Ciston}, \citenamefont {Taniguchi}, \citenamefont {Watanabe},\
  and\ \citenamefont {Bediako}}]{BediakoNM21}%
  \BibitemOpen
  \bibfield  {author} {\bibinfo {author} {\bibfnamefont {N.~P.}\ \bibnamefont
  {Kazmierczak}}, \bibinfo {author} {\bibfnamefont {M.}~\bibnamefont
  {Van~Winkle}}, \bibinfo {author} {\bibfnamefont {C.}~\bibnamefont {Ophus}},
  \bibinfo {author} {\bibfnamefont {K.~C.}\ \bibnamefont {Bustillo}}, \bibinfo
  {author} {\bibfnamefont {S.}~\bibnamefont {Carr}}, \bibinfo {author}
  {\bibfnamefont {H.~G.}\ \bibnamefont {Brown}}, \bibinfo {author}
  {\bibfnamefont {J.}~\bibnamefont {Ciston}}, \bibinfo {author} {\bibfnamefont
  {T.}~\bibnamefont {Taniguchi}}, \bibinfo {author} {\bibfnamefont
  {K.}~\bibnamefont {Watanabe}},\ and\ \bibinfo {author} {\bibfnamefont
  {D.~K.}\ \bibnamefont {Bediako}},\ }\bibfield  {title} {\bibinfo {title}
  {Strain fields in twisted bilayer graphene},\ }\href
  {https://doi.org/10.1038/s41563-021-00973-w} {\bibfield  {journal} {\bibinfo
  {journal} {Nat. Mater.}\ }\textbf {\bibinfo {volume} {20}},\ \bibinfo {pages}
  {956} (\bibinfo {year} {2021})}\BibitemShut {NoStop}%
\bibitem [{\citenamefont {Ezzi}\ \emph {et~al.}(2024)\citenamefont {Ezzi},
  \citenamefont {Pallewela}, \citenamefont {De~Beule}, \citenamefont {Mele},\
  and\ \citenamefont {Adam}}]{ShaffiquePRL24}%
  \BibitemOpen
  \bibfield  {author} {\bibinfo {author} {\bibfnamefont {M.~M.~A.}\
  \bibnamefont {Ezzi}}, \bibinfo {author} {\bibfnamefont {G.~N.}\ \bibnamefont
  {Pallewela}}, \bibinfo {author} {\bibfnamefont {C.}~\bibnamefont {De~Beule}},
  \bibinfo {author} {\bibfnamefont {E.~J.}\ \bibnamefont {Mele}},\ and\
  \bibinfo {author} {\bibfnamefont {S.}~\bibnamefont {Adam}},\ }\bibfield
  {title} {\bibinfo {title} {Analytical model for atomic relaxation in twisted
  moir\'e materials},\ }\href {https://doi.org/10.1103/PhysRevLett.133.266201}
  {\bibfield  {journal} {\bibinfo  {journal} {Phys. Rev. Lett.}\ }\textbf
  {\bibinfo {volume} {133}},\ \bibinfo {pages} {266201} (\bibinfo {year}
  {2024})}\BibitemShut {NoStop}%
\bibitem [{\citenamefont {Kang}\ and\ \citenamefont
  {Vafek}(2025)}]{KangLatticeRelax2025}%
  \BibitemOpen
  \bibfield  {author} {\bibinfo {author} {\bibfnamefont {J.}~\bibnamefont
  {Kang}}\ and\ \bibinfo {author} {\bibfnamefont {O.}~\bibnamefont {Vafek}},\
  }\bibfield  {title} {\bibinfo {title} {Numerical data for lattice
  relaxation},\ }\href {https://doi.org/10.17605/OSF.IO/USVKC}
  {10.17605/OSF.IO/USVKC} (\bibinfo {year} {2025})\BibitemShut {NoStop}%
\bibitem [{\citenamefont {Sinner}\ \emph {et~al.}(2023)\citenamefont {Sinner},
  \citenamefont {Pantale\'on},\ and\ \citenamefont {Guinea}}]{PacoPRL23}%
  \BibitemOpen
  \bibfield  {author} {\bibinfo {author} {\bibfnamefont {A.}~\bibnamefont
  {Sinner}}, \bibinfo {author} {\bibfnamefont {P.~A.}\ \bibnamefont
  {Pantale\'on}},\ and\ \bibinfo {author} {\bibfnamefont {F.}~\bibnamefont
  {Guinea}},\ }\bibfield  {title} {\bibinfo {title} {Strain-induced quasi-1d
  channels in twisted moir\'e lattices},\ }\href
  {https://doi.org/10.1103/PhysRevLett.131.166402} {\bibfield  {journal}
  {\bibinfo  {journal} {Phys. Rev. Lett.}\ }\textbf {\bibinfo {volume} {131}},\
  \bibinfo {pages} {166402} (\bibinfo {year} {2023})}\BibitemShut {NoStop}%
\bibitem [{\citenamefont {Guinea}\ and\ \citenamefont
  {Walet}(2019)}]{Guinea19}%
  \BibitemOpen
  \bibfield  {author} {\bibinfo {author} {\bibfnamefont {F.}~\bibnamefont
  {Guinea}}\ and\ \bibinfo {author} {\bibfnamefont {N.~R.}\ \bibnamefont
  {Walet}},\ }\bibfield  {title} {\bibinfo {title} {Continuum models for
  twisted bilayer graphene: Effect of lattice deformation and hopping
  parameters},\ }\href {https://doi.org/10.1103/PhysRevB.99.205134} {\bibfield
  {journal} {\bibinfo  {journal} {Phys. Rev. B}\ }\textbf {\bibinfo {volume}
  {99}},\ \bibinfo {pages} {205134} (\bibinfo {year} {2019})}\BibitemShut
  {NoStop}%
\bibitem [{\citenamefont {Cantele}\ \emph {et~al.}(2020)\citenamefont
  {Cantele}, \citenamefont {Alf\`e}, \citenamefont {Conte}, \citenamefont
  {Cataudella}, \citenamefont {Ninno},\ and\ \citenamefont
  {Lucignano}}]{Lucignano20}%
  \BibitemOpen
  \bibfield  {author} {\bibinfo {author} {\bibfnamefont {G.}~\bibnamefont
  {Cantele}}, \bibinfo {author} {\bibfnamefont {D.}~\bibnamefont {Alf\`e}},
  \bibinfo {author} {\bibfnamefont {F.}~\bibnamefont {Conte}}, \bibinfo
  {author} {\bibfnamefont {V.}~\bibnamefont {Cataudella}}, \bibinfo {author}
  {\bibfnamefont {D.}~\bibnamefont {Ninno}},\ and\ \bibinfo {author}
  {\bibfnamefont {P.}~\bibnamefont {Lucignano}},\ }\bibfield  {title} {\bibinfo
  {title} {Structural relaxation and low-energy properties of twisted bilayer
  graphene},\ }\href {https://doi.org/10.1103/PhysRevResearch.2.043127}
  {\bibfield  {journal} {\bibinfo  {journal} {Phys. Rev. Res.}\ }\textbf
  {\bibinfo {volume} {2}},\ \bibinfo {pages} {043127} (\bibinfo {year}
  {2020})}\BibitemShut {NoStop}%
\bibitem [{\citenamefont {Leconte}\ \emph {et~al.}(2022)\citenamefont
  {Leconte}, \citenamefont {Javvaji}, \citenamefont {An}, \citenamefont
  {Samudrala},\ and\ \citenamefont {Jung}}]{JungPRB22}%
  \BibitemOpen
  \bibfield  {author} {\bibinfo {author} {\bibfnamefont {N.}~\bibnamefont
  {Leconte}}, \bibinfo {author} {\bibfnamefont {S.}~\bibnamefont {Javvaji}},
  \bibinfo {author} {\bibfnamefont {J.}~\bibnamefont {An}}, \bibinfo {author}
  {\bibfnamefont {A.}~\bibnamefont {Samudrala}},\ and\ \bibinfo {author}
  {\bibfnamefont {J.}~\bibnamefont {Jung}},\ }\bibfield  {title} {\bibinfo
  {title} {Relaxation effects in twisted bilayer graphene: A multiscale
  approach},\ }\href {https://doi.org/10.1103/PhysRevB.106.115410} {\bibfield
  {journal} {\bibinfo  {journal} {Phys. Rev. B}\ }\textbf {\bibinfo {volume}
  {106}},\ \bibinfo {pages} {115410} (\bibinfo {year} {2022})}\BibitemShut
  {NoStop}%
\bibitem [{\citenamefont {Gargiulo}\ and\ \citenamefont
  {Yazyev}(2018)}]{Yazyev17}%
  \BibitemOpen
  \bibfield  {author} {\bibinfo {author} {\bibfnamefont {F.}~\bibnamefont
  {Gargiulo}}\ and\ \bibinfo {author} {\bibfnamefont {O.~V.}\ \bibnamefont
  {Yazyev}},\ }\bibfield  {title} {\bibinfo {title} {Structural and electronic
  transformation in low-angle twisted bilayer graphene},\ }\href@noop {}
  {\bibfield  {journal} {\bibinfo  {journal} {2D Materials}\ }\textbf {\bibinfo
  {volume} {5}} (\bibinfo {year} {2018})}\BibitemShut {NoStop}%
\bibitem [{\citenamefont {Cazeaux}\ \emph {et~al.}(2023)\citenamefont
  {Cazeaux}, \citenamefont {Clark}, \citenamefont {Engelke}, \citenamefont
  {Kim},\ and\ \citenamefont {Luskin}}]{Cazeaux2023}%
  \BibitemOpen
  \bibfield  {author} {\bibinfo {author} {\bibfnamefont {P.}~\bibnamefont
  {Cazeaux}}, \bibinfo {author} {\bibfnamefont {D.}~\bibnamefont {Clark}},
  \bibinfo {author} {\bibfnamefont {R.}~\bibnamefont {Engelke}}, \bibinfo
  {author} {\bibfnamefont {P.}~\bibnamefont {Kim}},\ and\ \bibinfo {author}
  {\bibfnamefont {M.}~\bibnamefont {Luskin}},\ }\bibfield  {title} {\bibinfo
  {title} {Relaxation and domain wall structure of bilayer moir{\'e} systems},\
  }\href {https://doi.org/10.1007/s10659-023-10013-0} {\bibfield  {journal}
  {\bibinfo  {journal} {Journal of Elasticity}\ }\textbf {\bibinfo {volume}
  {154}},\ \bibinfo {pages} {443} (\bibinfo {year} {2023})}\BibitemShut
  {NoStop}%
\bibitem [{\citenamefont {Wang}\ \emph
  {et~al.}(2023{\natexlab{b}})\citenamefont {Wang}, \citenamefont {Khosravi},
  \citenamefont {Silva}, \citenamefont {Fabrizio}, \citenamefont {Vanossi},\
  and\ \citenamefont {Tosatti}}]{TosattiPRB23}%
  \BibitemOpen
  \bibfield  {author} {\bibinfo {author} {\bibfnamefont {J.}~\bibnamefont
  {Wang}}, \bibinfo {author} {\bibfnamefont {A.}~\bibnamefont {Khosravi}},
  \bibinfo {author} {\bibfnamefont {A.}~\bibnamefont {Silva}}, \bibinfo
  {author} {\bibfnamefont {M.}~\bibnamefont {Fabrizio}}, \bibinfo {author}
  {\bibfnamefont {A.}~\bibnamefont {Vanossi}},\ and\ \bibinfo {author}
  {\bibfnamefont {E.}~\bibnamefont {Tosatti}},\ }\bibfield  {title} {\bibinfo
  {title} {Bending stiffness collapse, buckling, topological bands of
  freestanding twisted bilayer graphene},\ }\href
  {https://doi.org/10.1103/PhysRevB.108.L081407} {\bibfield  {journal}
  {\bibinfo  {journal} {Phys. Rev. B}\ }\textbf {\bibinfo {volume} {108}},\
  \bibinfo {pages} {L081407} (\bibinfo {year}
  {2023}{\natexlab{b}})}\BibitemShut {NoStop}%
\bibitem [{\citenamefont {Ochoa}(2019)}]{OchoaPRB19}%
  \BibitemOpen
  \bibfield  {author} {\bibinfo {author} {\bibfnamefont {H.}~\bibnamefont
  {Ochoa}},\ }\bibfield  {title} {\bibinfo {title} {Moir\'e-pattern
  fluctuations and electron-phason coupling in twisted bilayer graphene},\
  }\href {https://doi.org/10.1103/PhysRevB.100.155426} {\bibfield  {journal}
  {\bibinfo  {journal} {Phys. Rev. B}\ }\textbf {\bibinfo {volume} {100}},\
  \bibinfo {pages} {155426} (\bibinfo {year} {2019})}\BibitemShut {NoStop}%
\bibitem [{\citenamefont {Xie}\ and\ \citenamefont {Liu}(2023)}]{LiuPRB23}%
  \BibitemOpen
  \bibfield  {author} {\bibinfo {author} {\bibfnamefont {B.}~\bibnamefont
  {Xie}}\ and\ \bibinfo {author} {\bibfnamefont {J.}~\bibnamefont {Liu}},\
  }\bibfield  {title} {\bibinfo {title} {Lattice distortions, moir\'e phonons,
  and relaxed electronic band structures in magic-angle twisted bilayer
  graphene},\ }\href {https://doi.org/10.1103/PhysRevB.108.094115} {\bibfield
  {journal} {\bibinfo  {journal} {Phys. Rev. B}\ }\textbf {\bibinfo {volume}
  {108}},\ \bibinfo {pages} {094115} (\bibinfo {year} {2023})}\BibitemShut
  {NoStop}%
\bibitem [{\citenamefont {Ceferino}\ and\ \citenamefont
  {Guinea}(2023)}]{Guinea23}%
  \BibitemOpen
  \bibfield  {author} {\bibinfo {author} {\bibfnamefont {A.}~\bibnamefont
  {Ceferino}}\ and\ \bibinfo {author} {\bibfnamefont {F.}~\bibnamefont
  {Guinea}},\ }\bibfield  {title} {\bibinfo {title} {Pseudomagnetic fields in
  fully relaxed twisted bilayer and trilayer graphene},\ }\href
  {https://api.semanticscholar.org/CorpusID:265457046} {\bibfield  {journal}
  {\bibinfo  {journal} {2D Materials}\ }\textbf {\bibinfo {volume} {11}}
  (\bibinfo {year} {2023})}\BibitemShut {NoStop}%
\bibitem [{\citenamefont {Whittaker}\ and\ \citenamefont
  {Watson}(1996)}]{WhittakerBook}%
  \BibitemOpen
  \bibfield  {author} {\bibinfo {author} {\bibfnamefont {E.~T.}\ \bibnamefont
  {Whittaker}}\ and\ \bibinfo {author} {\bibfnamefont {G.~N.}\ \bibnamefont
  {Watson}},\ }\bibinfo {title} {A course of modern analysis}\ (\bibinfo
  {publisher} {Cambridge University Press},\ \bibinfo {year} {1996})\ pp.\
  \bibinfo {pages} {91--94}\BibitemShut {NoStop}%
\bibitem [{\citenamefont {Herzog-Arbeitman}\ \emph {et~al.}(2024)\citenamefont
  {Herzog-Arbeitman}, \citenamefont {Yu}, \citenamefont {C\u{a}lug\u{a}ru},
  \citenamefont {Hu}, \citenamefont {Regnault}, \citenamefont {Vafek},
  \citenamefont {Kang},\ and\ \citenamefont {Bernevig}}]{JonahHFarXiv24}%
  \BibitemOpen
  \bibfield  {author} {\bibinfo {author} {\bibfnamefont {J.}~\bibnamefont
  {Herzog-Arbeitman}}, \bibinfo {author} {\bibfnamefont {J.}~\bibnamefont
  {Yu}}, \bibinfo {author} {\bibfnamefont {D.}~\bibnamefont
  {C\u{a}lug\u{a}ru}}, \bibinfo {author} {\bibfnamefont {H.}~\bibnamefont
  {Hu}}, \bibinfo {author} {\bibfnamefont {N.}~\bibnamefont {Regnault}},
  \bibinfo {author} {\bibfnamefont {O.}~\bibnamefont {Vafek}}, \bibinfo
  {author} {\bibfnamefont {J.}~\bibnamefont {Kang}},\ and\ \bibinfo {author}
  {\bibfnamefont {B.~A.}\ \bibnamefont {Bernevig}},\ }\href
  {https://arxiv.org/abs/2405.13880} {\bibinfo {title} {Heavy fermions as an
  efficient representation of atomistic strain and relaxation in twisted
  bilayer graphene}} (\bibinfo {year} {2024}),\ \Eprint
  {https://arxiv.org/abs/2405.13880} {arXiv:2405.13880 [cond-mat.mes-hall]}
  \BibitemShut {NoStop}%
\bibitem [{\citenamefont {Herzog-Arbeitman}\ \emph {et~al.}(2025)\citenamefont
  {Herzog-Arbeitman}, \citenamefont {C\u{a}lug\u{a}ru}, \citenamefont {Hu},
  \citenamefont {Yu}, \citenamefont {Regnault}, \citenamefont {Kang},
  \citenamefont {Bernevig},\ and\ \citenamefont {Vafek}}]{JonahIKSarXiv25}%
  \BibitemOpen
  \bibfield  {author} {\bibinfo {author} {\bibfnamefont {J.}~\bibnamefont
  {Herzog-Arbeitman}}, \bibinfo {author} {\bibfnamefont {D.}~\bibnamefont
  {C\u{a}lug\u{a}ru}}, \bibinfo {author} {\bibfnamefont {H.}~\bibnamefont
  {Hu}}, \bibinfo {author} {\bibfnamefont {J.}~\bibnamefont {Yu}}, \bibinfo
  {author} {\bibfnamefont {N.}~\bibnamefont {Regnault}}, \bibinfo {author}
  {\bibfnamefont {J.}~\bibnamefont {Kang}}, \bibinfo {author} {\bibfnamefont
  {B.~A.}\ \bibnamefont {Bernevig}},\ and\ \bibinfo {author} {\bibfnamefont
  {O.}~\bibnamefont {Vafek}},\ }\href {https://arxiv.org/abs/2502.08700}
  {\bibinfo {title} {Kekul\'e spiral order from strained topological heavy
  fermions}} (\bibinfo {year} {2025}),\ \Eprint
  {https://arxiv.org/abs/2502.08700} {arXiv:2502.08700 [cond-mat.str-el]}
  \BibitemShut {NoStop}%
\bibitem [{\citenamefont {Uchida}\ \emph {et~al.}(2014)\citenamefont {Uchida},
  \citenamefont {Furuya}, \citenamefont {Iwata},\ and\ \citenamefont
  {Oshiyama}}]{OshiyamaPRB14}%
  \BibitemOpen
  \bibfield  {author} {\bibinfo {author} {\bibfnamefont {K.}~\bibnamefont
  {Uchida}}, \bibinfo {author} {\bibfnamefont {S.}~\bibnamefont {Furuya}},
  \bibinfo {author} {\bibfnamefont {J.-I.}\ \bibnamefont {Iwata}},\ and\
  \bibinfo {author} {\bibfnamefont {A.}~\bibnamefont {Oshiyama}},\ }\bibfield
  {title} {\bibinfo {title} {Atomic corrugation and electron localization due
  to moir\'e patterns in twisted bilayer graphenes},\ }\href
  {https://doi.org/10.1103/PhysRevB.90.155451} {\bibfield  {journal} {\bibinfo
  {journal} {Phys. Rev. B}\ }\textbf {\bibinfo {volume} {90}},\ \bibinfo
  {pages} {155451} (\bibinfo {year} {2014})}\BibitemShut {NoStop}%
\end{thebibliography}%

\appendix

\begin{widetext}
\section{Comparison with the experiments}
In this section, we compare the experiments reporting the lattice relaxation with the calculated relaxation from two different theoretical models proposed in Ref.~\cite{KoshinoPRB17} and \cite{KaxirasPRB18}. Instead of using the closed-form solution for lattice relaxation, we numerically solve the self-consistent equation (Eq.~\ref{Eqn:LatticeUqRelax}) using the iteration method to obtain the lattice relaxation for these two theoretical models. 

Fig.~\ref{Fig:ThetaRSPAB} in the main text illustrates the relaxation-induced twist, defined as $\theta_R = \half (\partial_x \delta U_y - \partial_y \delta U_x)$, at the SP and AB stacking points, for both the experiment and the calculation from two theoretical models. The experimental measurement is more consistent with the model proposed in Ref.~\cite{KaxirasPRB18}. As a consequence, this model is chosen to derive the closed form of the lattice relaxation with and without external heterostrain. In addition, this model is also used in upcoming works to study the electronic continuum model and the strain-induced change of the electron spectrum. In this section, we compare other quantities of the relaxation at high symmetry points. Based on the comparison of these quantities, the same conclusion is drawn that the model proposed in Ref.~\cite{KaxirasPRB18} is more consistent with the experimental measurements. 

The derivative of the lattice relaxation is a rank-$2$ tensor $\epsilon_{\mu\nu}(\fvec x) = \partial_{\mu} \delta U_{\nu}$ that contains four independent components. The antisymmetric part gives the local twist $\theta_R = \half (\epsilon_{xy} - \epsilon_{yx})$  as defined in the main text. The symmetric part $M^s_{\mu\nu}(\fvec x) = \half (\epsilon_{\mu\nu} + \epsilon_{\nu\mu})$ contains three independent parameters, and can be diagonalized by an orthogonal matrix. 
\begin{align}
    M^s(\fvec x) = R^T(\phi'(\fvec x)) \begin{pmatrix} \epsilon_{max}(\fvec x) & 0\\ 0 &   \epsilon_{min}(\fvec x) \end{pmatrix} R(\phi'(\fvec x)) \  ,
\end{align}
where $R(\phi')$ is the $2\times 2$ matrix describing the rotation along $\hat z$ axis with the rotation angle of $\phi'$, and both $\epsilon_{max}$ and $\epsilon_{min}$ are the magnitudes of strain along two principle axes. Note that all the parameters $\epsilon_{max}$, $\epsilon_{min}$, and $\phi'$ are position dependent. In addition to $\theta_R(\fvec x)$, Ref.~\cite{BediakoNM21} has introduced several quantities to describe the lattice relaxation. 
\begin{align}
    \gamma_{max} = \epsilon_{max} - \epsilon_{min} \ ,  \quad \theta_P(\fvec x) = - \phi(\fvec x) \ , \quad s_{xy} = \half \epsilon_{yx} \ , \quad s_{xy} = \half \epsilon_{yx} \quad \Longrightarrow \quad \theta_R = s_{yx} - s_{xy}  \ ,
\end{align}
where $\theta_P$ is the principle angle giving the orientation of the two principle axes of the strain matrix. 
\begin{figure}[htb] 
	\centering
	\subfigure[\label{FigS:RelaxComp:GammaSP}]{\includegraphics[width=0.32\columnwidth]{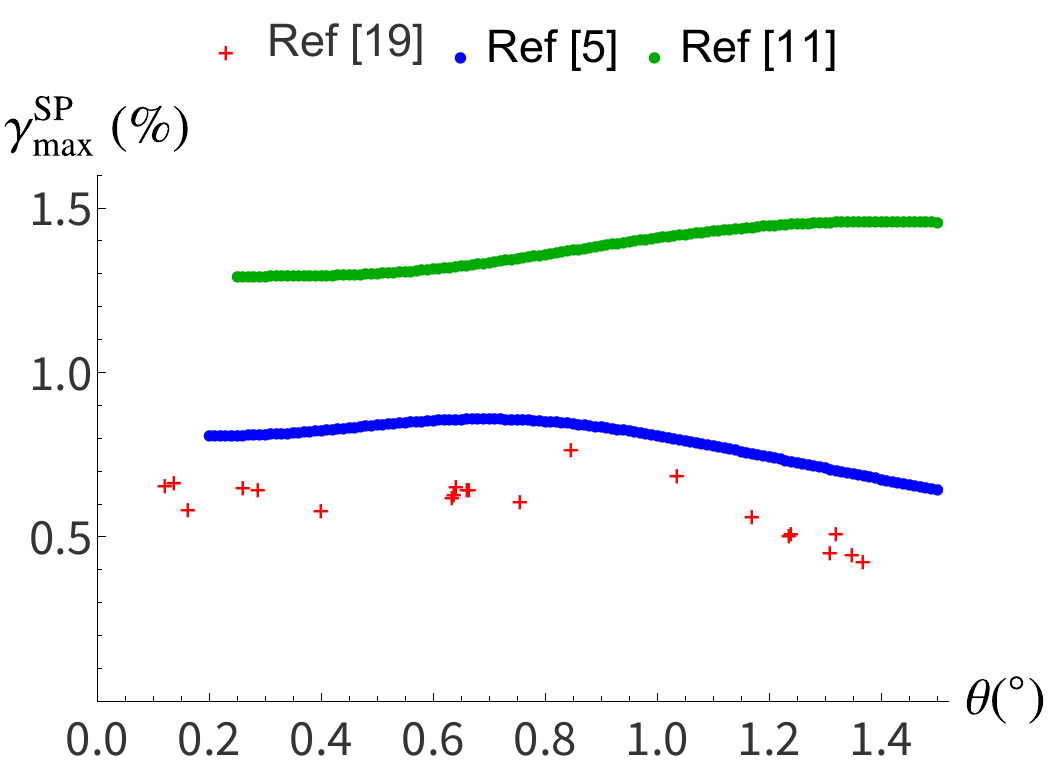}}
    \subfigure[\label{FigS:RelaxComp:SxySP}]{\includegraphics[width=0.32\columnwidth]{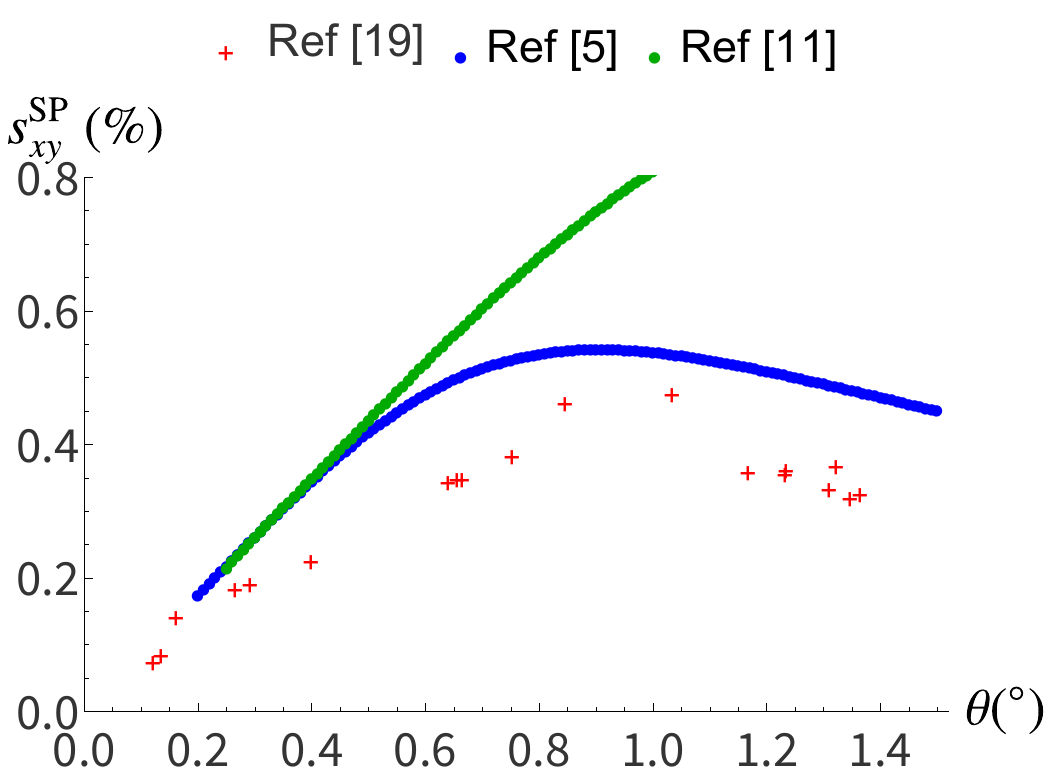}}
    \subfigure[\label{FigS:RelaxComp:SyxSP}]{\includegraphics[width=0.32\columnwidth]{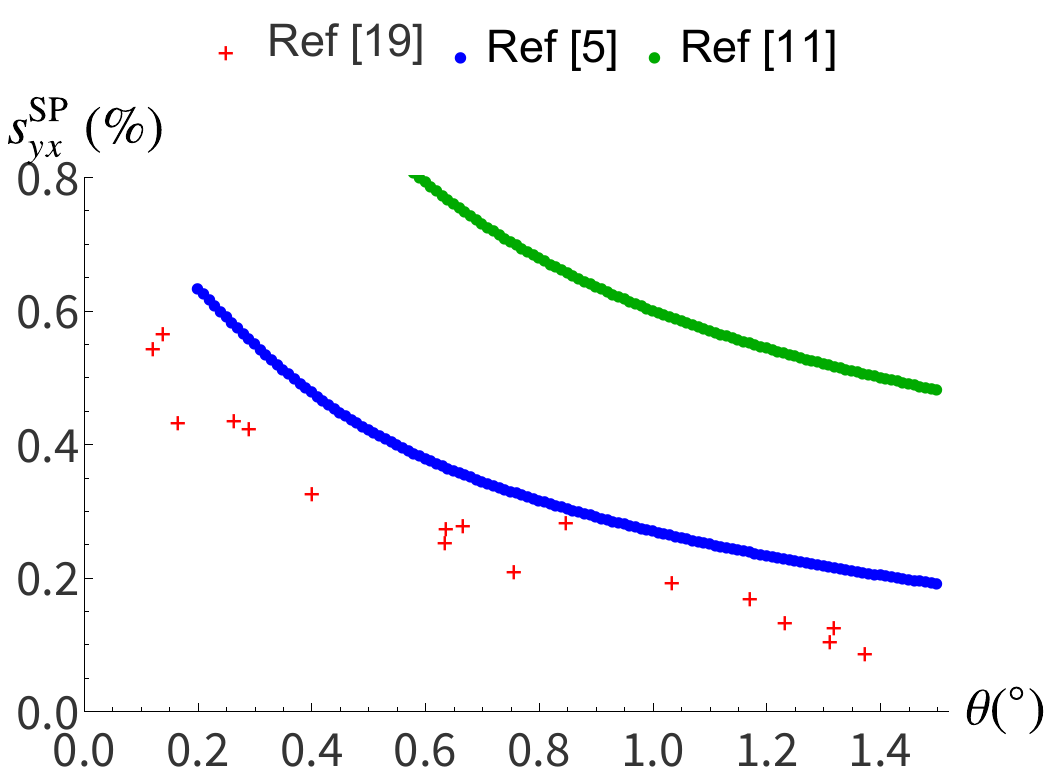}}
	\caption{Various relaxation parameters of (a) $\gamma_{max}$, (b) $s_{xy}$, and (c) $s_{yx}$ at SP stacking ($\fvec x = \half \fvec L_1$) when the heterostrain is absent. Three different datasets are included: the experimental measurements (red), the calculation of the model proposed in Ref.~\cite{KoshinoPRB17} (green), and the calculation of the model proposed in Ref.~\cite{KaxirasPRB18} (blue). The experimental measurements are more consistent with the model proposed in Ref.~\cite{KaxirasPRB18}.}
	\label{FigS:RelaxComp}
\end{figure}
Fig.~\ref{FigS:RelaxComp} illustrates $\gamma_{max}$, $s_{xy}$, and $s_{yx}$ at SP stacking for the experimental measurements, and for the two models proposed in Ref.~\cite{KoshinoPRB17} and \cite{KaxirasPRB18}, showing that the experimental measurements are consistent with the model proposed in Ref.~\cite{KaxirasPRB18}.

\section{Branch points}
\label{SecS:BP}
\begin{figure}[htb] 
	\centering
	\includegraphics[scale=0.5]{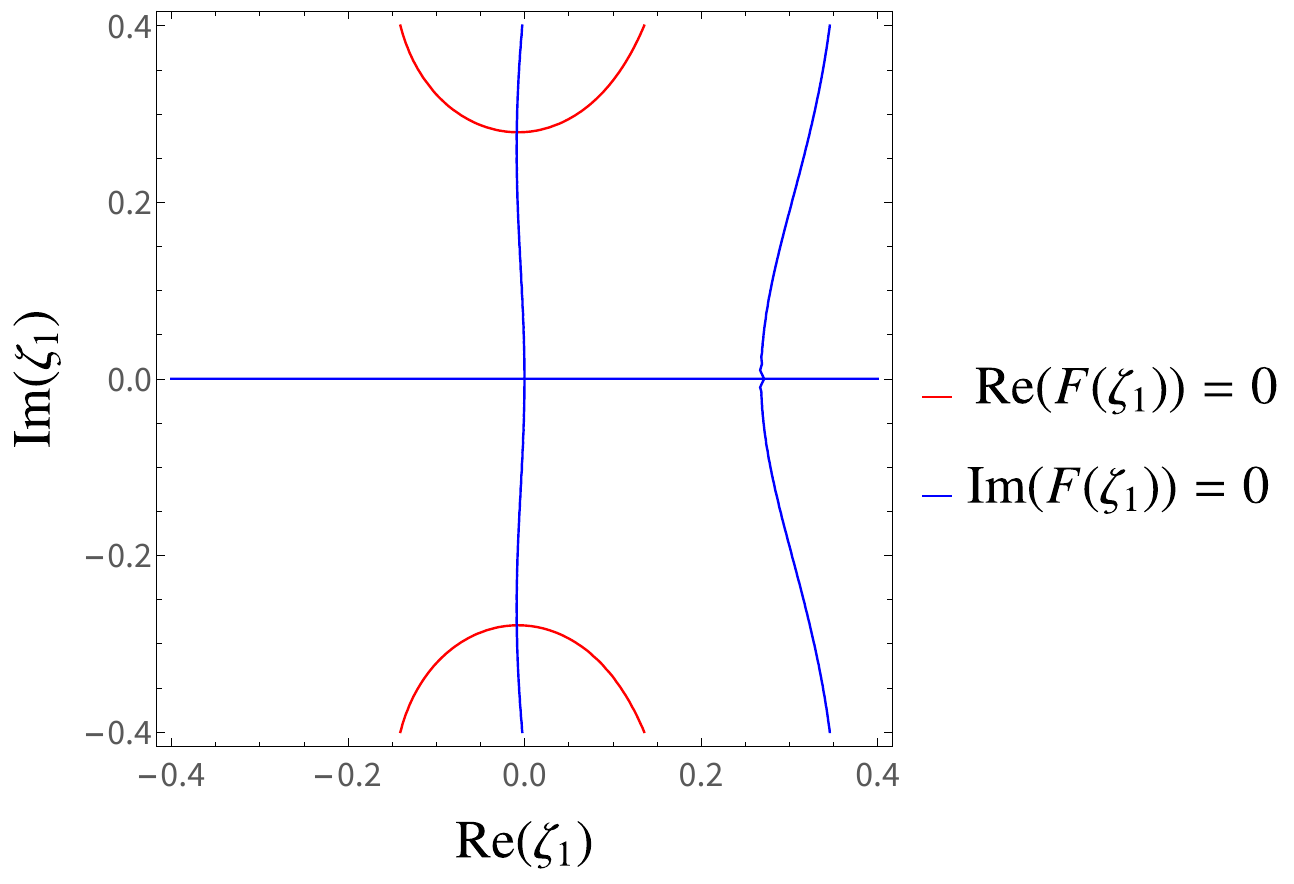}
	\caption{The contour plot of (red) $\text{Re}(F(\zeta_1)) = 0$ and (blue) $\text{Im}(F(\zeta_1)) = 0$ on the complex $\zeta_1$ plane, where $F(\zeta_1)$ is defined in Eq.~\ref{EqnS:FFunc}. The branch points $\lambda_{\text{BP}}$ are given by Eq.~\ref{EqnS:fEqual0}, where $\zeta_{1,c}$ is given by $F(\zeta_{1,c}) = 0$, the two intersection points of these two colored curves. Since these two points are the crossings, instead of the touching, of these two curves, the branch points are stable against any small perturbation of the parameters of Eq.~\ref{EqnS:gFunc}. The magnitude of $\lambda_{\text{BP}}$ is the radius of convergence.}
	\label{FigS:BranchFZeroPt}
\end{figure}

As discussed in Sec.~\ref{Sec:AnalyticalNoStrain}, $\zeta_1$ can be obtained by solving Eq.~\ref{Eqn:Zeta1ExactEqn}. The equation can be rewritten as
\begin{align}
    & 0 = f(\lambda,\ \zeta_1) = \zeta_1 + \lambda g(\zeta_1) \quad \mbox{with} \\
    & g(\zeta_1) = \frac{\rmd}{\rmd \zeta_1} \left( \frac{c_2}{c_1} \big( J_1(3\zeta_1) \big)^2 + \sum_{n = - \infty}^{\infty} \left( J_n(2\zeta_1) \big( J_{n+1}(\zeta_1) \big)^2  + \frac{c_3}{c_1} J_n(4 \zeta_1) \left( J_{n+2}(2\zeta_1) \right)^2 \right) \right) \ . \label{EqnS:gFunc}
\end{align}
In principle, for fixed $\lambda$, the nonlinear equation $f(\lambda, \zeta_1) = 0$ contains multiple solutions for $\zeta_1$.  The physical one is given by the solution that approaches $0$ as $\lambda \rightarrow 0$, and therefore, is well-defined at least for small $\lambda$. In addition, the solution contains branch points in the complex $\lambda$ plane, whose magnitude gives the radius of convergence. $g(\zeta_1)$ is a highly nonlinear function, and therefore approximation methods are needed to derive the explicit expression of $\zeta_1$ in terms of $\lambda$. The closed form of $\zeta_1$ in Eq.~\ref{Eqn:Zeta1Appr} is obtained by expanding $g(\zeta_1)$ to the order of $\mO(\zeta_1^2)$, which works very well for real $\lambda$ and real $\zeta_1$. However, due to the different convergence behavior of $J_n(x)$ with the complex argument $x$, relatively larger errors are introduced for complex $\lambda$ and thus for complex $\zeta_1$. Consequently, Eq.~\ref{Eqn:Zeta1Appr} cannot provide precise values of the branch points and the resulting radius of convergence. In this section, we provide a different method to obtain the branch points without explicitly expressing $\zeta_1$ in terms of $\lambda$. 

Note that at the branch points,
\begin{align}
    f(\lambda_{BP},\ \zeta_{1,{BP}}) = 0 \quad \Longrightarrow \quad \lambda_{BP} = - \zeta_{1, {BP}}/g(\zeta_{1,{BP}}) \label{EqnS:fEqual0}
\end{align}
Expanding the equation $f(\lambda, \zeta_1) = 0$ around the branch points $(\lambda_{BP}, \zeta_{1, BP})$, we obtain
\begin{align}
    0 & \approx f(\lambda_{BP},\ \zeta_{1,BP}) + (\lambda - \lambda_{BP}) \left. \partial_{\lambda} f(\lambda, \zeta_{1,BP}) \right|_{\lambda = \lambda_{BP}} +  (\zeta_1 - \zeta_{1, BP}) \left. \partial_{\zeta_1} f(\lambda_{BP}, \zeta_{1}) \right|_{\zeta_1 = \zeta_{1, BP}} \nonumber \\
    & \quad + \half  \left.  \left( (\lambda - \lambda_{BP})^2   \partial^2_{\lambda}  + 2 (\lambda - \lambda_{BP}) (\zeta_1 - \zeta_{1, BP}) \partial_{\lambda} \partial_{\zeta_1}    +  (\zeta_1 - \zeta_{1, BP})^2   \partial^2_{\zeta_1}  \right) f(\lambda,\ \zeta_{1})\right|_{\lambda = \lambda_{BP}, \zeta_1 = \zeta_{1, BP}} \nonumber \\
    & \quad + O((\lambda - \lambda_{BP})^3, (\zeta_1 - \zeta_{1,BP})^3) \ .
\end{align}
Substituting $f(\lambda, \zeta_1) = \zeta_1 + \lambda g(\zeta_1)$ and $f(\lambda_{BP},\ \zeta_{1,BP}) = 0$ into the above equation, we find
\begin{align}
    0 & \approx g(\zeta_{1,BP}) (\lambda - \lambda_{BP}) + (1 + \lambda_{BP} g'(\zeta_{1,BP})) (\zeta_1 - \zeta_{1,BP})  \nonumber \\
    & \qquad + \half \left( 2 (\lambda - \lambda_{BP}) (\zeta_1 - \zeta_{1, BP}) g'(\zeta_{1,BP}) + (\zeta_1 - \zeta_{1, BP})^2 \lambda_{BP} g''(\zeta_{1,BP}) \right)  \  ,
\end{align}
that gives the approximate solution of $\zeta_1$ when close to branch points. Around the branch points, $\zeta_1$ cannot be uniquely solved, therefore
\[  1 + \lambda_{BP} g'(\zeta_{1,BP}) = 0 \quad \mbox{and} \quad g(\zeta_{1,BP}) \neq 0 \qquad \Longrightarrow \quad g(\zeta_{1,BP}) - \zeta_{1,BP} g'(\zeta_{1,BP}) = 0 \quad \mbox{and} \quad  g(\zeta_{1,BP}) \neq 0 \ , \]
where in the last step, we substituted $\lambda_{BP}$ from the relation $\zeta_{1,BP} + \lambda_{BP} g(\zeta_{1,BP}) = 0$. Introducing
\begin{align}
     F(\zeta_1) = g(\zeta_1) - \zeta_1 g'(\zeta_1)   \   , \label{EqnS:FFunc}
\end{align}
with $g(\zeta_1)$ defined in Eq.~\ref{EqnS:gFunc}. The zero points of the function $F$ give $\zeta_{1, BP}$ at the branch points. Since the explicit formula of $g(\zeta_1)$ has already been given in Eq.~\ref{EqnS:gFunc}, $F(\zeta_1)$ can also be explicitly expressed as a nonlinear function. The numerical value of complex $\zeta_{1,BP}$ can be accurately obtained by various numerical methods. Consequently, $\lambda_{BP}$ can be obtained by Eq.~\ref{EqnS:fEqual0}. Fig.~\ref{FigS:BranchFZeroPt} shows two colored curves where the real part (in red) and the imaginary part (in blue) of $F$ vanish respectively on the complex $\zeta_1$ plane. The crossing points of these two colored curves give $\zeta_{1, BP}$. As illustrated in the plot, the two crossing points are stable against any small change of the parameters in the function $f(\lambda, \zeta_1)$, and thus the existence of these two branch points is robust.

\section{Improvement of the analytical formula for unstrained lattice relaxation}
\label{SecS:Improve}
\begin{figure}[htb] 
	\centering
    \subfigure[\label{FigS:LatticeRelaxa:Inner}]{\includegraphics[width=0.49\columnwidth]{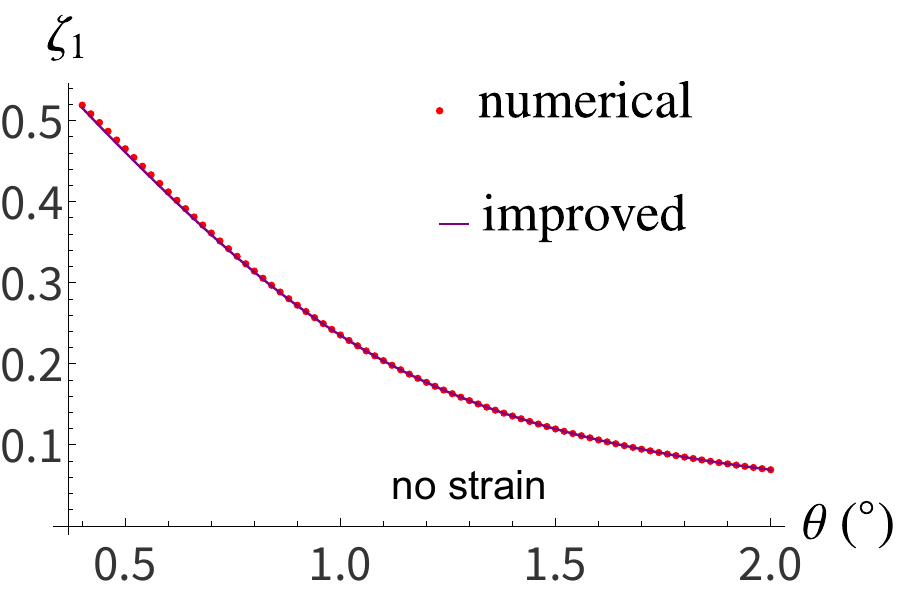}}
    \subfigure[\label{FigS:LatticeRelaxa:Outer}]{\includegraphics[width=0.49\columnwidth]{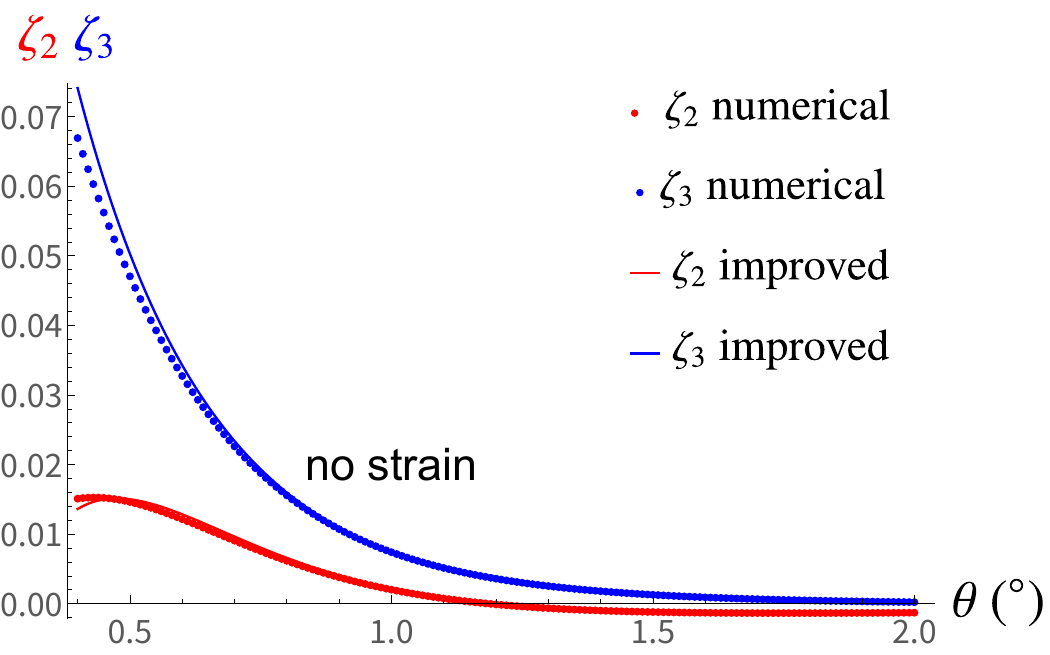}}
	\caption{Similar to Fig.~\ref{Fig:LatRelaxNoStrain}, (left) $\zeta_1$, (right red) $\zeta_2$ and (right blue) $\zeta_3$ for the lattice relaxation on the innermost, the 2nd, and the 3rd shells in the absence of external heterostrain. Dots are the solutions obtained by numerically solving Eq.\ref{Eqn:LatticeUqRelax}. The colored curves are the solutions obtained by solving $3\times 3$ matrix equation in Eq.~\ref{EqnS:ZetaMatrixEqn}, which are much more complicated but also more accurate compared with Eq.~\ref{Eqn:Zeta1Appr},  \ref{Eqn:Zeta2Appr}, and \ref{Eqn:Zeta3Appr}. }
	\label{FigS:LatticeRelax}
\end{figure}
In this section, we explicitly work out the improved formula for $\zeta_i$ ($i =1$, $2$, and $3$) that works well for the twist angle $\theta > 0.4^{\circ}$. Note that the lattice relaxation  can  be approximated as
\begin{align}
    \delta \fvec U(\fvec x) & \approx \frac2{|\fvec G_1|^2}\sum_{a = 1}^3 \left(  \zeta_1 \fvec G_a \sin(\fvec g_a \cdot \fvec x) + \zeta_2 (\fvec G_a - \fvec G_{a+1}) \sin((\fvec g_a - \fvec g_{a+1}) \cdot \fvec x) + 2\zeta_3 \fvec G_a \sin(2\fvec g_a \cdot \fvec x)  \right)
\end{align}
where we have introduced the variable $\fvec g_4 = \fvec g_1$ and $\fvec G_4 = \fvec G_1$ for notational convenience. Correspondingly, the elastic energy can be simplified as
\begin{align}
    U_E/A_{tot} = \text{const} + \frac32 \mathcal{G} \theta^2 \left( \zeta_1^2 + 9 \zeta_2^2 + 16 \zeta_3^2 \right) \ , 
\end{align}

As shown in Fig.~\ref{FigS:LatticeRelax}, $|\zeta_{2,3}| \lesssim 0.1$ for twist angle $\theta > 0.4^{\circ}$. Therefore, the interlayer adhesion energy can be approximated in the following form
\begin{equation}
    \frac{U_B}{3 c_1 A_{tot}} \approx F_1(\zeta_1) + \left( G_2(\zeta_1) \zeta_2 + G_3(\zeta_1) \zeta_3 \right) + \left( H_{22}(\zeta_1) \zeta_2^2 + H_{23}(\zeta_1) \zeta_2 \zeta_3 + H_{33}(\zeta_1) \zeta_3^2 \right)  \ ,   \label{Eqn:UbForm}
\end{equation}
where the cubic and higher order terms of $\zeta_{2,3}$ are neglected. For the relaxation at the inner shell, we write $\zeta_1 = \zeta_1^{(0)} + \delta \zeta_1$, where $\zeta_1^{(0)}$, given by Eq.~\ref{Eqn:Zeta1Appr}, is obtained by applying the following approximation
\begin{align}
    F_1(\zeta_1) \approx F_0(\zeta_1) = - \zeta_1 + \frac{\zeta_1^2}4 + \zeta_1^3 + \frac{c_2}{c_1} \frac94 \zeta_1^2 + \frac{c_3}{c_1} 2 \zeta_1^2  \label{Eqn:ApproxF}
\end{align}
and then solve the equation
\begin{align}
    \lambda^{-1} \zeta_1^{(0)} + F_0'(\zeta_1^{(0)}) = 0 \   .  \label{Eqn:minzeta10}
\end{align}
As discussed in the main text, $\zeta_1 \approx \zeta_1^{(0)}$ for $\theta > 0.4^{\circ}$ with the difference $|\delta \zeta_1| < 0.05$ revealed by full numerical calculation.


Without making the approximation in Eq.~(\ref{Eqn:ApproxF}), we obtain three equations for $\zeta_1$, $\zeta_2$, and $\zeta_3$ by minimizing the total energy $U_E + U_B$.
\begin{align}
    & \lambda^{-1} \zeta_1 + F_1'(\zeta_1) + G_2'(\zeta_1) \zeta_2 + G_3'(\zeta_1) \zeta_3 + \big( H_{22}'(\zeta_1) \zeta_2^2 + H'_{33}(\zeta_1) \zeta_3^2 + H_{23}'(\zeta_1) \zeta_2 \zeta_3 \big) = 0 \label{Eqn:minzeta1} \\
    & 9 \lambda^{-1} \zeta_2 + G_2(\zeta_1) + 2H_{22}(\zeta_1) \zeta_2 + H_{23}(\zeta_1) \zeta_3 = 0  \label{Eqn:minzeta2} \\
    & 16 \lambda^{-1} \zeta_3 + G_3(\zeta_1) + H_{23}(\zeta_1) \zeta_2 + 2H_{33}(\zeta_1) \zeta_3 = 0   \label{Eqn:minzeta3}
\end{align}
The above three coupled nonlinear equations are difficult to solve. We introduce
\begin{equation}
    \delta F(\zeta_1) = F_1(\zeta_1) - F_0(\zeta_1)
\end{equation}
and notice that
\begin{align}
    F_1'(\zeta_1) & = \left( F_1'(\zeta_1) - F_1'(\zeta_1^{(0)}) \right) + \left( F_1'(\zeta_1^{(0)}) - F'_0(\zeta_1^{(0)}) \right) + F_0'(\zeta_1^{(0)}) \nonumber \\
    & \approx F_1''(\zeta_1^{(0)}) \delta \zeta_1 + \delta F'(\zeta_1^{(0)}) + F_0'(\zeta_1^{(0)}) + O\left( \big( \delta \zeta_1 \big)^2 \right) \\
    G_i(\zeta_1) & \approx G_i(\zeta_1^{(0)}) + G_i'(\zeta_1^{(0)}) \delta \zeta_1 + O(\big(\delta \zeta_1\big)^2) \qquad \mbox{with } i = 2 \mbox{ and } 3
\end{align}
Thus, expanding Eq.~(\ref{Eqn:minzeta1})--(\ref{Eqn:minzeta3}) to the linear order of $\delta \zeta_1$, $\zeta_2$, and $\zeta_3$,  and using the formula in Eq.~(\ref{Eqn:minzeta10}), we obtain
\begin{align}
    & \left( F_1''(\zeta_1^{(0)}) + \lambda^{-1} \right) \delta \zeta_1 + G_2(\zeta_1^{(0)}) \zeta_2 + G_3(\zeta_1^{(0)}) \zeta_3 = - \delta F'(\zeta_1^{(0)})   \\
    & G_2'(\zeta_1^{(0)}) \delta \zeta_1 + \left( 9 \lambda^{-1} + 2 H_{22}(\zeta_1^{(0)}) \right) \zeta_2 + H_{23}(\zeta_1^{(0)}) \zeta_3 = - G_2(\zeta_1^{(0)}) \\
     & G_3'(\zeta_1^{(0)}) \delta \zeta_1 +  H_{23}(\zeta_1^{(0)}) \zeta_2 + \left( 16 \lambda^{-1} + 2 H_{33}(\zeta_1^{(0)}) \right) \zeta_3 = - G_3(\zeta_1^{(0)})
\end{align}
Written in the matrix form, 
\begin{align}
    \begin{pmatrix}
        F_1''(\zeta_1^{(0)}) + \lambda^{-1} & G_2'(\zeta_1^{(0)}) & G_3'(\zeta_1^{(0)}) \\
        G_2'(\zeta_1^{(0)}) & 2 H_{22}(\zeta_1^{(0)}) + 9 \lambda^{-1}   & H_{23}(\zeta_1^{(0)}) \\
        G_3'(\zeta_1^{(0)}) & H_{23}(\zeta_1^{(0)}) & 2 H_{33}(\zeta_1^{(0)}) + 16 \lambda^{-1}
    \end{pmatrix} 
    \begin{pmatrix}  \delta \zeta_1 \\ \zeta_2 \\ \zeta_3  \end{pmatrix}  = - \begin{pmatrix}  \delta F'(\zeta_1^{(0)}) \\ G_2(\zeta_1^{(0)}) \\ G_3(\zeta_1^{(0)})  \end{pmatrix}   \label{EqnS:ZetaMatrixEqn}  \ . 
\end{align}
$\delta \zeta_1$, $\zeta_2$, and $\zeta_3$ are obtained by solving the above matrix equation. 

The formula of $F_1$, $G_i$, and $H_{ij}$ ($i$, $j = 2$, $3$) can be derived but are quite complicated. For example,
\begin{align}
    F_1(\zeta_1) & = \frac{c_2}{c_1} \big( J_1(3\zeta_1) \big)^2 + \sum_{n = - \infty}^{\infty} \left( J_n(2\zeta_1) \big( J_{n+1}(\zeta_1) \big)^2  + \frac{c_3}{c_1} J_n(4 \zeta_1) \left( J_{n+2}(2\zeta_1) \right)^2 \right) 
\end{align}
Instead of presenting their exact expressions, here, we expand $F_1$ to the order of $\zeta_1^5$ or $\frac{c_{2,3}}{c_1} \zeta_1^4$, $G_i$ to the order of $\zeta_1^4$ or $\frac{c_{2,3}}{c_1} \zeta_1^3$, and $H_{ij}$ to the order of $\zeta_1^3$ or $\frac{c_{2,3}}{c_1} \zeta_1^2$. We found that the solution of Eq.~\ref{EqnS:ZetaMatrixEqn} agrees with the numerical calculation when $\theta \geq 0.4^{\circ}$. 

Their approximated formula are listed below,
\begin{align}
    F_1(\zeta_1) & \approx  \left( - \zeta_1 + \frac{\zeta_1^2}4 + \zeta_1^3 - \frac3{16}\zeta_1^4 - \frac{79}{192}\zeta_1^5  \right) + \frac{c_2}{c_1} \left( \frac94 \zeta_1^2 - \frac{81}{16} \zeta_1^4 \right) +  \frac{c_3}{c_1} \left( 2 \zeta_1^2 - 2 \zeta_1^3 - \frac{77}{12}\zeta_1^4  \right) \\
    G_2(\zeta_1) & \approx  \left( - \frac32 \zeta_1 + \frac32  \zeta_1^2 + \frac{21}{16} \zeta_1^3 - \frac{19}{16}\zeta_1^4 \right) +  \frac{c_2}{c_1} \left(  -3  + \frac{27}2 \zeta_1^2 - \frac{27}4 \zeta_1^3 \right) +  \frac{c_3}{c_1} \left( 6 \zeta_1 + 12 \zeta_1^2 - 33 \zeta_1^3 \right) \\
    G_3(\zeta_1) & \approx \left( - 2 \zeta_1 + \frac{17}{12} \zeta_1^3 - \frac{11}{48} \zeta_1^4 \right) +  \frac{c_2}{c_1} \frac{27}2 \zeta_1^2 + \frac{c_3}{c_1} \left( -4 + 22 \zeta_1^2 - 8 \zeta_1^3 \right) \\
    H_{22}(\zeta_1) & \approx \left( \frac92 \zeta_1 - \frac9{16} \zeta_1^2 - \frac{153}{32} \zeta_1^3  \right) + \frac{c_2}{c_1} \left( \frac94 - \frac{243}8 \zeta_1^2  \right) + \frac{c_3}{c_1} \left( 18 \zeta_1 - \frac{45}2 \zeta_1^2\right) \\
    H_{23}(\zeta_1) & \approx \left( \frac92 \zeta_1 + \frac32 \zeta_1^2 - 7 \zeta_1^3  \right) + \frac{c_2}{c_1} \left( 27 \zeta_1 - \frac{81}4 \zeta_1^2   \right) + \frac{c_3}{c_1} \left( 12 \zeta_1 - 48 \zeta_1^2 \right) \\
    H_{33}(\zeta_1) & \approx \left( 7 \zeta_1 - \frac14 \zeta_1^2 - \frac{35}6 \zeta_1^3  \right) - \frac{c_2}{c_1} \frac{81}4 \zeta_1^2 + \frac{c_3}{c_1} \left( 4 - 56 \zeta_1^2  \right) 
\end{align}
Substituting the above approximated formula into the matrix equation in Eq.~\ref{EqnS:ZetaMatrixEqn}, we obtain the improved formula for $\zeta_1$, $\zeta_2$, and $\zeta_3$, with their results plotted in Fig.~\ref{FigS:LatticeRelax}.

\section{Impact of external heterostrain}
In this section, we derive the equation for $\delta \fvec W_{\fvec G}$, especially the expression of $B(\fvec G, \{ \fvec W_{\fvec G'} \})$ in Eq.~\ref{Eqn:BDef}. Notice that Eq.~\ref{Eqn:LatticeUqRelax} is still valid in the presence of external heterostrain. As shown in Eq.~\ref{Eqn:MMat} -- \ref{Eqn:M2}, the matrix $M_{\mu\nu}(\fvec g_{\fvec G})= \sum_{i = 0}^2 M^{(i)}_{\fvec G}$ is decomposed into the sum of three parts ordered by the powers of $\delta \fvec g$ and gives the first term of Eq.~\ref{Eqn:WEqn}. In addition, since $\delta \fvec U(\fvec x) = 2\sum_{\fvec G \in \mathbb{H}} \fvec W_{\fvec G} \sin(\fvec g_{\fvec G} \cdot \fvec x)$, we have
\begin{align}
    \frac{\partial \delta  U_{\nu}(\fvec x)}{\partial W_{\fvec G, \mu}} = 2\delta_{\mu\nu} \sin(\fvec g_{\fvec G} \cdot \fvec x).
\end{align}
The Fourier components of $\sin(\fvec G' \cdot \fvec U^-(\fvec x))$ can be expressed as
\begin{align}
    \sin(\fvec G' \cdot \fvec U^-(\fvec x)) = \sum_{\fvec g} f_{\fvec  G'}(\fvec g, \{\fvec W_{\fvec G}\} ) \sin(\fvec g \cdot \fvec x) \quad \Longrightarrow \quad f_{\fvec  G'}(\fvec g, \{\fvec W_{\fvec G}\} ) = A_{tot}^{-1} \int \rmd^2 \fvec x\ \sin(\fvec G' \cdot \fvec U^-(\fvec x)) \sin(\fvec g \cdot \fvec x).
\end{align}
Then, the right hand side of Eq.~\ref{Eqn:LatticeUqRelax} can be written as
\begin{align}
    & \sum_{\fvec G'} V_{\fvec G'} f_{\fvec  G'}(\fvec g_{\fvec G}, \{\fvec W \} ) G'_{\mu} = \frac1{A_{tot}} \sum_{\fvec G'} V_{\fvec G'} \int \rmd^2 \fvec x\ \sin(\fvec G' \cdot \fvec U^-(\fvec x)) \sin(\fvec g_{\fvec G} \cdot \fvec x) G'_{\mu} \nonumber \\
    = & - \frac1{2 A_{tot}} \frac{\partial}{\pd W_{\fvec G, \mu}} \sum_{\fvec G'} V_{\fvec G'} \int \rmd^2 \fvec x\ \cos(\fvec G' \cdot \fvec U^-(\fvec x)) = - \half  \frac{\partial}{\pd W_{\fvec G, \mu}}  B(\fvec G, \{ \fvec W \} )\\
\mbox{where} \quad    & B(\fvec G, \{ \fvec W \} )  = \frac1{A_{tot}} \int \rmd^2 \fvec x\ \cos(\fvec G \cdot \fvec U^-(\fvec x)) =  \frac1{A_{tot}} \int \rmd^2 \fvec x\ \half \left( e^{i \fvec g_{\fvec G} \cdot \fvec x} e^{i \sum_{\fvec G' \in \mathbb{H}} 2\fvec G \cdot W_{\fvec G'} \sin(\fvec g_{\fvec G'} \cdot \fvec x)} + c.c \right) \nonumber \\
      & = \frac1{A_{tot}} \int \rmd^2 \fvec x\ e^{i \fvec g_{\fvec G} \cdot \fvec x} \prod_{\fvec G' \in \mathbb{H}} \left( \sum_n J_n(2 \fvec G \cdot \fvec W_{\fvec G'}) e^{i n \fvec g_{\fvec G'} \cdot \fvec x}  \right) \  .
\end{align}
This gives the definition of $B(\fvec G,\ \{ \fvec W \})$, same as Eq.~\ref{Eqn:BDef}.

As explained in the main text, $\fvec W_{\fvec G_a}$ can be approximately obtained by solving the matrix equation $(\mM + \mT)_{a\mu, b\nu} \delta W_{\fvec G_b, \nu} = - u_{a\mu}$, where the $\mM$ matrix is given by Eq.~\ref{Eqn:MMat}. 
At large twist angle $\theta \gtrsim 1.4^{\circ}$, $\lambda \approx \zeta_1 \lesssim 0.1$, and thus we can neglect $\mT$ matrix. For other cases, especially when $\theta \lesssim 1^{\circ}$, $\lambda$ becomes sizable, and thus we achieve a higher accuracy when we do not neglect the $\mT$ matrix. We thus expand the inverse matrix using the power series in $\mT$,
\begin{align}
    (\mM + \mT)^{-1} = \mM^{-1} - \mM^{-1} T \mM^{-1} + \mM^{-1} \mT \mM^{-1} T \mM^{-1} + O(\mT^3)  \   .
\end{align}
where the obtained $\delta \fvec W$ at different truncation orders are illustrated in Fig~\ref{Fig:DWSeries} for $\theta = 1.05^{\circ}$ and Fig.~\ref{FigS:DWSeriesTwist14} for $\theta = 1.4^{\circ}$.

\begin{figure*}[t] 
	\centering
	\subfigure[\label{FigS:DWSeriesTwist14:G1xEps02}]{\includegraphics[width=0.48\columnwidth]{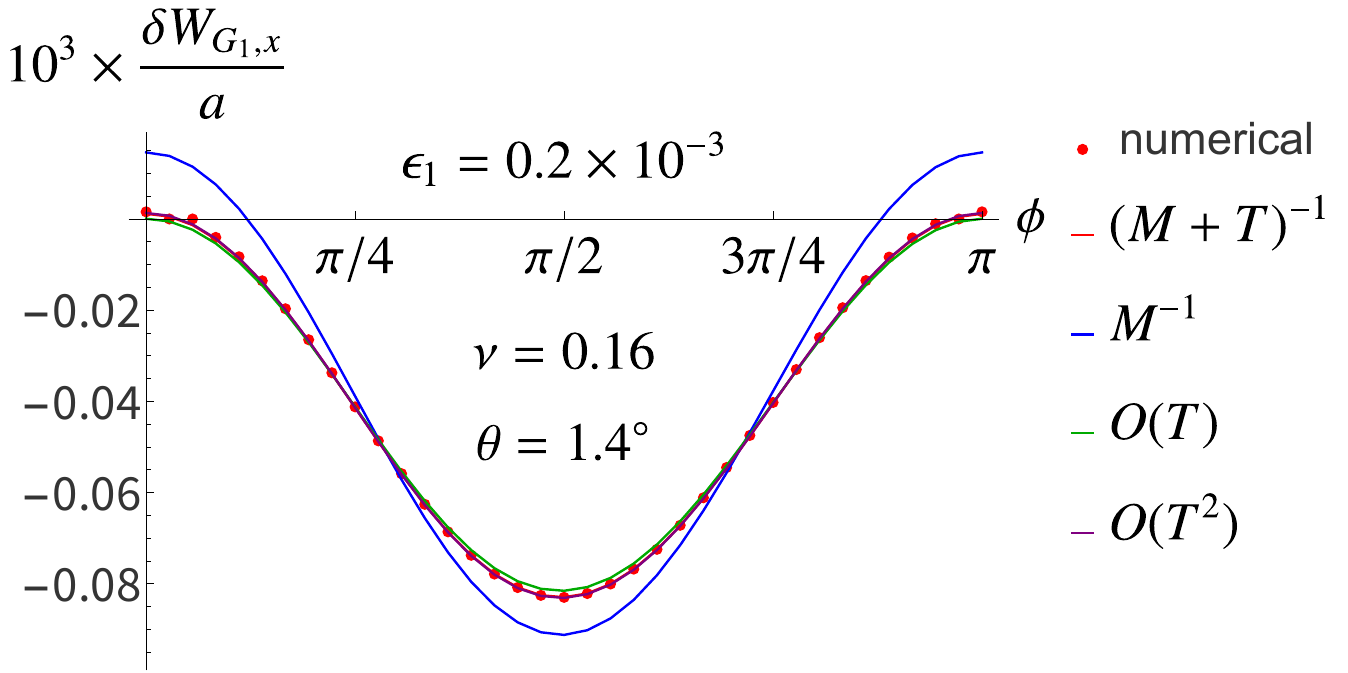}}
    \subfigure[\label{FigS:DWSeriesTwist14:G1yEps02}]{\includegraphics[width=0.48\columnwidth]{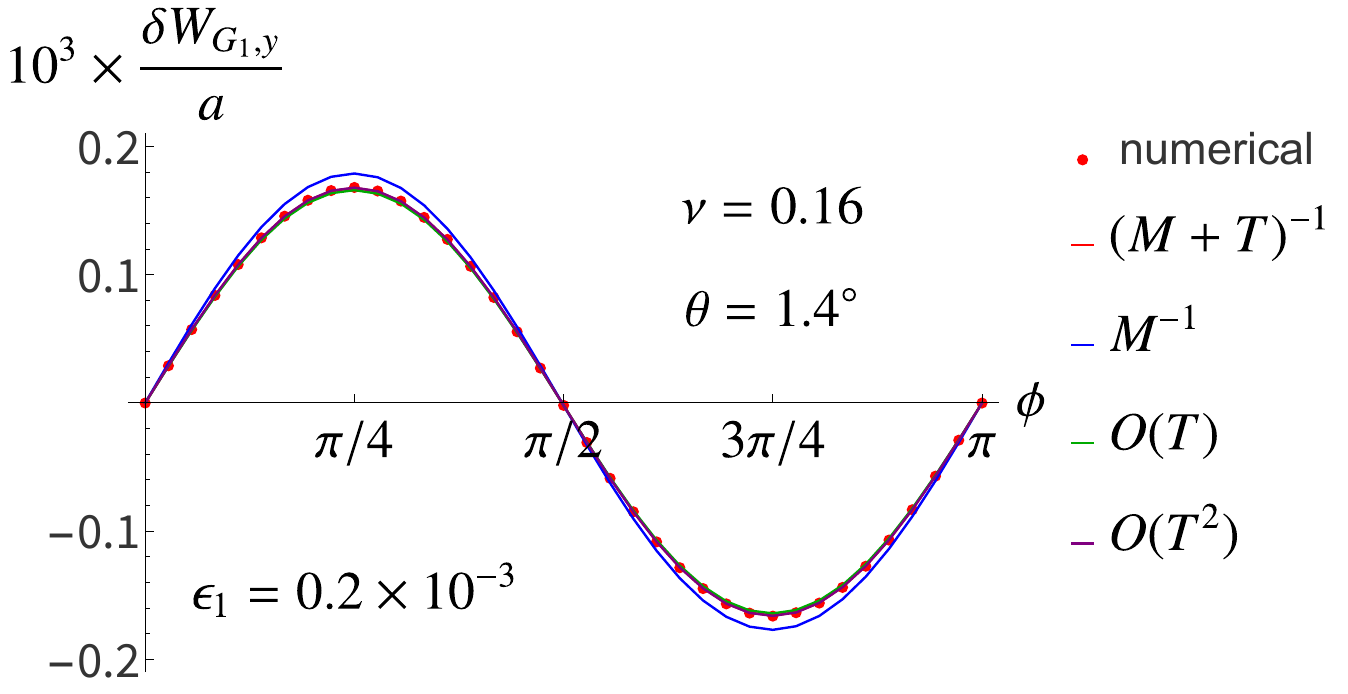}}
    \subfigure[\label{FigS:DWSeriesTwist14:G1xEps50}]{\includegraphics[width=0.48\columnwidth]{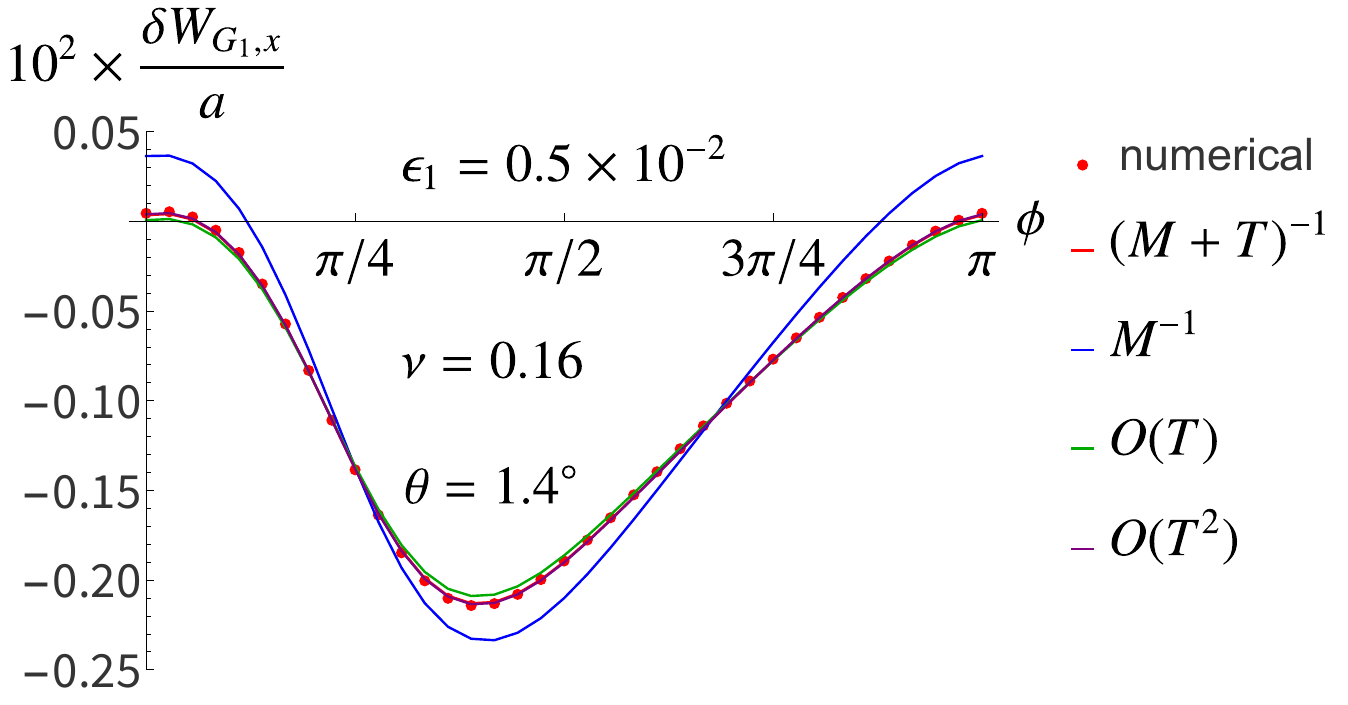}}
    \subfigure[\label{FigS:DWSeriesTwist14:G1yEps50}]{\includegraphics[width=0.48\columnwidth]{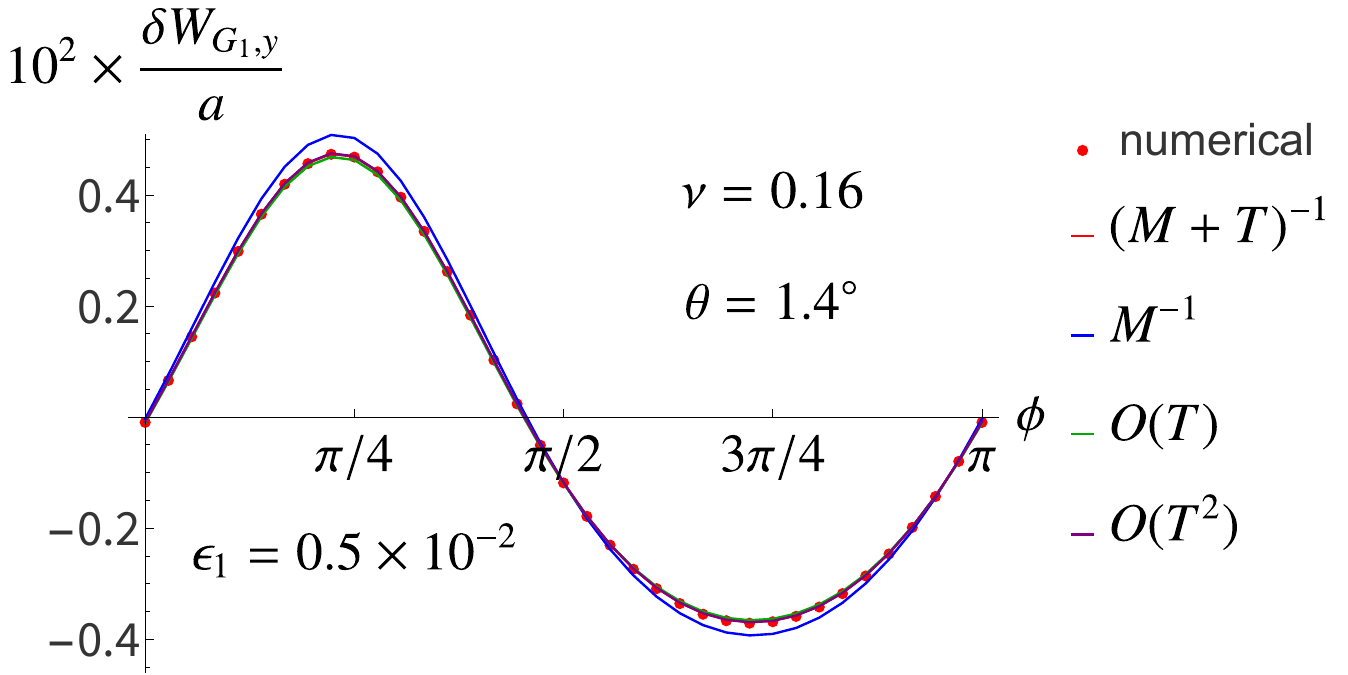}}
    \subfigure[\label{FigS:DWSeriesTwist14:G1xEps100}]{\includegraphics[width=0.48\columnwidth]{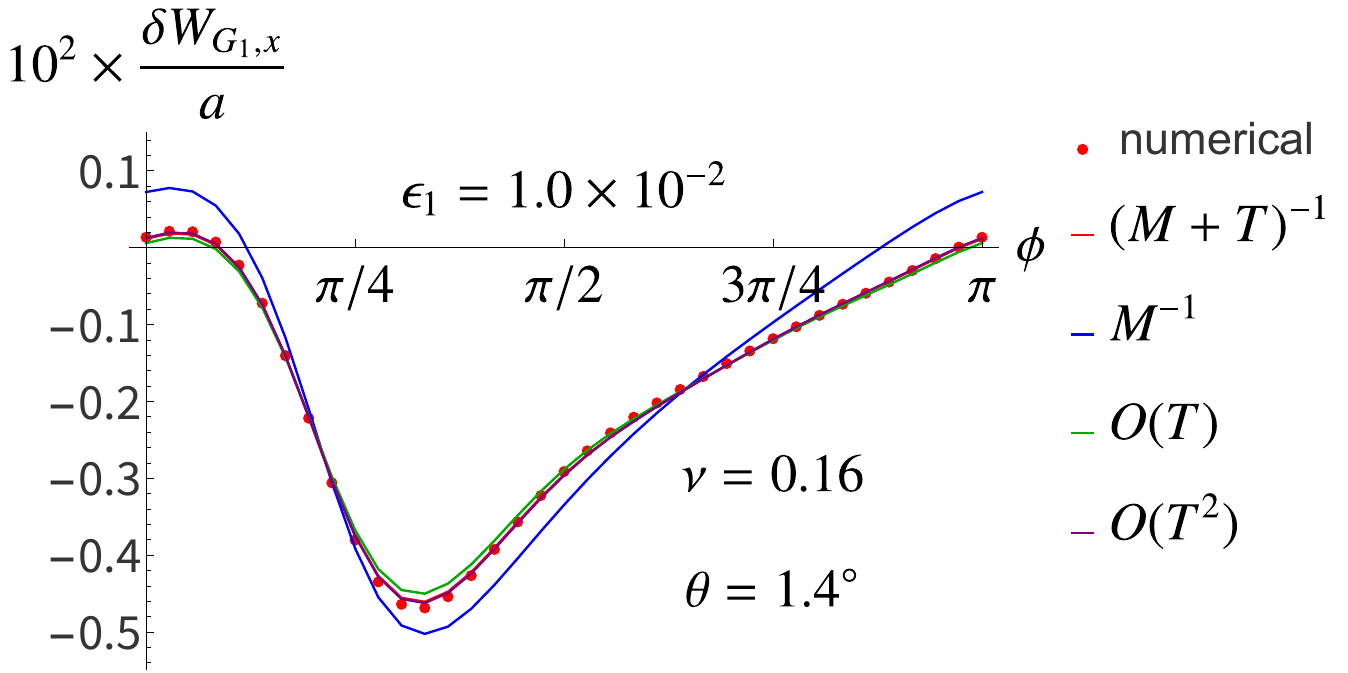}}
    \subfigure[\label{FigS:DWSeriesTwist14:G1yEps100}]{\includegraphics[width=0.48\columnwidth]{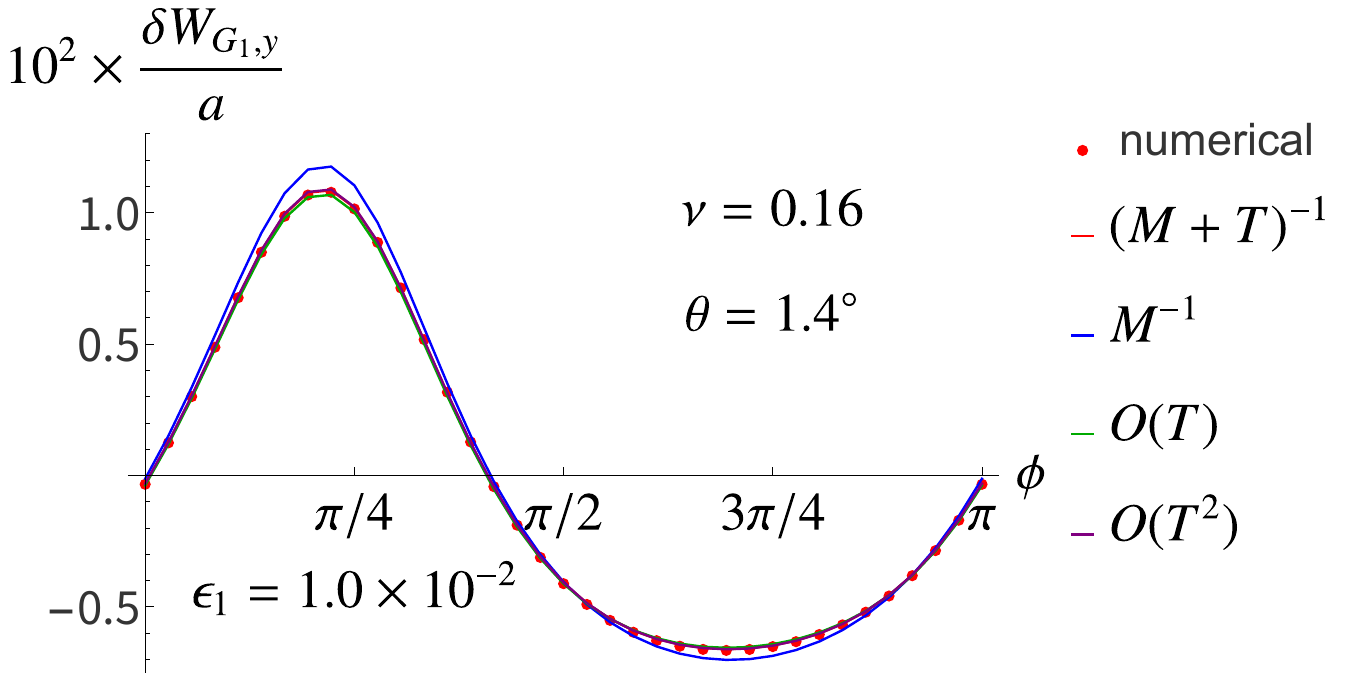}}
	\caption{Similar to Fig.~\ref{Fig:DWSeries}, we plot $\delta \fvec W_{\fvec G_1}$ as a function of $\phi$ for the twist angle $\theta = 1.4^{\circ}$. The truncation at $\mathcal{O}(\mT)$ is almost identical to $(\mM + \mT)^{-1}$. Compared with $\lambda \approx 0.256$ at $\theta = 1.05^{\circ}$ and the truncation order at $\mathcal{O}(\mT^2)$, smaller $\lambda \approx 0.1$ at $\theta =1.4^{\circ}$ leads to lower truncation order of the power series expansion in Eq.~\ref{Eqn:InverseMpT}.}
	\label{FigS:DWSeriesTwist14}
\end{figure*}


\subsection{Dominance of the first shell for the heterostrained lattice relaxation}
This subsection focuses on the estimates of $\delta \fvec W_{\fvec G}$ with small strain. We will argue that $\delta \fvec W_{\fvec G}$ for $\fvec G$ on the first shell is much larger than other $\delta \fvec W_{\fvec G}$s. For notational convenience, we introduce the variable, 
\begin{align}
    \epsilon_{max} & = \max(|\epsilon_1|,\ |\epsilon_2|) \quad \Longrightarrow \quad |\delta \fvec g| \sim \epsilon_{max} |\fvec G| 
\end{align}
When strain is small, $\epsilon_{max} \ll \theta$, and thus $|\delta \fvec g_{\fvec G}| \ll |\fvec g_{\fvec G}^{(0)}|$. Also, we generalize the variable $u_{a\mu}$ introduced in Eq.~\ref{Eqn:uForm} as
\begin{align}
    u_{\fvec G, \mu} & = \sum_{\nu} \left(  M_{\fvec G}^{(1)} + M_{\fvec G}^{(2)} \right)_{\mu\nu} W^{(0)}_{\fvec G, \nu} \  .
\end{align}
with the subscript $\fvec G$ for all reciprocal lattice vectors. Consequently, $\delta W$ can be written in the form of 
\begin{align}
    \delta W_{\fvec G, \mu} = - \sum_{\fvec G', \nu} \big( M + T \big)^{-1}_{\fvec G,\mu, \fvec G',\nu} u_{\fvec G',\nu} \ . \label{EqnS:DWEqn}
\end{align}

As the first step, we estimate the magnitude of each term in Eq.~\ref{EqnS:DWEqn}. From Eq.~\ref{Eqn:MMat} --\ref{Eqn:M2}, and note that $\mK \sim \mG$, it is found
\begin{align}
    \left| M^{(0)}_{\fvec G \mu, \fvec G \nu} \right| & \sim \mG |\fvec g_{\fvec G}^{(0)}|^2 \sim \mathcal{G} \theta^2 |\fvec G|^2 \ , \quad 
    \left| M^{(1)}_{\fvec G \mu, \fvec G \nu} \right| \sim \mathcal{G} \left| \delta \fvec g_{\fvec G} \right| \left| \fvec g^{(0)}_{\fvec G} \right| \sim \mathcal{G} \epsilon_{max} \theta |\fvec G|^2 \ , \quad \left| M^{(2)}_{\fvec G \mu, \fvec G \nu} \right| \sim \mathcal{G} \epsilon_{max}^2 |\fvec G|^2 \ . \label{EqnS:M012Appr} 
\end{align}
As a consequence, for small strain,
\begin{align}
    & \left| M^{(2)}_{\fvec G \mu, \fvec G \nu} \right| \ll \left| M^{(1)}_{\fvec G \mu, \fvec G \nu} \right| \ll \left| M^{(0)}_{\fvec G \mu, \fvec G \nu} \right| \quad  \Longrightarrow \quad \left| M_{\fvec G \mu, \fvec G \nu} \right| \sim \left| M^{(0)}_{\fvec G \mu, \fvec G \nu} \right| \sim \mathcal{G} \theta^2 |\fvec G|^2 \ .  \label{Eqn:MjHierarchy} 
\end{align}
On the other hand, the upper bound of the $T$ matrix elements can be estimated as
\begin{align}
    \left| T_{\fvec G \mu, \fvec G' \nu} \right| \lesssim c_1 |\fvec G_1|^2 = \lambda \mathcal{G} \theta^2 |\fvec G_1|^2 \ .  \label{Eqn:TAppr}
\end{align}
Clearly, when $\lambda \ll 1$, the above $T$ matrix elements are much smaller than the magnitude of the $M$ matrix elements, leading to the expansion of $(M + T)^{-1}$,
\begin{align}
    (M + T)^{-1} \sim M^{-1} - M^{-1} T M^{-1} + \mO(T^2)\ . \label{Eqn:InverseMpT2}
\end{align}

In addition, due to the hierarchy structure of $M^{(0, 1, 2)}$ in Eq.~\ref{Eqn:MjHierarchy}, for $\fvec G$ at the $j^{th}$ shell, 
\begin{align}
    & |u_{\fvec G\mu}| \sim \left| \big( M^{(1)}_{\fvec G} \big)_{\mu\nu} W^{(0)}_{\fvec G\nu} \right|  \sim \mathcal{G}\epsilon_{max} \theta |\fvec G|^2 \left| \fvec W^{(0)}_{\fvec G} \right|   \sim   \mathcal{G} \epsilon_{max} \theta |\zeta_j| |\fvec G|^3/|\fvec G_1|^2    \ ,  \label{Eqn:uGAppr}
\end{align} 
where Eq.~\ref{Eqn:W0} is used in the last step. 

As a result, $\{ \fvec u_{\fvec G} \}$ obtains the same hierarchy structure as $\zeta_j$ for the unstrained lattice relaxation. Especially, $|\fvec u_{\fvec G_a}|$ (for $a = 1$, $2$, and $3$) are much larger than other $\fvec u_{\fvec G}$ on outer $\fvec G$-shells. Thus, for $\fvec G$ on the first shell, 
\begin{align}
    |\delta W_{\fvec G_a \mu}| & \sim  \big( M^{-1} \big)_{\fvec G_a \mu, \fvec G_a \nu} | u_{\fvec G_a \nu} | \sim \frac{\epsilon_{max} \zeta_1}{\theta |\fvec G_1|} \  . \label{Eqn:W1InnerEstimate}
\end{align}
For $\fvec G$ on the $j^{th}$ $\fvec G$-shell with $j > 1$, using Eq.~\ref{Eqn:InverseMpT2}, we obtain 
\begin{align}
    \delta W_{\fvec G, \mu} & \approx - \big( M^{-1} \big)_{\fvec G\mu,  \fvec G \nu} u_{\fvec G \nu} +  \sum_{a = 1}^3 \sum_{\nu \rho \sigma}  \big( M^{-1} \big)_{\fvec G \mu, \fvec G\nu} T_{\fvec G \nu, \fvec G_a \rho} \big( M^{-1} \big)_{\fvec G_a \rho, \fvec G_a \sigma}  u_{\fvec G_a \sigma}  \ , \label{Eqn:W1OuterCount}
\end{align}
where $\nu$, $\rho$, and $\sigma$ are in-plane spatial indices. Combining Eq.~\ref{EqnS:M012Appr}, \ref{Eqn:TAppr}, and \ref{Eqn:uGAppr}, and noticing $|\fvec G| \sim |\fvec G_1|$ for $\fvec G$ at the first several momentum shells,  we conclude that
\begin{align}
    &  |\big( M^{-1} \big)_{\fvec G\mu,  \fvec G\nu} \fvec u_{\fvec G \nu}| \sim \frac{\epsilon_{max} \zeta_j}{\theta |\fvec G_1|}  \\
    &  |\big( M^{-1} \big)_{\fvec G \mu, \fvec G\nu} T_{\fvec G \nu, \fvec G'\rho} \big( M^{-1} \big)_{\fvec G' \rho, \fvec G'\sigma}  \fvec u_{\fvec G' \sigma} | \sim  \lambda \frac{\epsilon_{max} \zeta_1}{\theta |\fvec G_1|}  \quad \Longrightarrow \quad |\delta W_{\fvec G, \mu}| \sim \frac{\epsilon_{max}}{\theta |\fvec G_1|} \max\left( |\zeta_j|,\  \lambda \zeta_1 \right) \  . \label{Eqn:W1OuterEstimate}
\end{align}
Because of the hierarchy structure of $\zeta_j$ for the unstrained case, and $|\zeta_1| \sim \lambda  \ll 1$, $\delta \fvec W_{\fvec G_a}$ estimated in Eq.~\ref{Eqn:W1InnerEstimate} is found to be much larger than $\delta \fvec W$ of the outer shell estimated in Eq.~\ref{Eqn:W1OuterEstimate}, allowing us to keep only $\delta \fvec W_{\fvec G_a}$ (with $a = 1$, $2$ and $3$) and obtaining Eq.~\ref{Eqn:W1FiniteMatrixEqn}.


\section{Symmetry constraints}
Most of the symmetries of the unstrained moire system, including the six-fold rotation $C_{6z}$ along $\hat z$ axis, and mirror reflections along $xz$ and $yz$ planes, are broken by the presence of external heterostrain. Consequently, these transformations no longer impose constraints on the explicit form of the strained lattice relaxation, but interestingly, can still relate the lattice relaxations at different heterostrain orientations. Such relations play a crucial role in understanding the change of the electronic spectrum by strain, which we plan to address in upcoming papers. In this section, we will derive these relations.

As shown in Eq.~\ref{Eqn:StrainMat}, the heterostrain is described by three parameters: $\epsilon_1$, $\epsilon_2$, and $\phi$. Therefore, in this subsection, the strain matrix is labeled as $S^{\epsilon}(\epsilon_1, \epsilon_2, \phi)$, and correspondingly, the reciprocal lattice vectors are labeled as
\begin{align}
    g_{\fvec G, \mu}\left(\epsilon_1, \epsilon_2, \phi \right) = \left( S^{\epsilon}(\epsilon_1, \epsilon_2, \phi) + i \theta \sigma^2  \right)_{\mu\nu} G_{\nu}  \ . \label{EqnS:StrainedgG}
\end{align}
Similarly, we label the lattice relaxation as $\delta \fvec U(\fvec x; \epsilon_1, \epsilon_2, \phi)$. Since $R(\phi) = -R(\phi + \pi)$,  by the definition of $S^{\epsilon}$ in Eq.~\ref{Eqn:StrainMat}, $S^{\epsilon}(\epsilon_1, \epsilon_2, \phi) = S^{\epsilon}(\epsilon_1, \epsilon_2, \phi + \pi)$, implying that the strain matrix and thus the lattice relaxation $\delta \fvec U$ are periodic functions of $\phi$ with the same periods of $\pi$. The same is true for $\fvec W_{\fvec G}$, the Fourier components of $\delta \fvec U$
\begin{align}
    \fvec W_{\fvec G}(\epsilon_1, \epsilon_2, \phi) =  \fvec W_{\fvec G}(\epsilon_1, \epsilon_2, \phi + \pi) \ . \label{EqnS:C2WG}
\end{align}

In addition, by Eq.~\ref{Eqn:StrainMat},
\begin{align}
    S^{\epsilon}(\epsilon_1, \epsilon_2, \phi + \phi') = R^T(\phi') S^{\epsilon}(\epsilon_1, \epsilon_2, \phi) R(\phi') \ . 
\end{align}
Since $R(\phi) = \exp(-i \phi \sigma^2)$ and thus commutes with $\sigma^2$, Eq.~\ref{Eqn:gGMap} gives
\begin{align}
    & \qquad g_{\fvec G, \mu}(\epsilon_1, \epsilon_2, \phi + \phi') = \big( R^T(\phi') \big)_{\mu\nu} g_{R(\phi') \fvec G, \nu}\left( \epsilon_1, \epsilon_2, \phi \right)  \ . 
\end{align}
The set of $\fvec G$ vectors, which is unaffected by the heterostrain, is invariant under $C_{6z} = R(\pi/3)$. By setting $\phi' = \pi/3$ in the above formula, for two heterostrains $S^{\epsilon}(\epsilon_1, \epsilon_2, \phi)$ and $S^{\epsilon}(\epsilon_1, \epsilon_2, \phi + \pi/3)$, we conclude that the two sets of the corresponding moire reciprocal lattice vectors are related by $C_{6z}$ rotation. Since $\fvec G' \cdot \fvec U^-(\fvec x) = \fvec g_{\fvec G'} \cdot \fvec x + \fvec G \cdot \delta \fvec U(\fvec x)$,
\begin{align}
    & \fvec G \cdot \fvec U^-(\fvec x; \epsilon_1, \epsilon_2, \phi + \frac{\pi}3)  =  \fvec g_{\fvec G}(\epsilon_1, \epsilon_2, \phi + \frac{\pi}3) \cdot x + \fvec G \cdot \delta \fvec U(\fvec x; \epsilon_1, \epsilon_2, \phi + \frac{\pi}3) \nonumber \\
     = & \fvec g_{C_{6z}\fvec G}(\epsilon_1, \epsilon_2, \phi) \cdot (C_{6z} \fvec x)  + (C_{6z} \fvec G) \cdot C_{6z} \delta \fvec U(\fvec x; \epsilon_1, \epsilon_2, \phi + \frac{\pi}3)  \ . 
\end{align}
Since both the elastic energy $U_E$ and the interlayer adhesion potential $U_B$ are invariant under $C_{6z}$, we conclude
\begin{align}
    & \quad C_{6z} \delta \fvec U(\fvec x; \epsilon_1, \epsilon_2, \phi + \frac{\pi}3) =  \delta \fvec U(C_{6z}\fvec x; \epsilon_1, \epsilon_2, \phi) \nonumber \\
     \Longrightarrow \quad & \delta U_{\mu}\left(\fvec x; \epsilon_1, \epsilon_2, \phi + \frac{\pi}3 \right)  = \left( R\left( \frac{\pi}3 \right) \right)_{\nu\mu} \delta U_{\nu}\left( R\left( \frac{\pi}3 \right) \fvec x; \epsilon_1, \epsilon_2, \phi \right), \label{Eqn:C6DU}
\end{align}
leading to their corresponding Fourier components related by 
\begin{align}
    & W_{\fvec G, \mu}\left( \epsilon_1, \epsilon_2, \phi + \frac{\pi}3 \right)  =  \left( R\left( \frac{\pi}3 \right) \right)_{\nu\mu} W_{C_{6z}\fvec G, \nu}(\epsilon_1, \epsilon_2, \phi) \  . \label{Eqn:C6W}
\end{align}
In particular, for $\fvec G$ at the innermost shell, we obtain
\begin{align}
     & W_{\fvec G_2, \mu}(\epsilon_1, \epsilon_2, \phi) = \left( R\left( \frac{2\pi}3 \right) \right)_{\mu\nu}  W_{\fvec G_1, \nu}\left( \epsilon_1, \epsilon_2, \phi + \frac{2\pi}3 \right),\label{EqnS:C6WG2} \\
     & W_{\fvec G_3, \mu}(\epsilon_1, \epsilon_2, \phi) = - \left( R\left( \frac{\pi}3 \right) \right)_{\mu\nu}  W_{\fvec G_1, \nu}\left( \epsilon_1, \epsilon_2, \phi + \frac{\pi}3 \right).\label{EqnS:C6WG3}
\end{align}
Combining Eq.~\ref{EqnS:C2WG}, \ref{EqnS:C6WG2}, and \ref{EqnS:C6WG3}, we conclude that $\fvec W_{\fvec G_1}(\epsilon_1, \epsilon_2, \phi)$ by varying $\phi$ from $0$ to $\pi$ already provides $\fvec W_{\fvec G_a}(\epsilon_1, \epsilon_2, \psi)$ ($a = 1$, $2$, and $3$) with arbitrary angle $\psi$.  

In addition, the strain matrix also satisfies the relation
\begin{align}
    & S^{\epsilon}\left( \epsilon_1, \epsilon_2, \frac{\pi}2 - \phi \right)  = \sigma^3 S^{\epsilon}(\epsilon_2, \epsilon_1, \phi) \sigma^3 = - \sigma^3 S^{\epsilon}(-\epsilon_2, -\epsilon_1, \phi) \sigma^3, \\
   \Longrightarrow \quad & g_{\fvec G, \mu}\left( \epsilon_1, \epsilon_2, \frac{\pi}2 - \phi \right) =  - \sigma^3_{\mu\nu} g_{\mathcal{R}_y\fvec G, \nu}(-\epsilon_2, -\epsilon_1, \phi) = \sigma^3_{\mu\nu} g_{-\mathcal{R}_y\fvec G, \nu}(-\epsilon_2, -\epsilon_1, \phi),  \label{Eqn:gMoireC2x}
\end{align}
where $\mathcal{R}_y$ is the mirror reflection along $xz$ plane, that transforms an arbitrary in-plane vector $\fvec v$ to $(\mathcal{R}_y \fvec v)_{\mu} = \sigma^3_{\mu\nu} v_{\nu}$. As shown in Fig.~\ref{Fig:Schematic}, the set of all $\fvec G$ vectors is invariant under both $\mathcal{R}_y$ and $C_{2z}\mR_y = -\mathcal{R}_y$. Thus, for two heterostrains described by $S^{\epsilon}(\epsilon_1, \epsilon_2, \frac{\pi}2 - \phi)$ and $S^{\epsilon}(-\epsilon_2, -\epsilon_1, \phi)$, Eq.~\ref{Eqn:gMoireC2x} demonstrates that the two sets of the corresponding moire reciprocal lattice vectors are related by the mirror reflection $\mR_y$. Notice that
\begin{align}
    & \quad \fvec g_{\fvec G}( \epsilon_1, \epsilon_2, \frac{\pi}2 - \phi) + \fvec G \cdot \delta \fvec U(\fvec x; \epsilon_1, \epsilon_2, \frac{\pi}2 - \phi)  = \fvec g_{\mR_y \fvec G}(- \epsilon_2, -\epsilon_1, \phi) \cdot (- \mR_y \fvec x) + (\mR_y \fvec G) \cdot \mR_y \delta \fvec U(\fvec x; \epsilon_1, \epsilon_2, \frac{\pi}2 - \phi).
\end{align}
Since both $U_E$ and $U_B$ are invariant under $\mR_y$, we conclude
\begin{align}
    & \mR_y \delta \fvec U(\fvec x; \epsilon_1, \epsilon_2, \frac{\pi}2 - \phi) = \delta\fvec U(-\mR_y \fvec x; -\epsilon_2, - \epsilon_1, \phi) \nonumber \\
    & \Longrightarrow \delta U_{\mu}\left(\fvec x; \epsilon_1, \epsilon_2, \frac{\pi}2 - \phi \right) = \sigma^3_{\mu\nu} \delta U_{\nu}(-\mathcal{R}_y \fvec x; -\epsilon_2, - \epsilon_1, \phi)  \label{Eqn:RyDU}
\end{align}
and the corresponding Fourier components satisfy
\begin{align}
    W_{\fvec G, \mu}\left( \epsilon_1, \epsilon_2, \frac{\pi}2 - \phi \right) = \sigma^3_{\mu\nu} W_{\mathcal{R}_y \fvec G, \nu}(-\epsilon_2, - \epsilon_1, \phi)  \ . \label{Eqn:RyW} 
\end{align}
Although Eqs.~\ref{Eqn:C6DU}, \ref{Eqn:C6W}, \ref{Eqn:RyDU}, \ref{Eqn:RyW} are derived with external heterostrain, they are also valid in the absence of heterostrain by setting $\epsilon_1 = \epsilon_2 = 0$. The obtained formulas give the constraints on $\delta \fvec U$ and the corresponding Fourier components $\fvec W$ for the unstrained system, as have been derived in Ref.~\cite{JKPRB23}.


\end{widetext}

\end{document}